\documentclass[format=acmsmall, review=false, screen=true]{acmart}

\usepackage{booktabs} 

\usepackage[linesnumbered,ruled]{algorithm2e} 

\SetAlFnt{\small}
\SetAlCapFnt{\small}
\SetAlCapNameFnt{\small}
\SetAlCapHSkip{0pt}
\IncMargin{-\parindent}

\setcopyright{acmcopyright}
\acmJournal{POMACS}
\acmYear{2018}
\acmVolume{2} \acmNumber{1} \acmArticle{9} \acmMonth{3} \acmPrice{$15.00$}
\acmDOI{10.1145/3179412}

\usepackage[normalem]{ulem}
\usepackage{graphicx}
\usepackage{amsmath}
\usepackage{soul}
\usepackage{subfig}
\usepackage{relsize}
\usepackage{multirow}
\usepackage{hhline}
\usepackage{dirtytalk}
\usepackage{xcolor}
\usepackage{soul}
\usepackage{longtable}
\usepackage{tabularx}
\usepackage{afterpage}
\usepackage{rotating}
\usepackage{perpage}
\usepackage[titletoc,toc,title]{appendix}

\newcommand{\techname}{ECI-Cache}

\received{January 2017}
\received[revised]{March 2017}
\received[accepted]{December 2017}

\begin{document}
\title{\techname{}: A High-\underline{E}ndurance and \underline{C}ost-Efficient\\\underline{I}/O Caching Scheme for Virtualized Platforms}

\author{Saba Ahmadian}
\orcid{1234-5678-9012-3456}
\affiliation{%
  \institution{Sharif Univerisity of Technology}
  \streetaddress{Azadi}
  \city{Tehran}
  \postcode{11155-11365}
  \country{Iran}}
\email{ahmadian@ce.sharif.edu}

\author{Onur Mutlu}
\orcid{1234-5678-9012-3456}
\affiliation{%
	\institution{ETH Z{\"u}rich}
	\country{Switzerland}
	}
\email{omutlu@ethz.ch}

\author{Hossein Asadi}
\orcid{1234-5678-9012-3456}
\authornote{Corresponding Author}
\affiliation{%
	\institution{Sharif Univerisity of Technology}
	\streetaddress{Azadi}
	\city{Tehran}
	\postcode{11155-11365}
	\country{Iran}}
\email{asadi@sharif.edu}

\begin{abstract}
\sloppypar
In recent years, high interest in using \emph{Virtual Machines} (VMs) in data centers and Cloud computing has significantly increased the demand for high-performance data storage systems. A straightforward approach to provide a high performance storage system is using \emph{Solid-State Drives} (SSDs). Inclusion of SSDs in storage systems, however, imposes significantly higher cost compared to \emph{Hard Disk Drives} (HDDs).
Recent studies suggest using SSDs as a caching layer for HDD-based storage subsystems in virtualization platforms.
Such studies neglect to address the endurance and cost of SSDs, which can significantly affect the efficiency of I/O caching. 
Moreover, previous studies  \emph{only} configure the cache size to provide the required performance level for each VM, while neglecting other important parameters such as cache \emph{write policy} and \emph{request type}, which can adversely affect both performance-per-cost and endurance. 

\sloppypar
In this paper, we present a new \emph{high-\uline{E}ndurance and \uline{C}ost-efficient \uline{I}/O Caching} (\techname{}) scheme for virtualized platforms, which can significantly improve both the \emph{performance-per-cost} and \emph{endurance} of storage subsystems  as opposed to previously proposed I/O caching schemes. 
Unlike traditional I/O caching schemes which allocate cache size \emph{only} based on \emph{reuse distance} of accesses, we propose a new metric, \emph{Useful Reuse Distance} (URD), which considers the \emph{request type} in reuse distance calculation, resulting in improved performance-per-cost and endurance for the SSD cache.
Via online characterization of workloads and using URD, \techname{} partitions the SSD cache across VMs and is able to dynamically adjust the cache size and write policy for each VM. To evaluate the proposed scheme, we have implemented \techname{} in an open source hypervisor, QEMU (version 2.8.0), on a server running the CentOS 7 operating system (kernel version 3.10.0-327). Experimental results show that our proposed scheme improves the performance, performance-per-cost, and endurance of the SSD cache by 17\%, 30\% and 65\%, respectively, compared to the state-of-the-art \emph{dynamic} cache partitioning scheme.
\end{abstract}

\maketitle

\renewcommand{\shortauthors}{S. Ahmadian et al.}

\section{Introduction}
\label{sec:introduction}
\begin{figure}[!h]
	\centering
	\includegraphics[scale=0.7]{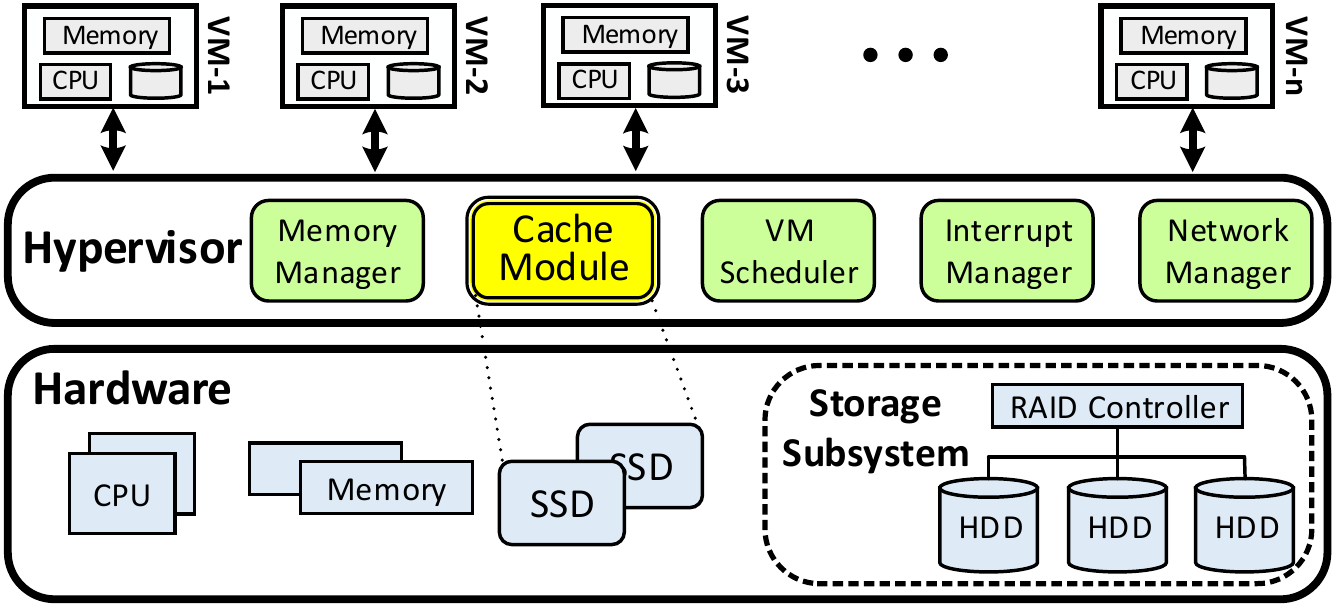}
	\caption{Example state-of-the-art virtualization platform.}
	\label{fig:vir}
\end{figure}

\sloppy
Virtualization is widely used in data centers and Cloud computing in order to improve the utilization of high-performance servers {\cite{virtualization}}.
Integrating various \emph{Virtual Machines} (VMs) running with different operating systems on a server provides more flexibility and higher resource utilization while delivering the desired performance for each VM. In addition, virtualization provides system isolation where each VM has access only to its own resources. In a virtualized platform, shown in Fig. \ref{fig:vir}, the resource allocation of each VM is managed by a hypervisor. By employing various modules such as a VM scheduler and a memory and network manager, the hypervisor orchestrates the sharing of resources between VMs according to their demand, in order to maximize the overall performance provided by the server (and this maximizes performance-per-cost by enabling the use of a smaller number of physical servers than one for each VM) \cite{uhlig2005intel,tickoo2010modeling}.
\sloppy
With increasing performance requirements of data-intensive applications in data centers, storage subsystems have become performance bottlenecks of computing systems. \emph{Hard Disk Drives} (HDDs), which are used as main media for data storage in storage systems, provide large capacity and low cost, but they suffer from low performance, particularly for random access workloads. The low performance of HDD-based storage systems can be avoided by employing high-performance storage devices such as \emph{Solid-State Drives} (SSDs). Compared to HDDs, SSDs provide higher performance due to their non-mechanical structure used to retrieve and store data. SSDs, however, impose up to 10X higher cost and support only a limited number of reliable writes \cite{cai2017error}, which makes the replacement of all HDDs by SSDs usually prohibitively expensive \cite{migratetossd,agrawal2008design,salkhordeh2015operating,ahmadian-ssd-rel-date}. 

\sloppy
In order to take advantage of the merits of both HDDs and SSDs, several studies from EMC$^2$, Facebook, FusionIO, Mercury, and VMware \cite{migratetossd,matthews2008intel,appuswamy2012integrating,kim2011hybridstore,nitro,azor,hystor,flashcache,mercury,fusionio,vfrm,emcvfcache,reca,meza2015large} employ high-performance SSDs as a caching layer for high-capacity HDDs in storage systems (as shown in Fig. \ref{fig:vir}). 
Applying such I/O caching on virtualization platforms requires a proper cache management scheme in order to achieve higher I/O performance.
In previous studies, such caches have been used as either a shared global cache \cite{emcvfcache, fusionio, vfrm} or a cache statically partitioned across VMs \cite{mercury}. 
The former scheme fails to provide a guaranteed minimum cache space for each VM. This is due to the fact that the entire cache is shared between all VMs and each VM can potentially use up an unbounded amount of the entire cache space, affecting the performance of the other VMs. The latter scheme \emph{statically} partitions SSD space between VMs where the partitioning is performed \emph{independently of} the characteristics of the workloads running on VMs. This scheme has two major shortcomings. 
First, the allocated cache space could be underutilized by the VM if there is low locality of reference in the workload access pattern. Second, since cache space partitioning and allocation are done statically offline, cache allocation for a new VM during runtime is \emph{not} practical using this scheme, which makes the scheme inflexible and the system underutilized. 
To alleviate the shortcomings of the two aforementioned schemes, \emph{partitioned I/O caching} has been proposed. Variants of this technique dynamically estimate and allocate cache space for each VM by estimating an efficient cache size for each VM \cite{scave, vcacheshare,centaur}. 
They do so by calculating the reuse distance of the workload, i.e., the maximum distance between two accesses to an identical address \cite{parda,shards,centaur} .
Unfortunately, such I/O caching schemes only focus on estimating cache size for VMs and neglect other key parameters, such as \emph{write policy} (i.e., how write requests are handled by the cache), \emph{request type} (i.e., read or write requests), and their corresponding impact on the workload \emph{reuse distance}, 
which greatly affects the performance-per-cost and endurance  of the SSD cache, as we show in this work. 


In this paper, we propose a new \emph{high-\uline{E}ndurance and \uline{C}ost-efficient \uline{I}/O caching} (\techname{}) scheme which can be used for virtualized platforms in large-scale data centers.  
\techname{} aims to improve both the performance-per-cost and endurance of the SSD cache by dynamically configuring 1) an efficient cache size to maximize the performance of the VMs and 2) an effective write policy that improves the endurance and performance-per-cost of each VM.
To this end, we propose a metric called  \emph{Useful Reuse Distance} (URD), which minimizes the cache space to allocate for each VM while maintaining the performance of the VM.
The main objective of URD is to reduce the allocated cache space for each VM. The reduced cache space is obtained by computing workloads reuse distance based on request type, \emph{without} considering unnecessary write accesses (i.e., writes to a block without any further read access).
Employing URD in our proposed I/O caching scheme maximizes the performance-per-cost and also enhances the endurance of the SSD cache by allocating much smaller cache space compared to state-of-the-art cache partitioning schemes. 
We also propose a detailed analysis of the effect of write policy on the performance and endurance of an SSD cache, clearly demonstrating the negative impact of having the same write policy for VMs with \emph{different} access patterns (as used in previous studies) on the IO performance and SSD endurance. 
To achieve a sufficiently high hit ratio, \techname{} \emph{dynamically} partitions the cache across VMs.

In the proposed scheme, we mainly focus on two approaches: 1) URD based per-VM cache size estimation and 2) per-VM effective write policy assignment, via online monitoring and analysis of IO requests for each VM.
In the first approach, we allocate much smaller cache space compared to previous studies for each VM, which results in improved performance-per-cost. In the second approach, we assign an effective write policy for each VM in order to improve the endurance of the I/O cache while minimizing the negative impact on performance. The integration of these two approaches enables ECI-Cache to partition and manage the SSD cache between VMs more effectively than prior mechanisms \cite{scave,vcacheshare,centaur}.

We have implemented \techname{} on QEMU (version 2.8.0) \cite{qemu}, an open source hypervisor  (on the CentOS 7 operating system, kernel version 3.10.0-327). We evaluate our scheme on an HP ProLiant DL380 Generation 5 (G5) server \cite{hp_g5} with four 146GB SAS 10K HP HDDs \cite{hp_hdd} (in RAID-5 configuration), a 128GB Samsung 850 Pro SSD \cite{samsung_ssd}, 16GB DDR2 memory, and 8 x 1.6GHz Intel(R) Xeon CPUs. We run more than fifteen workloads from the SNIA MSR traces \cite{msr} on VMs.
Experimental results show that \techname{} 1) improves performance by 17\% and performance-per-cost by 30\% compared to the state-of-the-art cache partitioning scheme \cite{centaur}, 2) reduces the number of writes committed to the SSD by 65\% compared to \cite{centaur}, thereby greatly improving the SSD lifetime. 

To our knowledge, we make the following contributions.
\begin{itemize}

	\item This paper is the first to differentiate the concept of reuse distance based on the \emph{type of each request}. We propose a new metric, \emph{Useful Reuse Distance} (URD), whose goal is to reduce the cost of the SSD cache by allocating a smaller cache size for each VM.
	
	\item By conducting extensive workload analyses, we demonstrate the importance of dynamically adjusting the cache \emph{write policy} on a per-VM basis, which no previous I/O caching policy explicitly takes into account.
	We use these analyses to develop a mechanism that can efficiently adjust both cache size and write policy on a per-VM basis.
	
	\item We propose \techname{}, which consists of two key novel components: 1) dynamic per-VM cache size estimation, and cache partitioning using the URD metric and 2) per-VM write policy to improve both system performance-per-cost and SSD cache lifetime.
	
	\item We implement \techname{} in QEMU, an open source hypervisor. Our extensive evaluations of \techname{} on a large number of diverse workloads show that \techname{}  significantly improves performance-per-cost over the best previous dynamic cache partitioning policy and reduces the number of writes to the SSD.
	
\end{itemize}
The rest of the paper is organized as follows. Sec. \ref{sec:related} discusses related work. In Sec. \ref{sec:example}, we provide an illustrative example and motivation. In Sec. \ref{sec:URD}, we propose the metric of URD. In Sec. \ref{sec:arch}, we present our proposed technique. Sec. \ref{sec:results} provides experimental setup and results. Finally, Sec. \ref{sec:conclusion} concludes the paper.

\section{Related Work}
\label{sec:related}

Previous studies on I/O caching in virtualization platforms investigate 1) the location of the cache or 2) the cache partitioning policy. The former set of works explores caching techniques based on where the I/O cache resides, whereas the latter examines mechanisms for sharing and partitioning of the cache across VMs. Based on the location of the SSD cache, three main alternatives for I/O caching in virtualization platforms have been introduced, as shown in Fig. \ref{fig:alternatives_caching}. 
We next describe possible schemes for I/O caching and discuss their advantages and shortcomings.
\subsection{VM-based I/O Caching}
In \emph{VM-based I/O caching} (Fig. \ref{fig:vmbased}), each VM in a virtualized platform has full control on a portion of the SSD cache. In this scheme, separate SSD slots are allocated for each VM and the cache management is conducted by VMs. 
Cache size adjustment, cache partitioning, and sharing techniques cannot be applied in this scheme. In order to employ a \emph{VM-based I/O caching} scheme in a virtualized platform, caching schemes presented in \cite{migratetossd,nitro,hystor,reca} can be applied on VMs and improve the IO performance of individual VMs, which is likely not efficient in virtualized platforms.
\subsection{Storage system-based I/O Caching}
In \emph{Storage system-based I/O caching} (Fig. \ref{fig:storagebased}), VMs and the hypervisor have no knowledge of the SSD cache, which
prohibits the advantages achieved by cache partitioning schemes.
Similar to \emph{VM-based I/O caching}, previous caching techniques, such as \cite{migratetossd,nitro,hystor,reca}, can be employed in storage systems, but such techniques cannot be managed in a virtualized platform.
\subsection{Hypervisor-based I/O Caching}
\begin{figure}[!t]
	\centering
	\subfloat[{VM-based}]{\includegraphics[width=.3\textwidth]{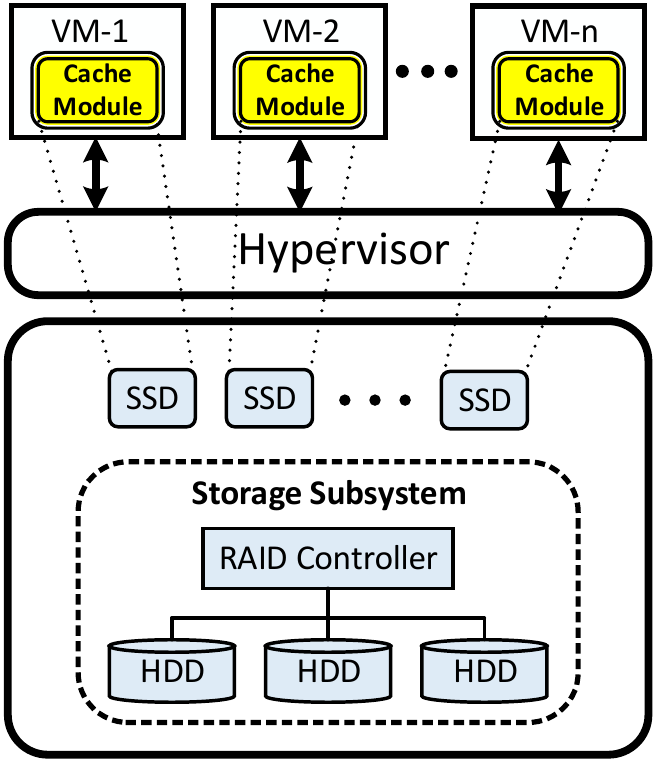}%
		\label{fig:vmbased}}
	\hfil
	\hspace{-.8pt}
	\subfloat[{Storage-based}]{\includegraphics[width=.3\textwidth]{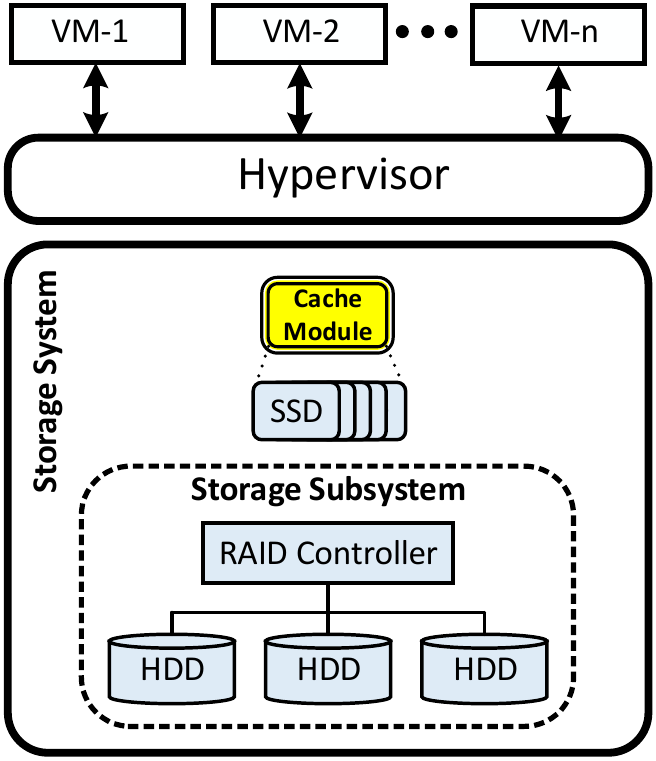}%
		\label{fig:storagebased}}
	\hfil
	\hspace{-.8pt}
	\subfloat[{Hypervisor-based}]{\includegraphics[width=.3\textwidth]{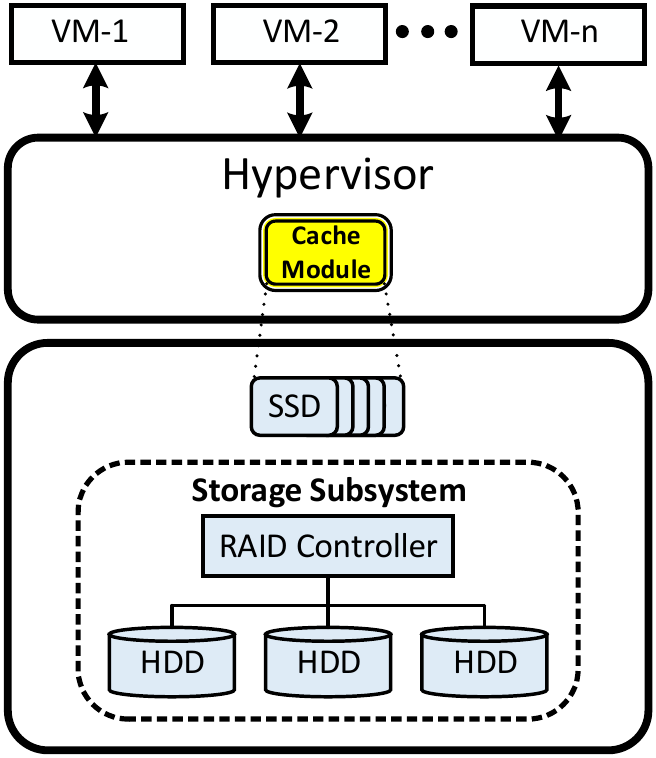}%
		\label{fig:hypervisorbased}}
	
	\caption{I/O caching using SSDs in virtualized platforms.}
	\label{fig:alternatives_caching}
\end{figure}

In \emph{Hypervisor-based I/O caching} (Fig. \ref{fig:hypervisorbased}), cache management is done by the hypervisor. Since the hypervisor has full knowledge about workloads running on VMs, it can perform efficient cache management and cache space partitioning across VMs. This type of I/O caching scheme has been proposed frequently in previous studies \cite{scave,vcacheshare,centaur}. These works mainly focus on cache management, sharing, and partitioning across VMs. From the partitioning perspective, \emph{Hypervisor-based I/O caching} schemes can be divided into two groups: global/static and dynamic cache partitioning. We next describe state-of-the-art \emph{hypervisor-based I/O caching} schemes. 

\subsubsection{Global Caching and Static Cache Partitioning}
Examples of global caching or static cache partitioning schemes include the EMC VFCache \cite{emcvfcache}, NetApp Mercury \cite{mercury}, Fusion-io ioTurbine \cite{fusionio}, and vFRM \cite{vfrm}. In global caching, each VM can use up the \emph{entire} SSD cache, thereby potentially adversely affecting the performance of the other VMs \cite{scave}. 
In static cache partitioning, SSD cache space is \emph{equally} partitioned across VMs based on the number of VMs in the platform, without
taking into account the data access and reuse patterns of each VM. Static cache partitioning is also unable to allocate cache space for newly-added VMs during online operation.

\subsubsection{Dynamic Cache Partitioning}
Dynamic cache partitioning schemes alleviate the shortcoming of global and static cache partitioning \cite{cloudcache,scave,vcacheshare,centaur,argon,janus}.
These techniques partition the SSD cache across VMs based on the cache space demand of each VM, and they are aware of data access and reuse pattern of VMs.
Argon \cite{argon} presents a storage server that partitions the memory cache across services based on their access patterns. This scheme allocates minimum cache space for each service to achieve a predefined fraction of hit ratio namely \emph{R-Value}. To estimate the required cache size for each service, Argon employs an online cache simulation mechanism and finds the fraction of accesses that are served by cache (namely I/O absorption ratio). 
Janus \cite{janus} partitions the flash tier between workloads at the filesystem level. Janus maximizes the total read accesses served from the flash tier by allocating the required space for each workload. The required space is estimated based on the ratio of read operations of each workload.
S-CAVE \cite{scave} is a hypervisor-based online I/O caching scheme that allocates cache space for each VM by dynamically estimating the working set size of workloads running on VMs. To minimize the possibility of data loss, S-CAVE uses the \emph{Write-Through} (WT) policy in cache configuration.
vCacheShare \cite{vcacheshare} is another hypervisor-based I/O caching scheme that dynamically partitions the SSD cache space across VMs. This scheme considers locality and reuse intensity (i.e., burstiness of cache hits) in order to estimate the required cache size for each VM.
vCacheShare reduces the number of writes in the SSD cache by using the \emph{Write Around} policy where write operations are \emph{always} directed to the storage subsystem. Such scheme shows improved performance only for read operations while it has no performance improvement for write operations.
Centaur \cite{centaur}, another online partitioning scheme for virtualized platforms, aims to maximize the IO performance of each VM as well as meeting QoS targets in the system. It employs \emph{Miss Ratio Curves} (MRCs) to estimate an efficient cache space allocation for each VM.
Centaur does \emph{not} consider the negative impact of write operations on SSD cache lifetime in 1) cache size estimation, or 2) write policy assignment.
Centaur employs the \emph{Write-Back} (WB) policy to maximize the IO performance without considering the impact of the WB policy on the number of writes into the SSD.  
CloudCache \cite{cloudcache} estimates each VM's cache size by considering \emph{Reuse Working Set Size} (RWSS), which captures temporal locality and also reduces the number of writes into the SSD cache. In addition, this scheme employs a VM migration mechanism to handle the performance demands of VMs in the Cloud.

{To summarize, among previous studies, S-CAVE {\cite{scave}}, vCacheShare {\cite{vcacheshare}}, Centaur {\cite{centaur}}, and CloudCache {\cite{cloudcache}} are the closest to our proposal. However, they only consider cache space partitioning and do not consider adaptive write policies. The cache size estimation scheme presented in S-CAVE, which is based on working set size estimation fails in cache size estimation for workloads with sequential access patterns and has become deprecated, as shown in {\cite{vcacheshare}}.
	vCacheShare and CloudCache perform cache size estimation based on reuse intensity. Such cache allocation schemes are based on assumptions that cannot be applied to the I/O cache in the storage subsystem, as demonstrated in {\cite{centaur}}.
	Reuse intensity based schemes are \emph{only} effective for workloads that are aligned with their size estimation schemes and would not be accurate compared to reuse distance based schemes such as {\cite{centaur}}.
	The state-of-the-art scheme is Centaur, which works based on MRCs and reuse distance analysis. This scheme does \emph{not} consider 1) the impact of request type on reuse distance calculation and 2) the impact of write policy on either endurance or performance.}

\section{Motivation and Illustrative Example}
\label{sec:example}
\begin{figure}[!h]
	\centering
	\includegraphics[scale=0.4]{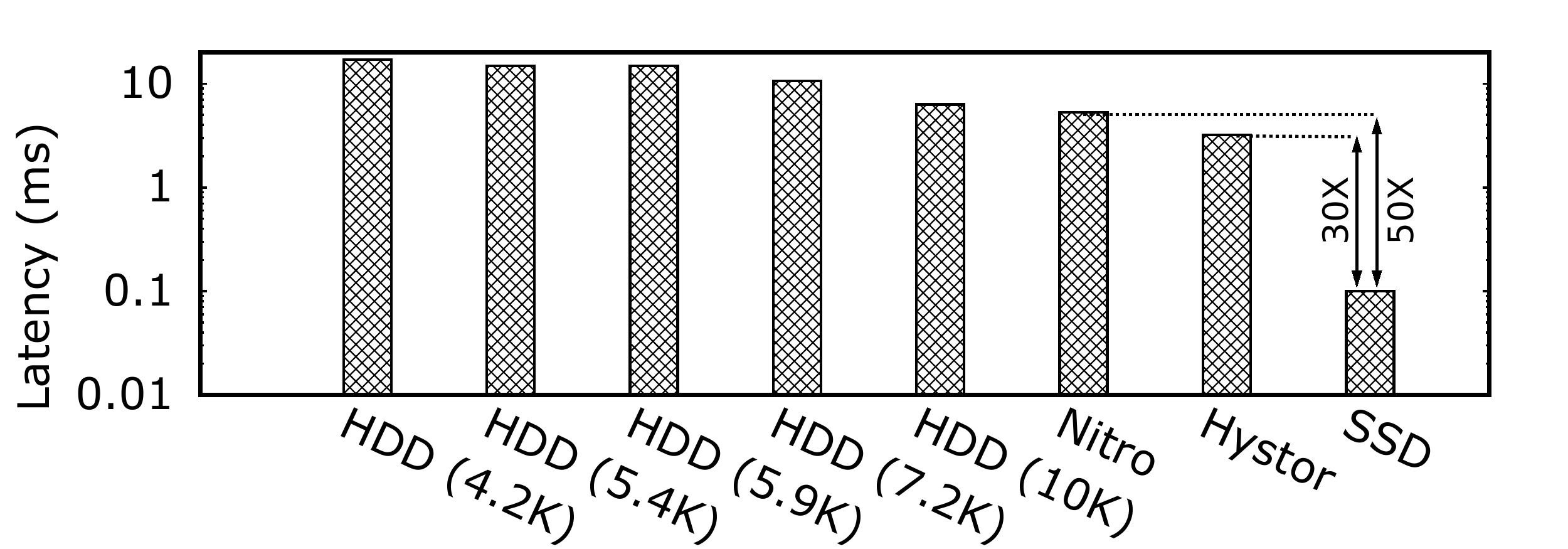}
	\caption{HDD and SSD access latency vs. I/O caching \cite{nitro,hystor}.}
	\label{fig:hdd-ssd-latency}
\end{figure}

The main purpose of employing a high-performance cache layer is to reduce the number of accesses to the disk subsystem. Theoretically, an ideal SSD cache layer would provide access latency equal to the SSD device access latency.
However, due to limited SSD size and imperfect write policy, the observed average latency of accesses to the SSD cache is much higher than the SSD device latency.
For example, we find that the access latency of the caching technique presented in \cite{nitro} is $~$50X higher than the raw access latency of the employed SSD device, as shown in Fig. \ref{fig:hdd-ssd-latency}.\footnote{All numbers in Fig. \ref{fig:hdd-ssd-latency} are based on the results reported in \cite{nitro,hystor}.}

The major parameters affecting IO cache performance are cache size, write policy, and replacement policy.
In this work, we mainly study the effect of cache size and write policy on the performance and endurance of an SSD cache. A commonly-used metric to compare the performance of different cache configurations is cache \emph{hit ratio}. To maximize the cache hit ratio, the cache size should be large enough to house the working set of the running workload. Similar to size, write policy can also affect the performance of the cache particularly for write-intensive workloads. There are three major write policies that are fundamentally different: 1) \emph{Write-Back} (WB), 2) \emph{Write-Through} (WT), and 3) \emph{Read-Only} (RO).
WB reduces the number of accesses to the storage subsystem by buffering temporal writes in the SSD cache and writing them back to the storage subsystem only after the buffered dirty blocks are evicted. WB can improve the performance of both read and write operations, but it suffers from low reliability since the SSD cache can become a single point of failure (i.e., buffered writes can get lost before being propagated to the storage subsystem, if the SSD fails).
WT policy buffers write operations but also transfers them to the storage subsystem at the same time. This policy improves the performance of \emph{only} read operations but provides a higher level of reliability by assuming that each written block is propagated immediately to the storage system. 
RO caches only the read operations and sends write operations directly to the storage subsystem without buffering them. Therefore, it is not able to improve the performance of write operations but it keeps them reliable. 

We conduct experiments to show the impact of cache size and cache write policy on IO performance and SSD endurance. To this end, we perform several experiments on a system with a 200GB HDD and 30GB SSD (our experimental setup is reported in Table \ref{table:mot_exp}). We employ EnhanceIO \cite{enhanceio} as an open source SSD caching scheme in the experiments.
To investigate the impact of write policy on performance and endurance, we run 30 workloads from Filebench on a fixed-size SSD cache with both the WB and RO policies.
We omit the results of the WT policy, since WT has the same endurance as WB and provides less performance than WB.

\begin{table}[!h]
	\centering
	\caption{Setup of the motivational experiments.}\label{table:mot_exp}
	\resizebox{0.8\textwidth}{!} {
		\begin{tabular}{ |l|l| }
			\hline
			\textbf{HW/SW} & \textbf{Description} \\\hline\hline
			Server & HP Proliant DL380 G5 \\  \hline
			CPU & 8x 1.6GHz Intel(R) Xeon \\  \hline
			Memory & 16 GB DDR2, Configured Clock Speed: 1600 MHz \\  \hline 
			HDD &  438GB: four SAS 10K HP‌ HDDs in RAID5 (partition size = 200GB) \\  \hline
			SSD &  128GB Samsung 850 Pro (partition size = 30GB) \\  \hline\hline
			OS & Centos 7 (Kernel version: 3.10.0-327) \\ \hline
			Filesystem & {ext3 (Buffer cache is disabled)} \cite{tweedie2000ext3} \\ \hline
			
		\end{tabular}
	}
\end{table}

\begin{figure*}[!h]
	\centering
	\subfloat[Fileserver]{\includegraphics[width=.25\textwidth]{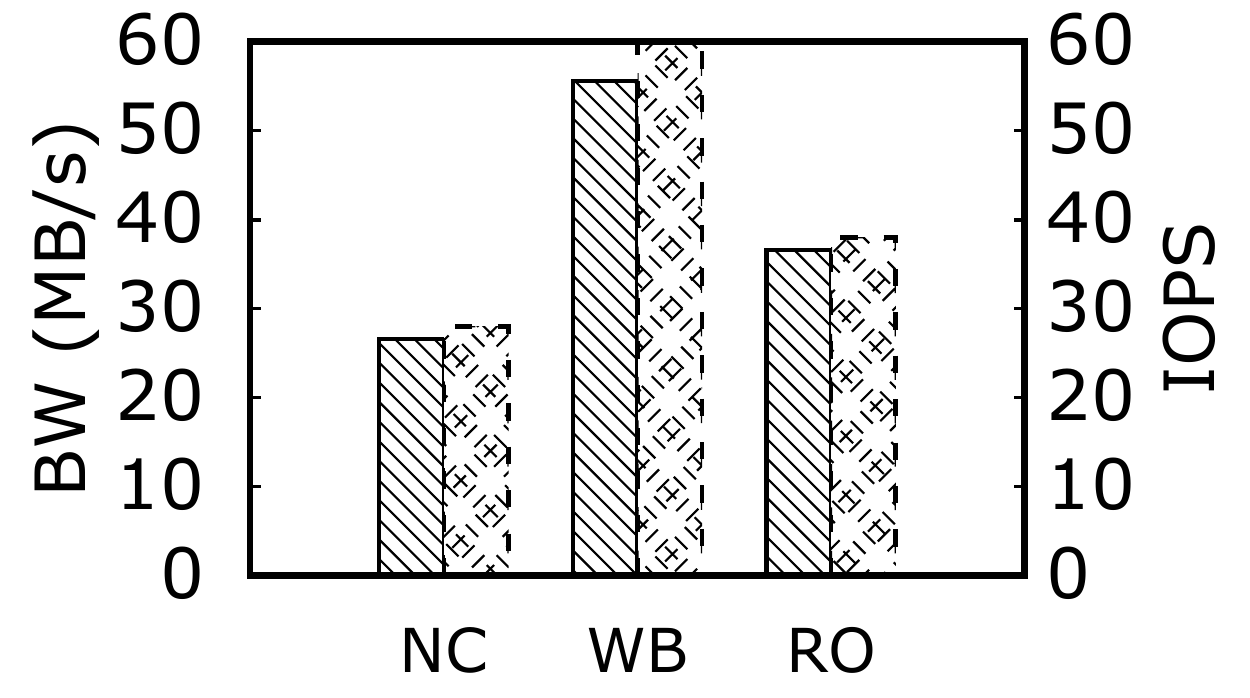}%
		\label{fig:fileserver}}
	\hfil
	\subfloat[RandomRW]{\includegraphics[width=.25\textwidth]{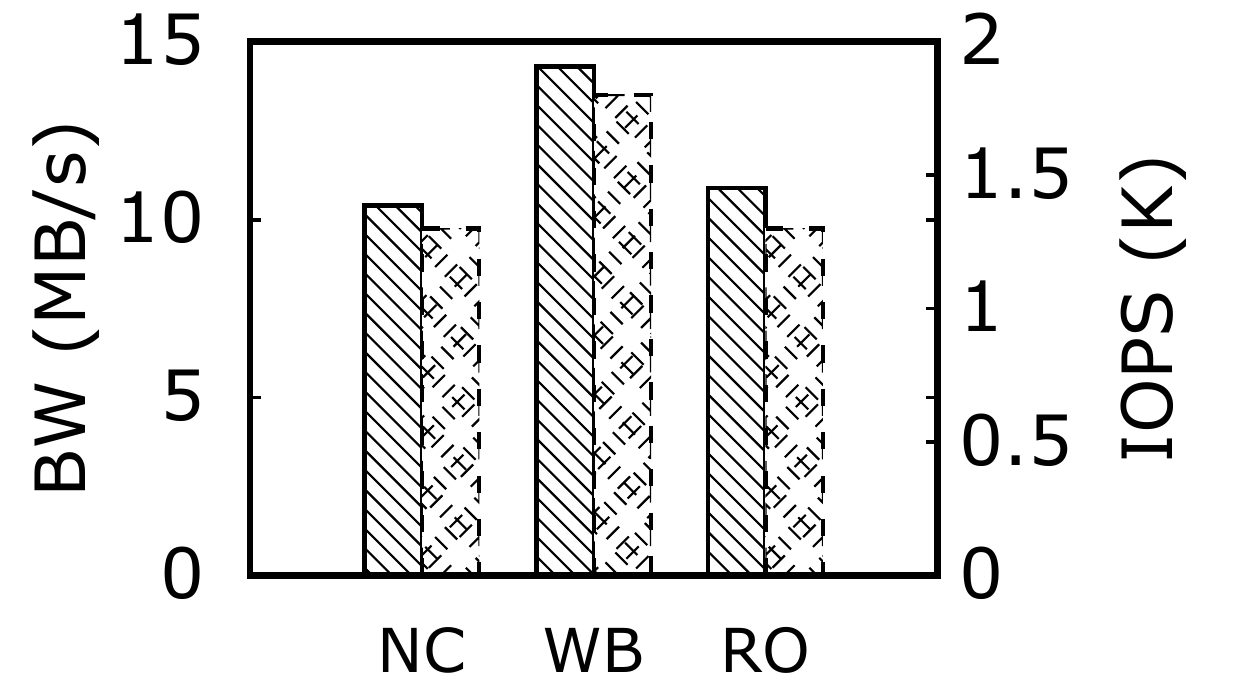}%
		\label{fig:randomrw}}
	\hfil
	\subfloat[Varmail]{\includegraphics[width=.25\textwidth]{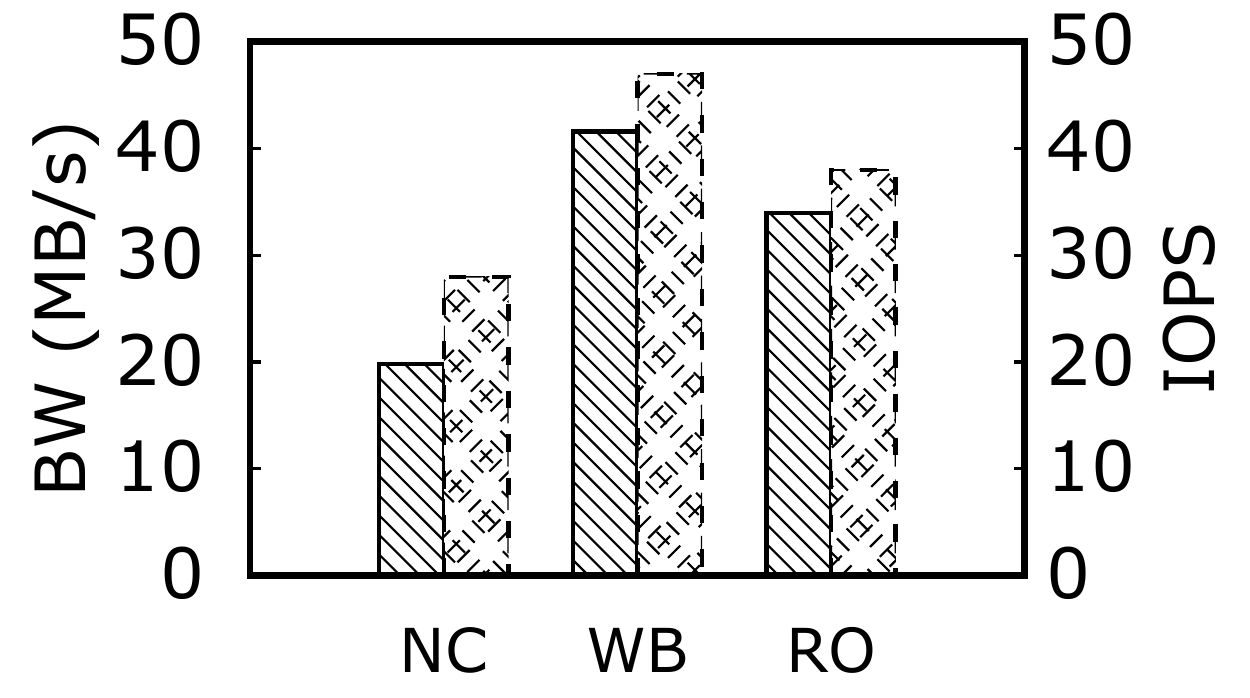}%
		\label{fig:varmail}}
	\hfil
	\subfloat[Webserver]{\includegraphics[width=.25\textwidth]{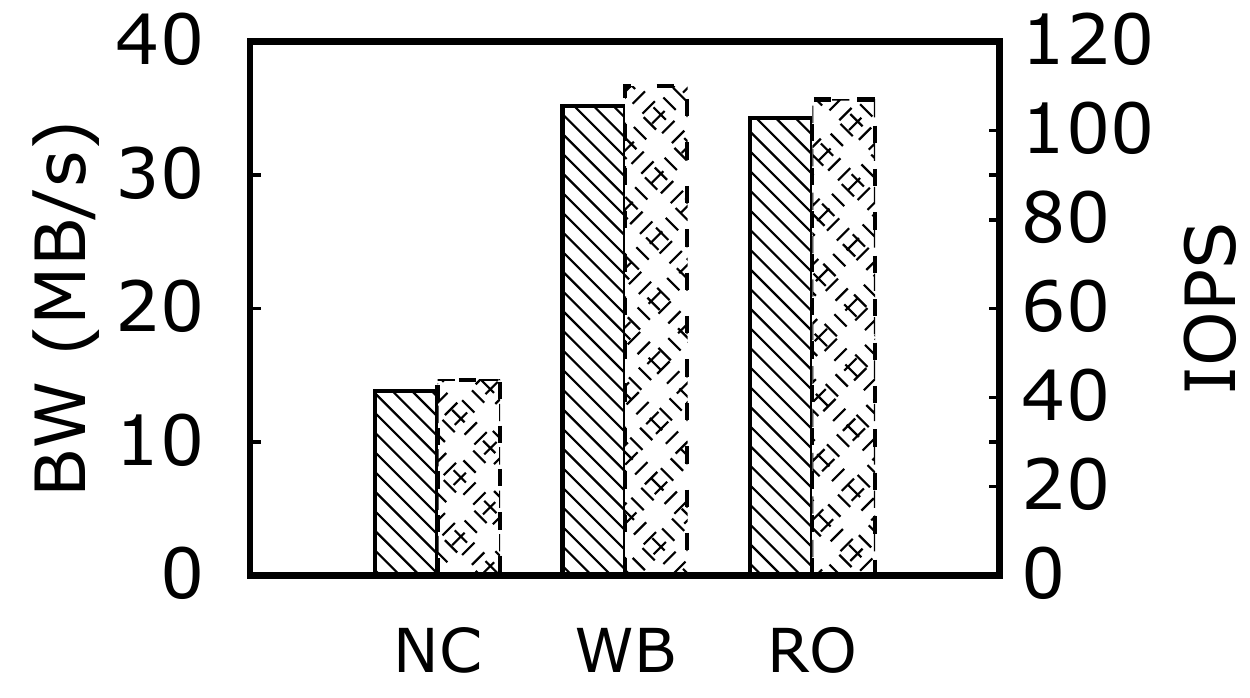}%
		\label{fig:webserver}}
	\hfil
	\subfloat[CopyFiles]{\includegraphics[width=.25\textwidth]{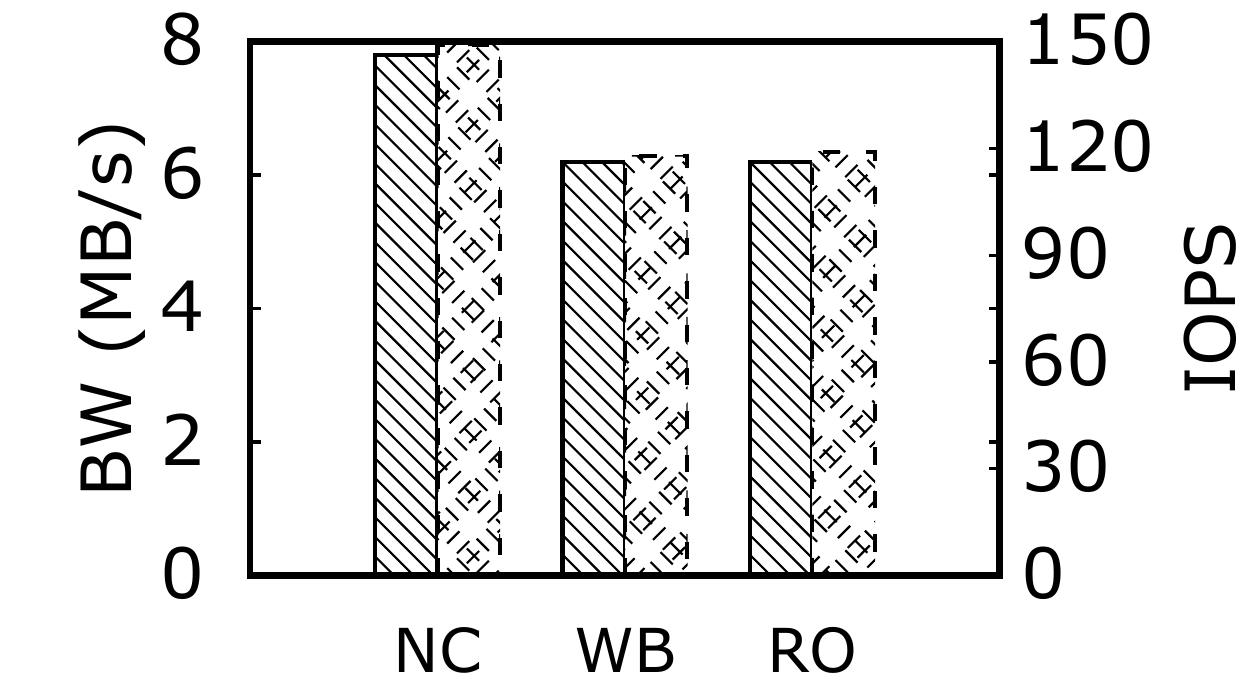}%
		\label{fig:copyfiles}}
	\hfil
	\subfloat[Webproxy]{\includegraphics[width=.25\textwidth]{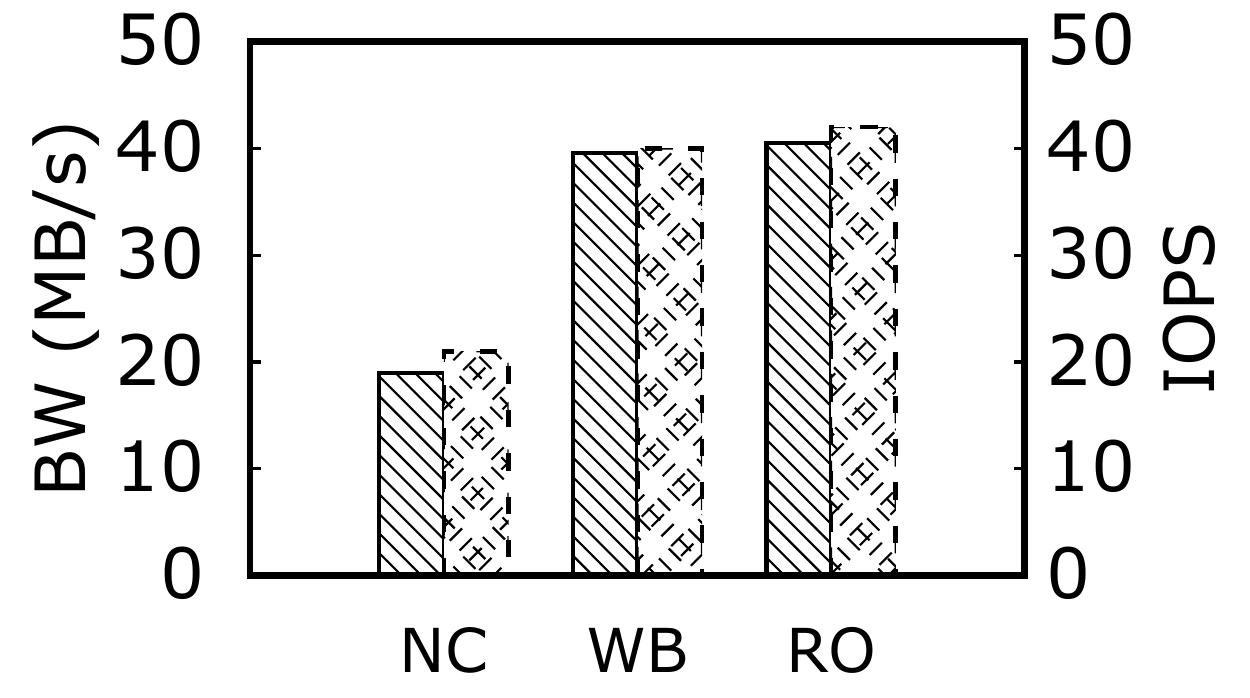}%
		\label{fig:webproxy}}
	\hfil
	\subfloat[Mongo]{\includegraphics[width=.25\textwidth]{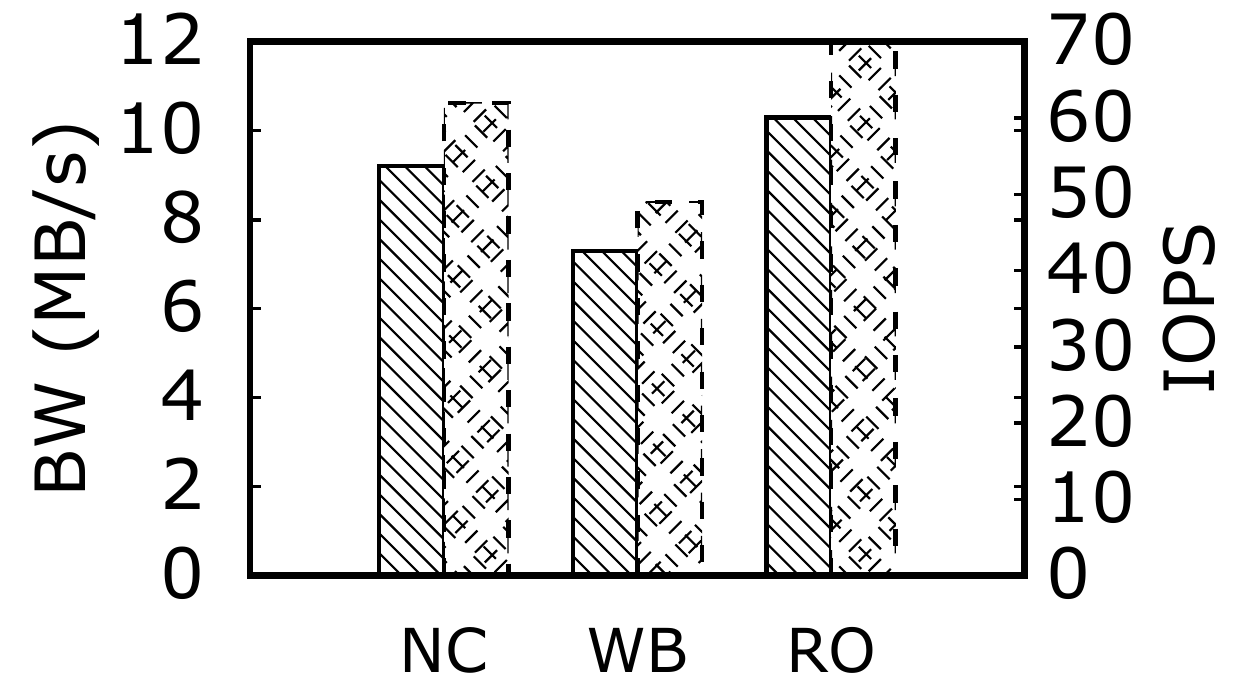}%
		\label{fig:mongo}}
	\hfil
	\subfloat[SingleStreamRead]{\includegraphics[width=.25\textwidth]{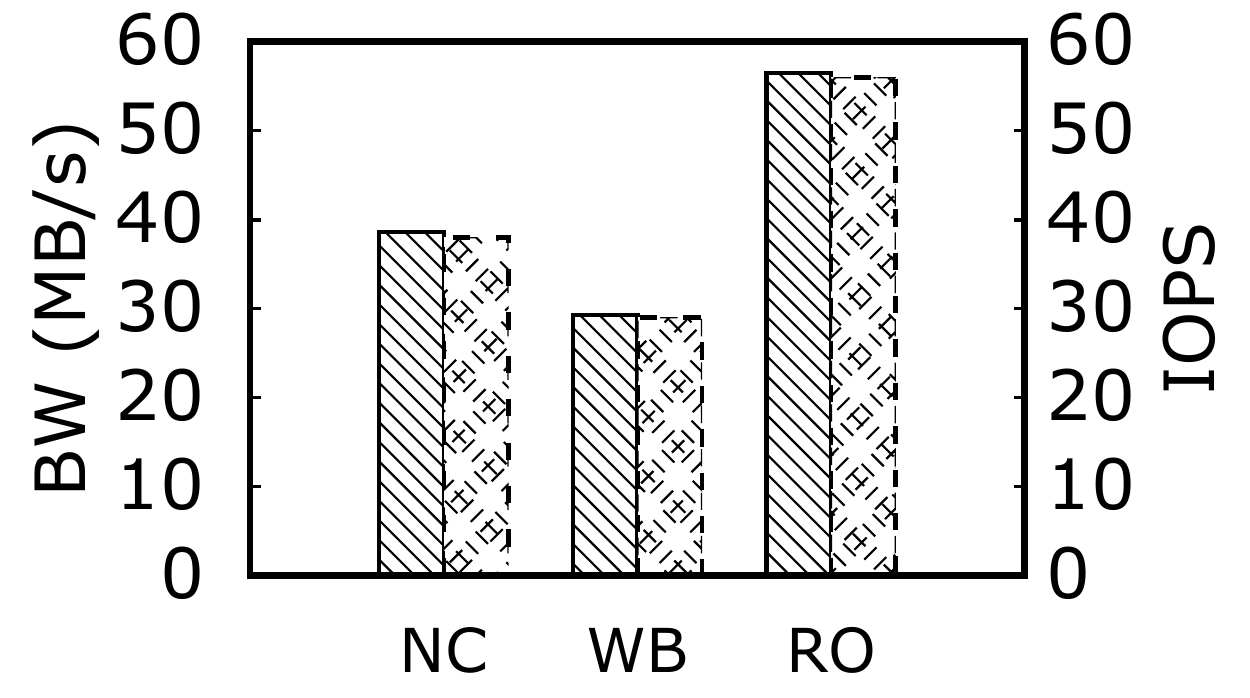}%
		\label{fig:singlestreamread}}
	\caption{Impact of write policy on the performance of workloads (NC: No Cache, WB: Write Back, RO: Read Only).}
	\label{fig:simple_exp_policy}
\end{figure*}

Fig. \ref{fig:simple_exp_policy} shows the impact of write policy on both \emph{Bandwidth} (i.e., the amount of data that is transmitted for a workload in one second) and \emph{I/O Per Second (IOPS)} of eight sample workloads (Fig. \ref{fig:fileserver} through Fig. \ref{fig:singlestreamread}).
We make five major observations: 1) the SSD cache has 2.4X performance improvement with the WB policy
in the \emph{Fileserver} workload where the RO cache has only 1.6X improvement on the IO performance of this workload (Fig. \ref{fig:fileserver}). 2) WB policy improves the IO performance of \emph{RandomRW} and \emph{Varmail} workloads, both over no caching (Fig. \ref{fig:randomrw} and Fig. \ref{fig:varmail}).
3) \emph{Webserver} and \emph{Webproxy} workloads achieve good and similar performance with both the WB and RO write policies (Fig. \ref{fig:webserver} and Fig. \ref{fig:webproxy}). 4) Employing the SSD cache  has negative impact on the performance of the \emph{CopyFiles} workload (Fig. \ref{fig:copyfiles}).
5) The RO write policy can significantly improve the performance of \emph{Mongo} and \emph{SingleStreamRead} workloads by 20\% and 48\%, respectively (Fig. \ref{fig:mongo} and Fig. \ref{fig:singlestreamread}).

Our main experimental conclusions are as follows: 
\begin{enumerate}
	\item In workloads such as \emph{Fileserver}, \emph{Varmail}, and \emph{SingleStreamRead}, only a specific cache configuration can improve IO performance.
	About 45\% of the workloads prefer the WB policy and about 33\% of workloads prefer the RO policy.
	Hence, it is necessary to employ a \emph{workload-aware write policy} for the IO cache. Random-access and write-intensive workloads prefer WB while random-access and read-intensive workloads can be satisfied with the RO policy.
	
	\item In 20\% of workloads, e.g., \emph{Webserver} and \emph{Webproxy}, both WB and RO write policies result in similar improvements. Hence, we can employ RO instead of WB in order to reduce the number of writes to the SSD cache. Such workloads that are random and read-intensive do not take advantage of buffering writes in the SSD cache.
	
	\item In workloads such as \emph{Mongo} and \emph{CopyFiles}, the SSD cache provides little performance improvement. Hence, one can allocate SSD cache space to other workloads that can benefit more from the available cache space. One can prevent the allocation of cache space for the workloads that do \emph{not} benefit from allocated cache, and hence reduce the number of unnecessary writes into the SSD cache. 
\end{enumerate}

\section{Useful Reuse Distance}
\label{sec:URD}
\emph{Traditional Reuse Distance} (TRD) schemes \cite{mattson, parda, shards,ding2003predicting,zhong2009program,fang2004reuse,ding2001reuse,berg2004statcache,loc-appr-conf} work based on the addresses of requests \emph{without} considering the types of the requests. We provide examples demonstrating the benefit of considering the \emph{request type} of the workloads in the calculation of reuse distance. The main objective of this analysis is to present the metric of \emph{Useful Reuse Distance} (URD), which enables assigning smaller amounts of cache space to the VMs while preserving I/O performance. 

\begin{figure}[!htb]
	\centering
	\subfloat[]{\includegraphics[width=.33\textwidth]{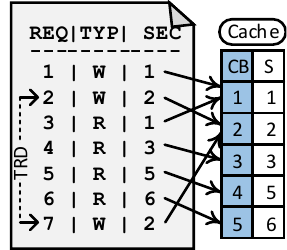}%
		\label{fig:trad_rd_1}}
	\hfil
	\subfloat[]{\includegraphics[width=.33\textwidth]{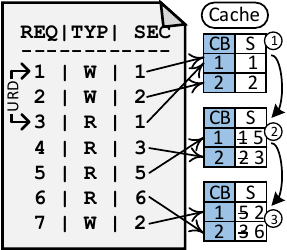}%
		\label{fig:enh_rd_1}}
	
	\caption{Comparison of cache size allocation (a) without  and (b) with considering request type (REQ: Request, TYP: Type, SEC: Sector, W: Write, R: Read, and CB: Cache Block).}
	\label{fig:example_rd}
\end{figure}

We examine a sample workload (shown in Fig. \ref{fig:example_rd}) and show how reuse distance analysis assigns cache size for the workload in two cases: 1) without considering request type (TRD) and 2) considering request type (URD).
In the sample workload given in Fig. \ref{fig:trad_rd_1}, the maximum reuse distance is due to the access of $Req_7$ to the second sector which was previously (five requests before) accessed by $Req_2$. Hence, the maximum TRD of the workload is equal to 4, and according to TRD, we should assign cache space equal to 5 blocks in order to maximize the hit ratio of this workload.
Fig. \ref{fig:trad_rd_1} also shows the contents of the allocated cache to this workload based on TRD. 
It can be seen that when we allocate cache space based on TRD, we reserve one block of cache (Block 2) to keep data that will be written (modified) by the next request ($Req_7$) without any read access to the block.

\begin{figure}[!t]
	\centering
	\includegraphics[scale=1.0]{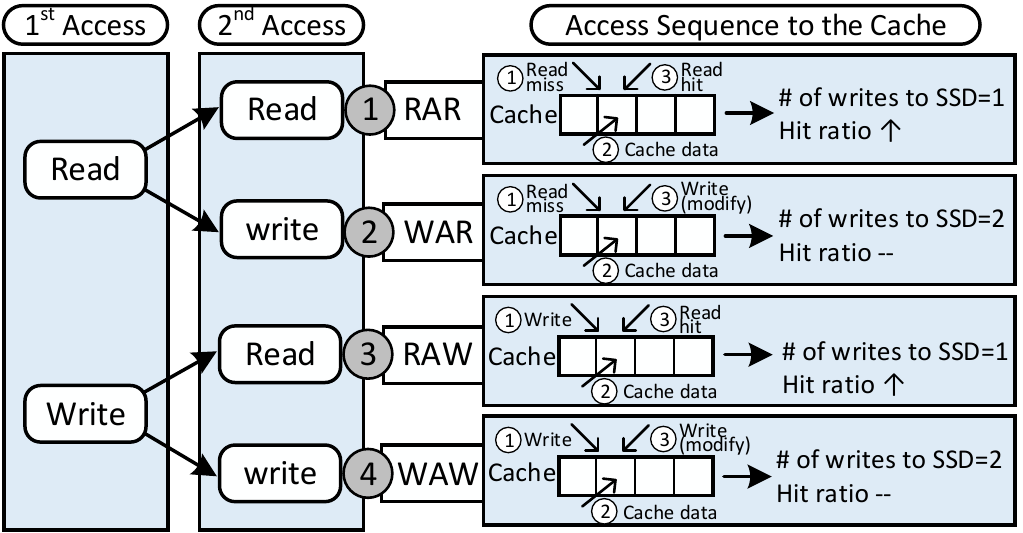}
	\caption{I/O access sequences.}
	\label{fig:io_class}
\end{figure}

\begin{figure}[!htb]
	\centering
	\subfloat[]{\includegraphics[width=.4\textwidth]{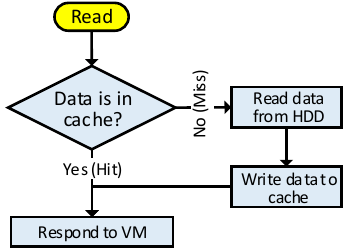}%
		\label{fig:read_hit_miss}}
	\hfil
	\subfloat[]{\includegraphics[width=.4\textwidth]{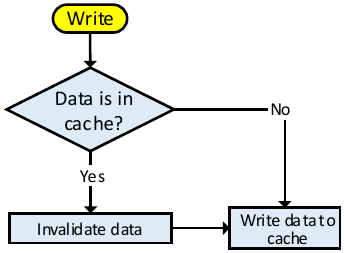}%
		\label{fig:write_hit_miss}}
	
	\caption{Flowchart of I/O cache access for  a (a) Read or (b) Write.}
	\label{fig:read_write_hit_miss}
\end{figure}

Here we classify the sequence of accesses in four groups based on their type (illustrated in Fig. \ref{fig:io_class}): 1) \emph{Read After Read} (RAR), 2) \emph{Write After Read} (WAR), 3) \emph{Read After Write} (RAW), and 4) \emph{Write After Write} (WAW). We show how data blocks of such accesses are stored {in both the WB and WT caches (i.e., allocate on write)}.
Fig. \ref{fig:read_write_hit_miss} shows the operation of the cache for both read and write requests \cite{netappflexcache}. The operation of cache is defined for two cases: 1) for read accesses, if the data is found in the cache we read data from cache. Otherwise, the data is read from the disk subsystem and is stored in the cache for further access. 2) Write operations are directly written to the cache and may modify the previously written data in the cache.
We define cache hit only for read requests.

\begin{enumerate}
	\item RAR: In the first read access, a cache miss fetches data from HDD to the cache. The second access reads the data from the cache. In this case, caching the data block of the first access improves the hit ratio at the cost of imposing one write access to the SSD.
	\item WAR: The first read access leads to fetching the data from HDD to the cache. The second access modifies the data in the cache \emph{without} any read access to this block. In this case, caching the data block of the first read access does \emph{not} improve the hit ratio but it comes at the cost of \emph{two writes} into the SSD.
	\item RAW: The first access writes the data to the cache. The second access reads the data from the cache. In this case, caching the data block of the first access increases the hit ratio of the cache at the cost of imposing one write access into the SSD.
	\item WAW: The first access writes the data to the cache. The second access modifies the data \emph{without} any read access. In this case, caching the data block of the first access does \emph{not} improve the hit ratio but comes at the cost of \emph{two writes} to the cache.
	
\end{enumerate}

We now show how we can allocate a smaller cache size to the sample workload (shown in Fig. \ref{fig:enh_rd_1}) by distinguishing between the four different types of access patterns we just described. We call this scheme \emph{Useful Reuse Distance} (URD) as it takes into account the \emph{request type} in calculating the reuse distance of the workload. URD only considers accesses to the \emph{referenced} data. It \emph{eliminates} WAW and WAR access patterns from the reuse distance calculation.
{It considers only the maximum $reuse~distance$ of RAR and RAW access patterns in reuse distance calculation.}
The maximum URD of the sample workload (Fig. \ref{fig:enh_rd_1}) is equal to 1, due to the read access of $Req_3$ to the first sector of disk which was previously (two request before) written by $Req_1$. In this case, we assign cache size equal to only two blocks. Fig. \ref{fig:enh_rd_1} shows the contents of the allocated cache space based on URD for the sample workload. 
It can be seen that by employing the concept of URD, we achieve hit ratio similar to the TRD scheme while reducing the allocated cache size.

{To summarize, in order to show how URD is able to allocate a smaller cache space compared to TRD and at the same time also achieve a similar hit ratio, we classify the workloads into two groups:}
\begin{enumerate}
	\item {Workloads where RAR and RAW (RA*) accesses are involved in the maximum reuse distance calculation.}
	\item {Workloads where WAR and WAW (WA*) accesses are involved in the maximum reuse distance calculation.}
\end{enumerate}
{These workloads are characterized with the following two equations, respectively.\footnote{RD: Reuse Distance.}
	\begin{equation} 
	\begin{cases}
	\scriptsize
	1: RD(WA*) \leq  RD(RA*) \to TRD \propto RD(RA*),\\ 
	~~~~URD \propto RD(RA*) \to TRD = URD  \\ 
	2: RD(WA*) >  RD(RA*) \to TRD \propto RD(WA*),\\ 
	~~~~URD \propto RD(RA*) \to TRD > URD
	\end{cases}
	\label{equ:conditions_RD}
	\end{equation}
}

{In the workloads of the first group (Eq. {\ref{equ:conditions_RD}}: part 1), both TRD and URD work similarly in cache size estimation. On the other hand, in the workloads of the second group  (Eq. {\ref{equ:conditions_RD}}: part 2), the URD of the workload is smaller than TRD and hence URD allocates a smaller cache size compared to TRD. This is because URD considers only the RA* accesses. The maximum reuse distance of RA* requests is smaller than the maximum reuse distance of WA* requests for the workloads in the second group and hence URD provides smaller maximum reuse distance and leads to the allocation of a smaller cache space. In this case, URD achieves a similar hit ratio while allocating a \emph{smaller} cache space compared to TRD.} 
\section{\techname{} Architecture}
\label{sec:arch}
\begin{figure}[!t]
	\centering
	\subfloat[Sequential accesses.]{\includegraphics[width=.4\textwidth]{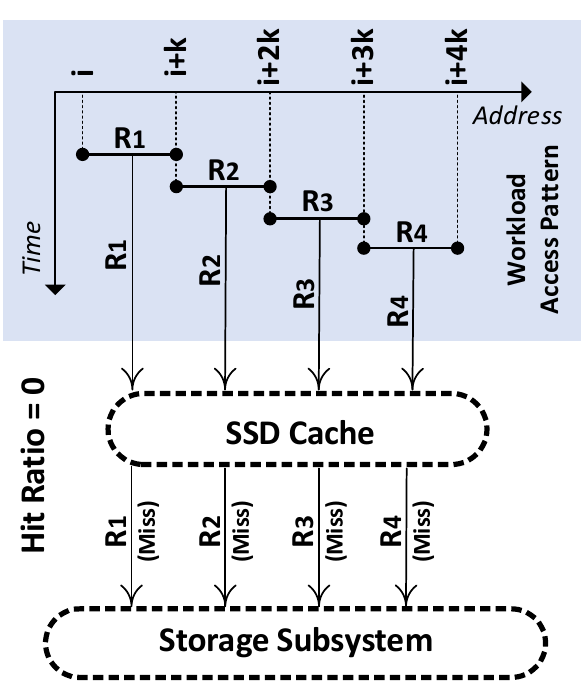}%
		\label{fig:seq_pattern}}
	\hfil
	\subfloat[Random accesses.]{\includegraphics[width=.4\textwidth]{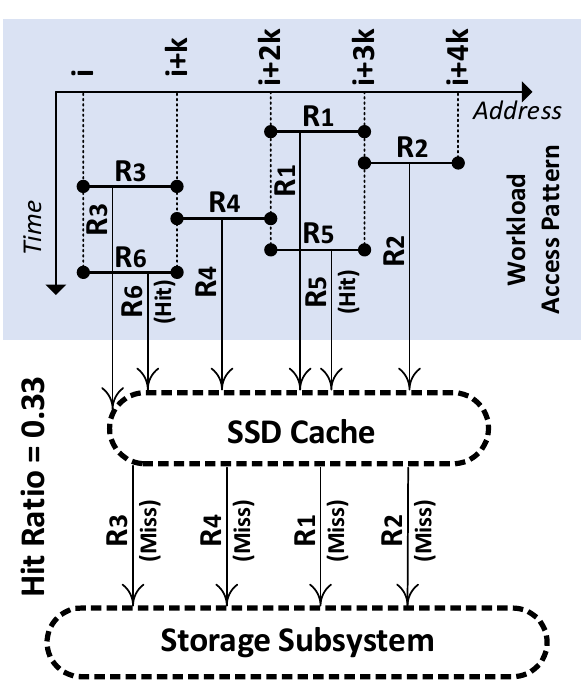}%
		\label{fig:rand_pattern}}
	\caption{Cache hits in sequential and random access patterns.}
	\label{fig:hit_pattern}
\end{figure}

In this section, we describe the architecture of  the \techname{}. \techname{} 1) collects and analyzes the access patterns of the VMs and 2) allocates an efficient and effective cache size and write policy to each VM. Fig. \ref{fig:\techname{}} provides an overview of the \techname{} architecture in the hypervisor of a virtualization platform. As shown in this figure, \techname{} consists of three major components: (1) \emph{Monitor}, (2) \emph{Analyzer}, and (3) \emph{Actuator}. \techname{} resides in the path of IO requests coming from VMs to the storage subsystem. \emph{Monitor} captures and collects information about the IO behavior of each VM. \emph{Analyzer} decides the cache size and write policy by characterizing the IO behavior of the corresponding VM. \emph{Actuator} realizes the decisions made by \emph{Analyzer} by allocating an efficient and effective cache space and write policy for each VM in the SSD cache. We describe each component in more detail:

\begin{enumerate}
	\item\emph{Monitor} receives all the IO requests coming from VMs and extracts important information such as \emph{VM Identification Number} (\emph{VM-ID}), \emph{request type}, \emph{destination address}, and \emph{request size} by using \emph{blktrace}, a block layer IO tracing tool {\cite{blktrace}} {that is available in the Linux kernel (version 2.6.16 and upper). \emph{Blktrace} receives event traces from the kernel and records the IO information.} We modified the source code of \emph{blktrace} to extract the owner of each request (i.e., the VM that sends the request) at the hypervisor level. Such modification helps us to classify the requests and discover the access patterns of the running workloads in different VMs. The extracted information is passed to \emph{Analyzer}.
	\item\emph{Analyzer} decides 1) the target destination of a given IO request, 2) an efficient cache size for each VM, and 3) the write policy of the I/O cache for each VM, based on 1) the information it receives from \emph{Monitor} and 2) a database it employs, called \emph{VM Info}. \emph{Analyzer} keeps information about each VM, such as cache size, write policy, workload characteristics, and the number of VMs running in the system in the \emph{VM Info} database.
	\item\emph{Actuator} is responsible for realizing the decisions made by \emph{Analyzer}. It allocates the decided cache space for each VM, configures the decided write policy, and also routes the IO requests to the SSD cache or the storage subsystem.
	\emph{Actuator} keeps logs for blocks stored in either the SSD cache or the storage subsystem in a table (namely \emph{Map Table}). This table is used for responding to future requests. 
\end{enumerate}
\begin{figure}[!t]
	\centering
	\includegraphics[scale=0.9]{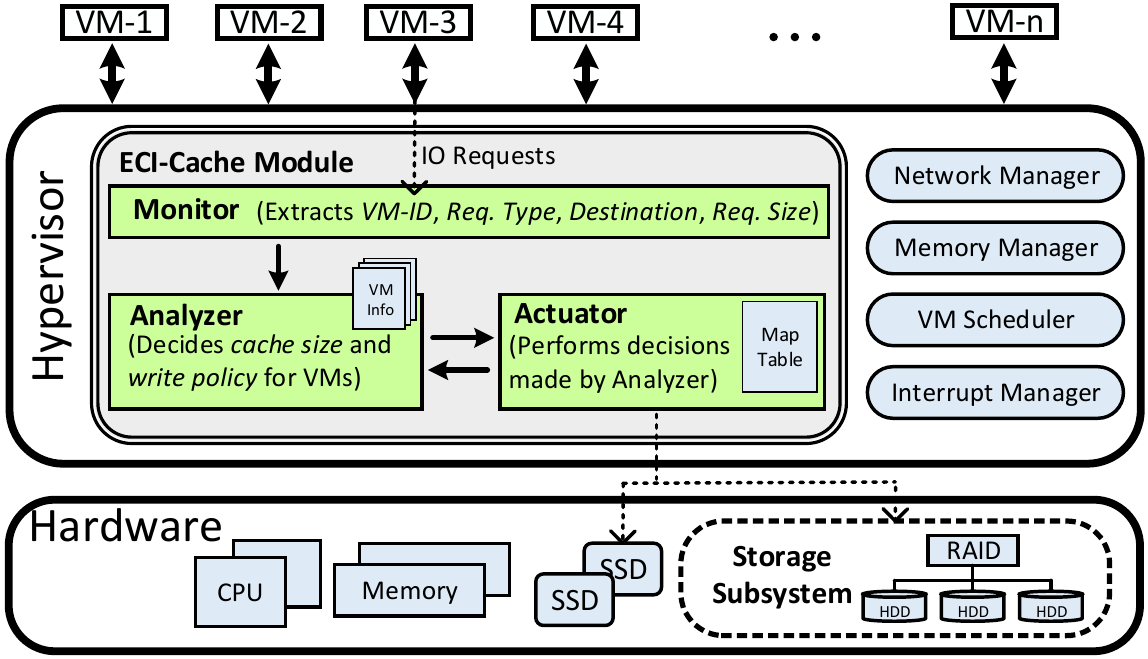}
	\caption{Architecture of \techname{}.}
	\label{fig:\techname{}}
\end{figure}
Typically, in a virtualization platform, there are several VMs running various workloads with different IO behavior. The hypervisor is responsible for partitioning the cache space efficiently between VMs.
In workloads with a sequential access pattern, there is little locality of reference and hence buffering data to capture future reuse demonstrates very poor hit ratio (Fig. \ref{fig:seq_pattern}). 
In workloads with a random access pattern, the probability of referencing the blocks that are accessed previously is significantly greater than that in workloads with a sequential access pattern and hence buffering data with a random access pattern improves the hit ratio (Fig. \ref{fig:rand_pattern}).
Hence, in our proposed architecture, cache space will be allocated to only VMs with random read or write access patterns. Sec. \ref{sec:size_estimation} describes our proposed algorithm for cache space partitioning across VMs via online characterization of the workloads. Sec \ref{sec:policy} describes the write policy assignment to each VM. 

\subsection{Efficient Cache Size Estimation}
\label{sec:size_estimation}
We propose an \techname{} size allocation algorithm, which aims to allocate an efficient cache size for each VM based on the reuse distances of the running workloads. Previously, in Sec. \ref{sec:URD}, we proposed the metric of URD and provided an example that showed the effect of considering request type on 1) the reuse distance of a workload and 2) the estimated cache size. We also demonstrated that employing URD instead of TRD‌ in cache size estimation preserves the performance of the workload while reducing the allocated cache size. Such a scheme allocates a much smaller cache space to the workloads and hence improves the performance-per-cost of each VM. It also reduces the number of unnecessary writes (due to WAW and WAR operations) into the SSD cache, thereby improving SSD lifetime. 

Periodically, \techname{} calculates the URD of the running workloads in VMs and then estimates an efficient cache size for each VM. {To provide the minimum latency for all VMs, we employ an optimization algorithm that meets the conditions of Eq. \ref{equ:conditions}. In this equation, $c_i$ is the allocated cache space for each $VM_i$, $C$ is the total SSD cache space, $N$ is the number of running VMs, and $h_i(c_i)$ denotes the achieved hit ratio for the running workload on $VM_i$ when we allocate cache space equal to $c_i$. In addition, $T_{hdd}$ and $T_{ssd}$ indicate the average read/write access latency to the HDD and the SSD, respectively.}

\begin{equation} 
\begin{cases}
LatencyVM_i=h_i(c_i) \times T_{ssd} + (1-h_i(c_i)) \times T_{hdd} \\
\text{Objective: }Minimize[\sum_{i=1}^{N}LatencyVM_i]\\
\text{Constraint1: }\sum_{i=1}^{N}c_{i} \leq C \\
\text{Constraint2: }0 \leq c_i \leq c_{{urd}_i}
\end{cases}
\label{equ:conditions}
\end{equation}

\indent According to the constraints in Eq. \ref{equ:conditions}, \techname{} partitions the cache space in a way that aims to \emph{minimize} the aggregate latency of all VMs.
Since allocating cache space for workloads with sequential access patterns achieves very poor hit ratio, \techname{} allocates cache space for VMs with random access patterns that have a high degree of access locality. 
{ECI-Cache estimates the true amount of cache space for each VM based on URD. The total SSD cache space should be greater than or equal to the sum of the cache sizes of all N VMs. Hence, we have two possible cases for the allocated cache size: (1) the sum of estimated cache sizes for all VMs using URD is \emph{less} than the total SSD cache space or (2) the sum of estimated cache sizes is \emph{greater} than the total SSD cache space. In the first case (i.e., when the SSD cache space is \emph{not limited}), ECI-Cache would allocate the exact estimated cache space for each VM and thus maximize the overall hit ratio. In the second case (i.e., when the SSD cache space is \emph{limited}), since we have shortage of cache space, it is necessary to \emph{recalculate} the cache size for each VM in order to fit each VM into the existing cache space.}

Algorithm \ref{alg:size} shows how \techname{} estimates and allocates cache space in a virtualized platform with $N$ VMs (Algorithm \ref{alg:size_detail} in the Appendix provides a more detailed version of Algorithm \ref{alg:size}). Initially, a previously defined minimum cache size ($c_{i_{min}}$) is allocated to each VM.\footnote{In the experiments, $c_{i_{min}}$ is set to $1,000$ blocks.} 
\begin{algorithm}
	\scriptsize
	\DontPrintSemicolon
	\caption{ECI-Cache size allocation algorithm.}
	\label{alg:size}
	\tcc{\textbf{Inputs:} Number of VMs: ($N$), SSD cache size: ($C$), HDD Delay: ($T_{HDD}$), SSD Delay: ($T_{SSD}$)}
	\tcc{\textbf{Output:} Efficient cache size for each VM: ($c_{eff}[1..N]$)}
	Sleep for $\Delta t$\; \label{lst:line:sleep}
	Extract the traces of the workloads running  on the VMs including 1) destination address, 2) request size, and 3) request type\; \label{lst:line:delay} \label{lst:line:extract}
	\For {$i = 1$ to $N$}
	{
		$URD[i]=calculateURD(VM[i])$\; \label{lst:line:get_urd}
		$size_{urd}[i]=calculateURDbasedSize(URD[i])$\; \label{lst:line:calc_urd_size}
		$csum+=size_{urd}[i]$\; \label{lst:line:csum}
	}
	\tcc{Check the feasibility of size estimation and minimize overall latency for estimated $size_{urd}[1..N]$}
	\If{csum $\leq C$}
	{\label{lst:line:if1}
		$c_{eff}[1..N]=size_{urd}[1..N]$\;\label{lst:line:feasible_alloc} 
	}\ElseIf{csum $> C$}
	{\label{lst:line:elseif1}
		Create hit ratio function of VMs ($H_i(c)$) based on reuse distance of the workloads.\;\label{lst:line:create_hit_func}
		$c_{eff}[1..N]=calculateEffSize(size_{urd}[1..N], C)$\; \label{lst:line:calc_eff_size}
	}
	$allocate(c_{eff}[1..N], VM[1..N])$\; \label{lst:line:allocate}
	\tcc{\newline Functions Declaration:\newline calculateURD}
	{
		\SetKwProg{Fn}{Function}{ is}{end}
		\Fn{$calculateURD$($VM$)}
		{
			\tcc{This function calls PARDA \cite{parda} which is modified to calculate URD (reuse distance only for RAR and RAW requests.)}
			return $URD$\;\label{lst:line:return_urd}
		}
	}
	{
		\SetKwProg{Fn}{Function}{ is}{end}
		\tcc{\newline calculateURDbasedSize}
		\Fn{$calculateURDbasedSize$($URD$)}
		{
			$size_{urd} = URD \times cacheBlkSize$\;\label{lst:line:calc_urd_based}
			return $size_{urd}$\;\label{lst:line:return_urd_based}
		}
	}
	{
		\SetKwProg{Fn}{Function}{ is}{end}
		\tcc{\newline calculateEffSize}
		\Fn{$calculateEffSize$($size_{urd}[1..N], C$)}
		{
			$initialSize = \{c_{min}, ..., c_{min}\} $\;\label{lst:line:init}
			$lowerBound = \{ c_{min}, ..., c_{min}\}$\;\label{lst:line:lowerbound}
			$upperBound = \{ size_{urd}[1], ..., size_{urd}[N]\}$\;\label{lst:line:upperbound}
			$weightVM = \{1, ..., 1\}$\;\label{lst:line:weight}
			\tcc{Here we use $fmincon$ function from MATLAB Optimization toolbox.} 
			$c_{eff}[1..N] = fmincon(ObjectiveFunction, initialSize, weightVM, C_{tot}, \{\}, \{\}$ $~~~~~~~~~~~~~, lowerBound, upperBound)$\;\label{lst:line:fmincon}
			return $c_{eff}[1..N]$\;\label{lst:line:return_eff}
		}
	}
	{
		\SetKwProg{Fn}{Function}{ is}{end}
		\tcc{\newline ObjectiveFunction}
		\Fn{{$ObjectiveFunction$()}}
		{
			\For{$~i = 1$ to $N$}
			{
				$h[i]=H_i(c[i])$\;\label{lst:line:hit_vm_i}
				$hsum+=h[i]$\;\label{lst:line:sum_hit}
				$csum+=c[i]$\;\label{lst:line:sum_size}
			}
			$diff = C - csum$\;\label{lst:line:diff}
			$Obj = diff + (hsum) \times T_{SSD}+(N - hsum) \times T_{HDD}$\;\label{lst:line:objjj}
			return $Obj$\;\label{lst:line:return_objjj}
		}
	}
\end{algorithm}

At specific time intervals ({$\Delta t$}), we separately extract the {IO} traces of the running workloads on the VMs {into text files} (line {\ref{lst:line:sleep}} and line {\ref{lst:line:extract}}).
{The information included in the IO traces for each request are: 1) the destination address, 2) size, 3) type, and 4) $VM_{ID}$ of the request. This information is extracted by the \emph{Monitor} part of the ECI-Cache.}

{In the next step, we use the collected traces to calculate the URD of the workloads using the {$calculateURD$} function (line {\ref{lst:line:get_urd}}). In the {$calculateURDbasedSize$} function (line {\ref{lst:line:calc_urd_size}}), based on the calculated reuse distances, we find the required cache space for each VM that maximizes the hit ratio. We check the feasibility of the estimated cache sizes to see if the sum of estimated cache sizes ({$csum$} which is calculated in line {\ref{lst:line:csum}}) is less than the total SSD cache capacity (line {\ref{lst:line:if1}}). When the condition in line {\ref{lst:line:if1}} is met and the sum of estimated cache spaces for all VMs is less than or equal to the total SSD‌ cache capacity, we call the cache space allocation {\say{$feasible$}}; otherwise (when the condition in line {\ref{lst:line:elseif1}} is met) we call the allocation {\say{$infeasible$}}. In case of in-feasibility (in line {\ref{lst:line:calc_eff_size}}), we need to recalculate the cache space allocation of each VM such that the SSD cache capacity is not exceeded. To do so, we run the {$calculateEffSize$} function that employs an optimized minimization algorithm ({\say{$fmincon$}}) {\cite{fmincon}} to find the most efficient set of cache space allocations that minimizes the aggregate latency of all VMs under the constraint that the total allocated cache space is less than the SSD‌ cache capacity.\footnote{To this end, we use the {\say{$fmincon$}} function from the MATLAB Optimization toolbox \cite{fmincon}.}
	The input of the minimization algorithm is (1) the existing SSD‌ cache capacity, (2) the hit ratio function of each VM (which will be described in Algorithm {\ref{alg:sample_hit_ratio}}) based on allocated cache space (H(c)) that has been extracted by analyzing the reuse distances of the workloads, and (3) the estimated cache sizes by the algorithm which cannot be fit into the existing SSD cache capacity. Finally, in line {\ref{lst:line:allocate}} we allocate efficient cache spaces for each VM.}

{Algorithm {\ref{alg:sample_hit_ratio}} shows the structure of the hit ratio function. $h(c_i)$ provides the hit ratio that can be obtained, if we assign a specific cache space ($c_i$) to $VM_i$.  This function is extracted from the output of the $calculateURD$ function (described in Algorithm \ref{alg:size}). 
	In the $calculateURD$ function, we extract the ratio of accesses with useful reuse distance $N$ ($URD_N$). Then, for each $c_i$ ($c_i=URD_N \times cacheBlkSize$), $h_i(c_i)$ is equal to ratio of accesses with useful reuse distance $N$.
	The specific cache space $c_i$ is calculated based on the different reuse distances whose hit ratio is the ratio of corresponding reuse distances.}
In each time interval, in case of infeasibility, we update the hit ratio function of each VM and feed it to the minimization algorithm. The minimization algorithm uses the hit ratio function of running VMs to minimize the sum of latency of all VMs, as calculated using Eq. {\ref{equ:conditions}}.

\begin{algorithm}
\scriptsize
	\DontPrintSemicolon
	\caption{{The structure of hit ratio function.}}
	\label{alg:sample_hit_ratio}
	{
	\SetKwProg{Fn}{Function}{ is}{end}
	\Fn{$H_i$($c$)}
	{
		\If{$0 \leq c<m_1$} {$h=h_1$}
		\ElseIf{$m_1 \leq c<m_2$} {$h=h_2$}
		\text{...}\;
		\ElseIf{$m_{k-1} \leq c<m_k$} {$h=h_k$}
		return $h$\;
	}
}
\end{algorithm}

\subsection{Write Policy Estimation}
\label{sec:policy}
In order to allocate the most efficient write policy for each VM, \techname{} analyzes the access patterns and also the request types of the running workloads on the VMs. We choose between RO and WB policies for each VM.
The key idea is to use 1) the RO policy for VMs with write operations without any further read access and 2) the WB policy for VMs with referenced write operations (i.e., write operations with further read access).
To minimize the number of unnecessary writes, we assign the RO policy to the caches of VMs with read-intensive access patterns (including RAR and RAW accesses). The RO policy improves the performance of read operations and increases the lifetime of the SSD. In addition, such a scheme is more reliable since it does \emph{not} buffer writes in the cache.

{As mentioned previously in Sec. \ref{sec:arch}, \emph{Analyzer} is responsible for the write policy assignment for each VM and it does so periodically (every $\Delta$t). {\emph{Analyzer}} checks the ratio of WAW and WAR operations (namely, $writeRatio$). If the $writeRatio$ of the running workload exceeds a defined threshold, we change the write policy to RO, to avoid storing a large number of written blocks in the cache. This is due to two reasons: 1) such a workload includes a large amount of writes and holding such writes in the SSD cache would likely not have a positive impact on the hit ratio, 2) caching such a  large amount of writes has a negative impact on the endurance of the SSD. We select the WB cache policy when the running workload on a VM includes a large fraction of \emph{RAW} accesses. In addition, we assign the RO policy to the caches with a larger fraction of \emph{WAW} and \emph{WAR} accesses.}

Algorithm \ref{alg:policy} shows how \techname{} assigns an efficient write policy for each VM's cache space. Initially, we assign the WB policy for a VM's cache space. Then, periodically, we analyze the behavior of the running workload and re-assess the write policy. 
In line \ref{lst:line:delay_2} of Algorithm \ref{alg:policy}, {after a period of time ($\Delta t$), we calculate the ratio of WAW and WAR requests ($writeRatio$) for $VM_i$} (line \ref{lst:line:readratio}). In line \ref{lst:line:pol_if}, we check whether the ratio of WAW‌ and WAR requests is greater than a threshold (namely $wThreshold$) or not. If the ratio of such requests is greater than $wThreshold$, we assign the RO policy for $VM_i$ (in line \ref{lst:line:set_eff}); otherwise the policy remains as WB.

\begin{algorithm}[!htb]
	\scriptsize
	\DontPrintSemicolon
	\caption{\techname{} write policy assignment algorithm.}
	\label{alg:policy}
	\tcc{\textbf{Inputs:} Number of VMs: ($N$)}
	\tcc{\textbf{Output:} Efficient cache policy for each VM: ($P_{i_{eff}}$)}
	$set(P_{i_{eff}}, VM_i) = WB$ \label{lst:line:set_wb_init} \tcc{Initialization}
	\For {$i = 1$ to $N$}
	{
		Sleep for $\Delta t$\; \label{lst:line:delay_2}
		$writeRatio = {{getNumOfWAW(VM_i)+getNumOfWAR(VM_i)} \over getNumOfReq(VM_i)}$ \label{lst:line:readratio}\;
		\If{$writeRatio \geq wThreshold$}
		{\label{lst:line:pol_if}
			$set(P_{i_{eff}}, VM_i) = RO$ \label{lst:line:set_eff}\;
		}
	}
\end{algorithm}


\section{Experimental Results}
\label{sec:results}
In this section, we provide comprehensive experiments to evaluate the effectiveness of the \techname{}.  
\subsection{Experimental Setup}
To evaluate our proposed scheme, we conduct  experiments on a real test platform, an HP ProLiant DL380 Generation 5 (G5) server \cite{hp_g5} with four 146GB SAS 10K HP HDDs \cite{hp_hdd} (in RAID-5 configuration), a 128GB Samsung 850 Pro SSD\footnote{{The capacity of selected SSD is \emph{larger} than the sum of  efficient cache spaces for the running VMs. Thus, there is no SSD cache capacity shortage in the experiments. As a result, employing an SSD with a larger size would not provide any improvement in the performance of VMs.}} \cite{samsung_ssd} used as the SSD cache, 16GB DDR2 memory from Hynix Semiconductor \cite{hynix_ram}, and 8 1.6GHz Intel(R) Xeon CPUs \cite{intel_cpu}. 
We run the QEMU hypervisor on the \emph{Centos 7} operating system (kernel version 3.10.0-327) and create different VMs running \emph{Ubuntu 15.04} and \emph{Centos 7} operating systems on the hypervisor. 
{The configuration of the device layer is in the default mode where the \emph{request merge} option in the device layer is enabled for a 128-entry device queue size.} 
We have integrated \techname{} with QEMU‌ to enable dynamic partitioning of the SSD‌ cache and allocation of an efficient cache space and  write policy for each VM. 
\subsection{Workloads}

We use MSR traces from SNIA \cite{msr}, comprising more than fifteen workloads, as real workload traces in our experiments. 
We  run \emph{Ubuntu 15.04} and \emph{Centos 7} operating systems on the VMs and allocate two virtual CPUs, 1GB memory, and 25GB of hard disk for each VM. The experiments are performed with 16 VMs. {Table {\ref{table:workload_def}} shows the workloads run on each VM. The SSD‌ cache is shared between VMs. ECI-Cache estimates the most efficient cache size for each VM and partitions the SSD cache space between the VMs based on the running workload's IO pattern and request type. We have also implemented the state-of-the-art IO caching scheme for virtualized platforms, Centaur {\cite{centaur}}, on our test platform. We run the same experiments with Centaur. Similar to ECI-Cache, Centaur works based on reuse distance but it does \emph{not} consider request type in reuse distance calculation. In addition, it does \emph{not} have any control on the write policy of the SSD cache for each VM.}
\begin{table*}[!h]
	\centering
	\caption{Information of running workloads on the VMs.}\label{table:workload_def}
	\vspace{-0.5em}
	\scriptsize
	\begin{tabular}{|m{0.08\textwidth}||m{0.035\textwidth}m{0.028\textwidth}m{0.028\textwidth}m{0.028\textwidth}m{0.028\textwidth}m{0.028\textwidth}m{0.028\textwidth}m{0.028\textwidth}m{0.032\textwidth}m{0.028\textwidth}m{0.028\textwidth}m{0.028\textwidth}m{0.028\textwidth}m{0.028\textwidth}m{0.031\textwidth}m{0.031\textwidth}|}
		\hline
		$VM_{ID}$& \begin{turn}{90}VM0\end{turn} & \begin{turn}{90}VM1\end{turn} & \begin{turn}{90}VM2\end{turn} & \begin{turn}{90}VM3 \end{turn}& \begin{turn}{90}VM4 \end{turn}& \begin{turn}{90}VM5 \end{turn}& \begin{turn}{90}VM6 \end{turn}& \begin{turn}{90}VM7\end{turn} & \begin{turn}{90}VM8 \end{turn}& \begin{turn}{90}VM9\end{turn} & \begin{turn}{90}VM10 \end{turn}& \begin{turn}{90}VM11\end{turn} & \begin{turn}{90}VM12 \end{turn}& \begin{turn}{90}VM13 \end{turn}& \begin{turn}{90}VM14 \end{turn}& \begin{turn}{90}VM15\end{turn}\\ \hline\hline
		Workload & \begin{turn}{90}wdev\underline{\hspace{.05in}}0 \end{turn}& \begin{turn}{90}web\underline{\hspace{.05in}}1\end{turn} & \begin{turn}{90}stg\underline{\hspace{.05in}}1\end{turn} &\begin{turn}{90} ts\underline{\hspace{.05in}}0 \end{turn}&\begin{turn}{90} hm\underline{\hspace{.05in}}1 \end{turn}&\begin{turn}{90} mds\underline{\hspace{.05in}}0 \end{turn}&\begin{turn}{90} proj\underline{\hspace{.05in}}0\end{turn} & \begin{turn}{90}prxy\underline{\hspace{.05in}}0 \end{turn}&\begin{turn}{90} rsrch\underline{\hspace{.05in}}0 \end{turn}&\begin{turn}{90} src1\underline{\hspace{.05in}}2 \end{turn}&\begin{turn}{90} prn\underline{\hspace{.05in}}1 \end{turn}&\begin{turn}{90} src2\underline{\hspace{.05in}}0 \end{turn}&\begin{turn}{90} web\underline{\hspace{.05in}}0 \end{turn}&\begin{turn}{90} usr\underline{\hspace{.05in}}0 \end{turn}&\begin{turn}{90} rsrch\underline{\hspace{.05in}}2 \end{turn}&\begin{turn}{90} mds\underline{\hspace{.05in}}1\end{turn}\\ \hline
		Run Time (min)& \begin{turn}{90}$1,140$ \end{turn}& \begin{turn}{90}$160$ \end{turn}&\begin{turn}{90} $2,190$\end{turn} &\begin{turn}{90} $1,800$ \end{turn}&\begin{turn}{90} $600$ \end{turn}&\begin{turn}{90} $1,210$ \end{turn}& \begin{turn}{90}$4,220$ \end{turn}&\begin{turn}{90} $12,510$ \end{turn}&\begin{turn}{90} $1,430$ \end{turn}&\begin{turn}{90} $1,900$ \end{turn}& \begin{turn}{90}$1,1230$ \end{turn}&\begin{turn}{90} $1,550$ \end{turn}&\begin{turn}{90} $2,020$ \end{turn}&\begin{turn}{90} $2,230$ \end{turn}&\begin{turn}{90} $200$ \end{turn}&\begin{turn}{90} $1,630$\end{turn}\\
		\hline
	\end{tabular}
\end{table*}
\subsection{Cache Allocation to Multiple VMs}
\label{sec:space_alloc}
{
	To show how ECI-Cache affects performance, performance-per-cost, and allocated cache space to VMs compared to Centaur, we conduct experiments in two conditions: 1) when the SSD cache capacity is limited, i.e., when the total SSD cache size is less than the sum of the estimated cache spaces for the VMs. In this case, the cache size estimation by ECI-Cache and Centaur may become infeasible and 2) when the SSD cache capacity is unlimited, i.e., the cache has enough space to allocate the required and efficient cache space for each VM.\footnote{The experiments of the second case are provided in Appendix \ref{sec:unlimited_ssd}.} In the experiments, VMs are run concurrently and cache space is partitioned across VMs for both ECI-Cache and Centaur schemes (we perform the same, yet separate and independent experiments for both schemes and then compare the results).}

{The experiment is performed on 16 running VMs. An initial cache space equal to $10,000$ cache blocks (block size is equal to 8KB) with WB policy is allocated to each VM. The total SSD cache capacity is 3 million blocks.}
{Cache space is calculated in 10-minute time intervals ($\Delta t$) for both ECI-Cache and Centaur. We select 10-min time intervals to reduce the time overhead of the URD calculation to less than 5\% (this trade-off has been obtained based on the reported URD‌ calculation overheads in Table {\ref{table:time_overhead}} in Appendix {\ref{sec:urd-overhead}}). Reducing the length of the time interval provides more accurate estimations but it also increases the time overhead of the URD‌ calculation.}

\begin{figure*}[!h]
	\centering
	
	\subfloat[]{\includegraphics[scale=0.2]{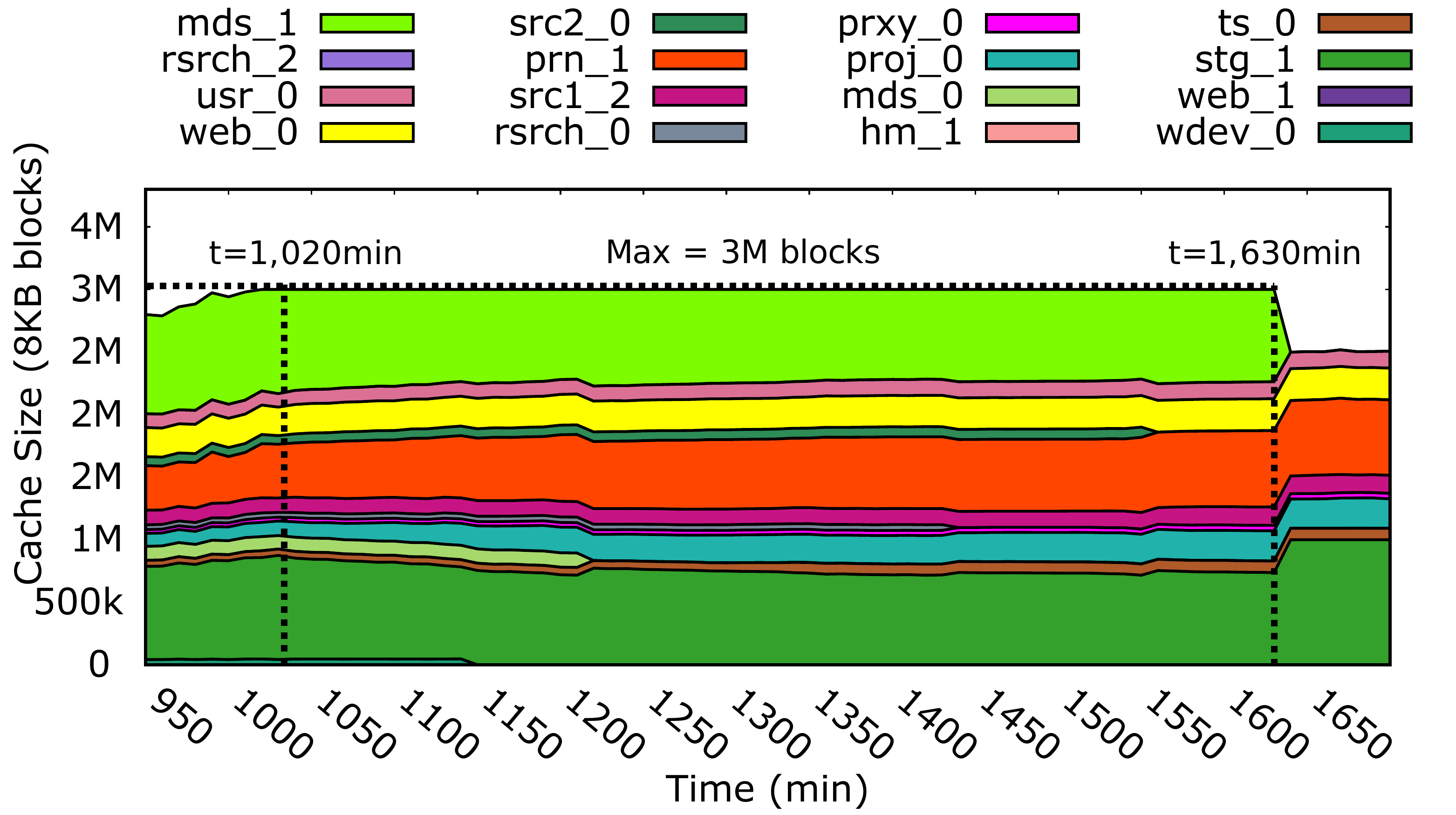}
		\label{fig:centaur_limited_cs}}
	\hfil
	\subfloat[]{\includegraphics[scale=0.2]{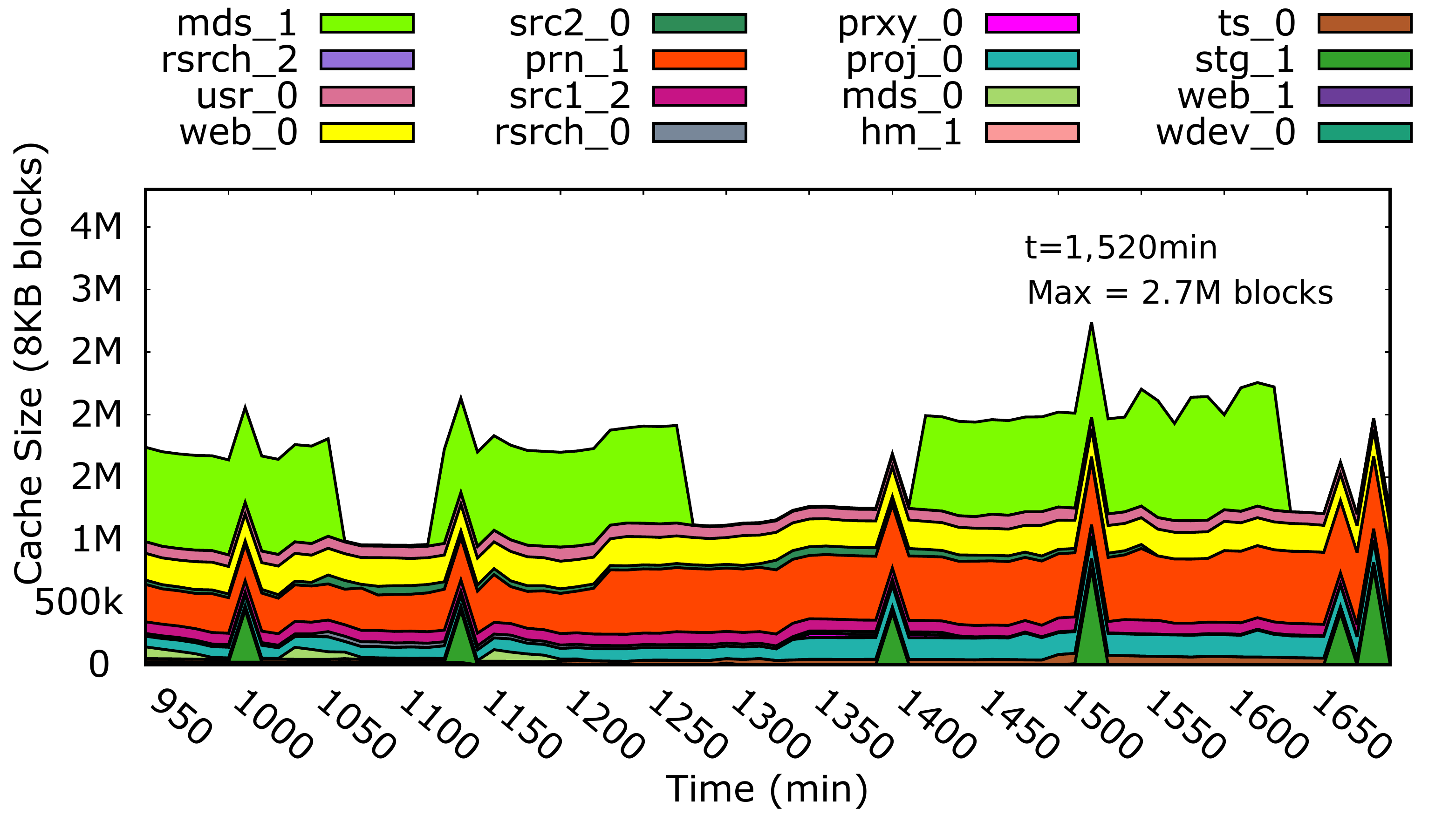}
		\label{fig:eci_limited_cs}}
	\hfil
	\caption{{Cache allocation for the VMs in infeasible state by (a) Centaur and (b) ECI-Cache.}}
	
\end{figure*}

{Fig. {\ref{fig:centaur_limited_cs}} and Fig. {\ref{fig:eci_limited_cs}} show how Centaur and ECI-Cache allocate cache spaces for the VMs in the time interval from $t=950~min$ to $t=1,700~min$ of the experiment when the SSD cache capacity is limited to 3 million blocks.
	We make two major observations: 1) Centaur becomes infeasible and reduces the allocated cache spaces of the VMs to fit in the existing SSD capacity and 2) ECI-Cache never becomes infeasible since the sum of  estimated cache spaces for the VMs is less than existing SSD capacity.
	This is because Centaur estimates a \emph{larger} cache space for each VM because it does \emph{not} consider the request type while ECI-Cache estimates much \emph{smaller} cache space because it \emph{does} consider the request type. When the estimated cache space becomes infeasible (in Centaur), Centaur employs an optimization algorithm to find efficient cache sizes which can be fit in the existing SSD cache capacity. To this end, Centaur allocates smaller cache space to each VM and hence achieves a smaller hit ratio than ECI-Cache. In these experiments, in-feasibility does not happen for ECI-Cache. However, when ECI-Cache becomes infeasible, Centaur would be infeasible, too. In such cases, both schemes should apply optimization algorithms to reduce the estimated cache spaces in order to fit in the existing SSD cache.\footnote{{In Section {\ref{sec:URD}}, we showed that the estimated cache space based on TRD‌ (Centaur)‌ would be greater than or equal to the estimated cache space by URD (ECI-Cache), and hence Centaur would reduce the cache size of each VM more than ECI-Cache would. In this case, the performance degradation in Centaur would be greater than performance degradation in ECI-Cache.}} We conclude that in infeasible cases, ECI-Cache provides greater hit ratio and better performance than Centaur.}
\begin{figure*}[!t]
	\centering
	\subfloat[VM0: wdev\underline{\hspace{.05in}}0]{\includegraphics[width=.25\textwidth]{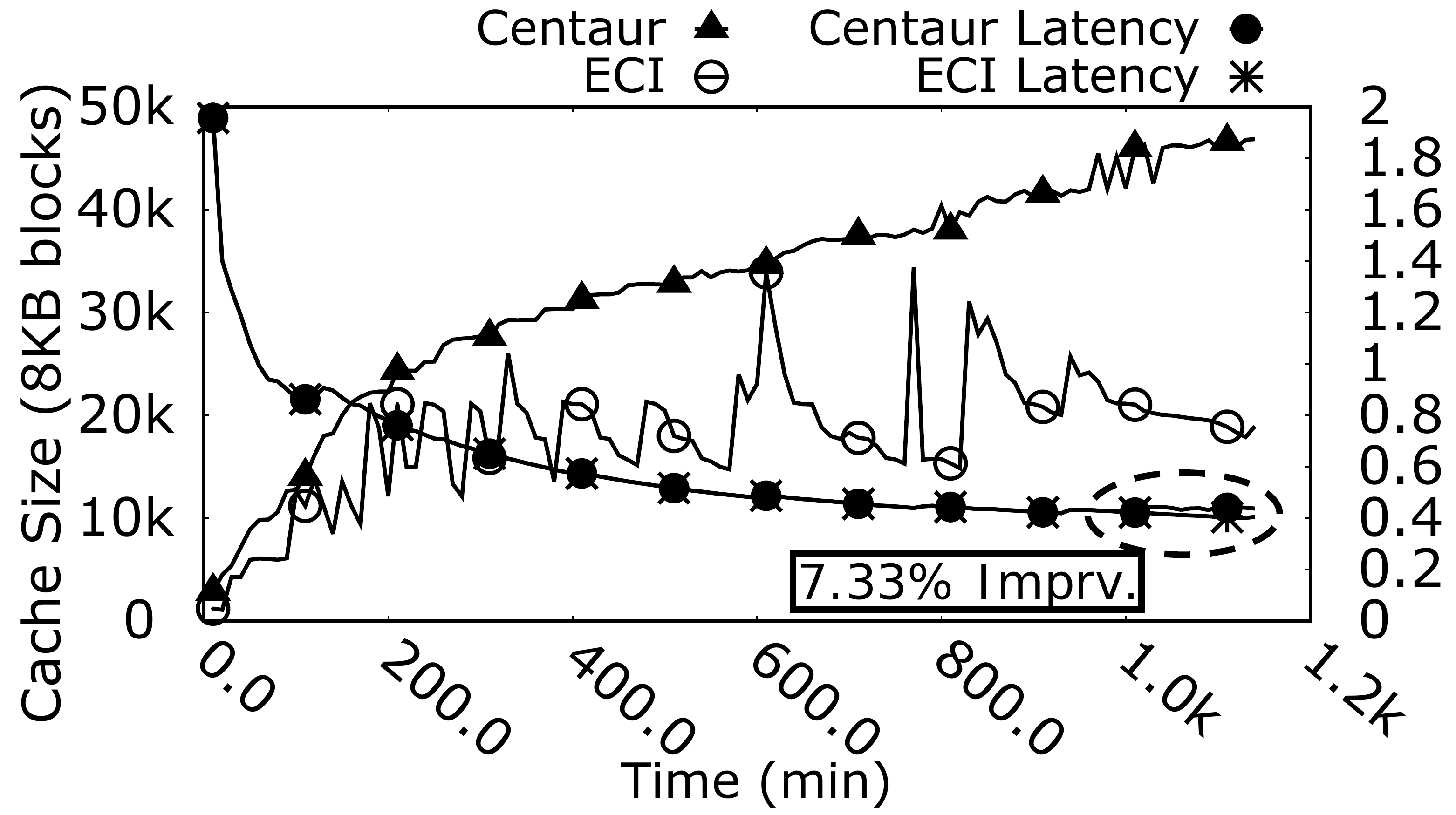}%
		\label{fig:wdev0}}
	\hfil
	\subfloat[VM1: web\underline{\hspace{.05in}}1]{\includegraphics[width=.25\textwidth]{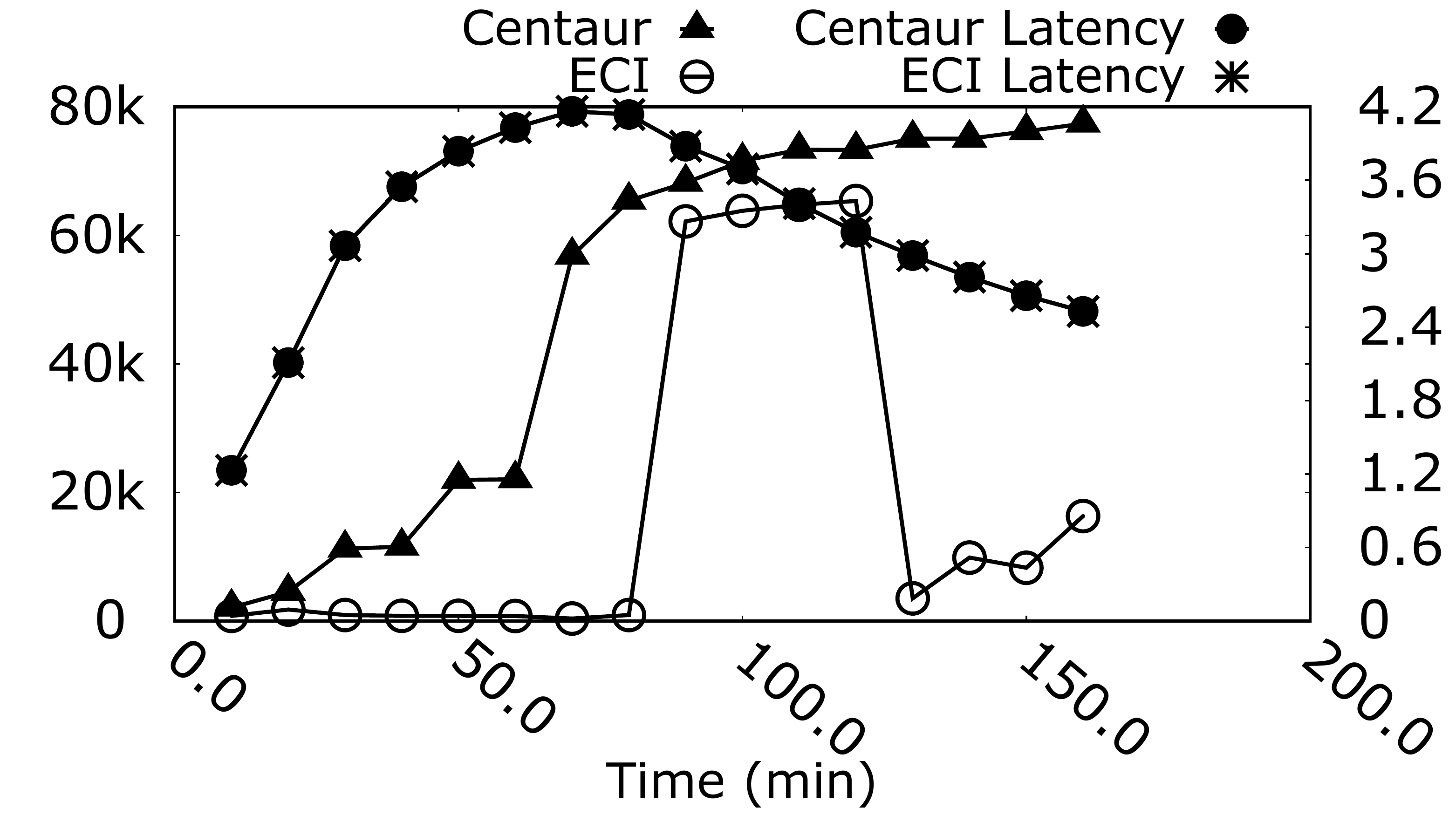}%
		\label{fig:web1}}
	\hfil
	\subfloat[VM2: stg\underline{\hspace{.05in}}1]{\includegraphics[width=.25\textwidth]{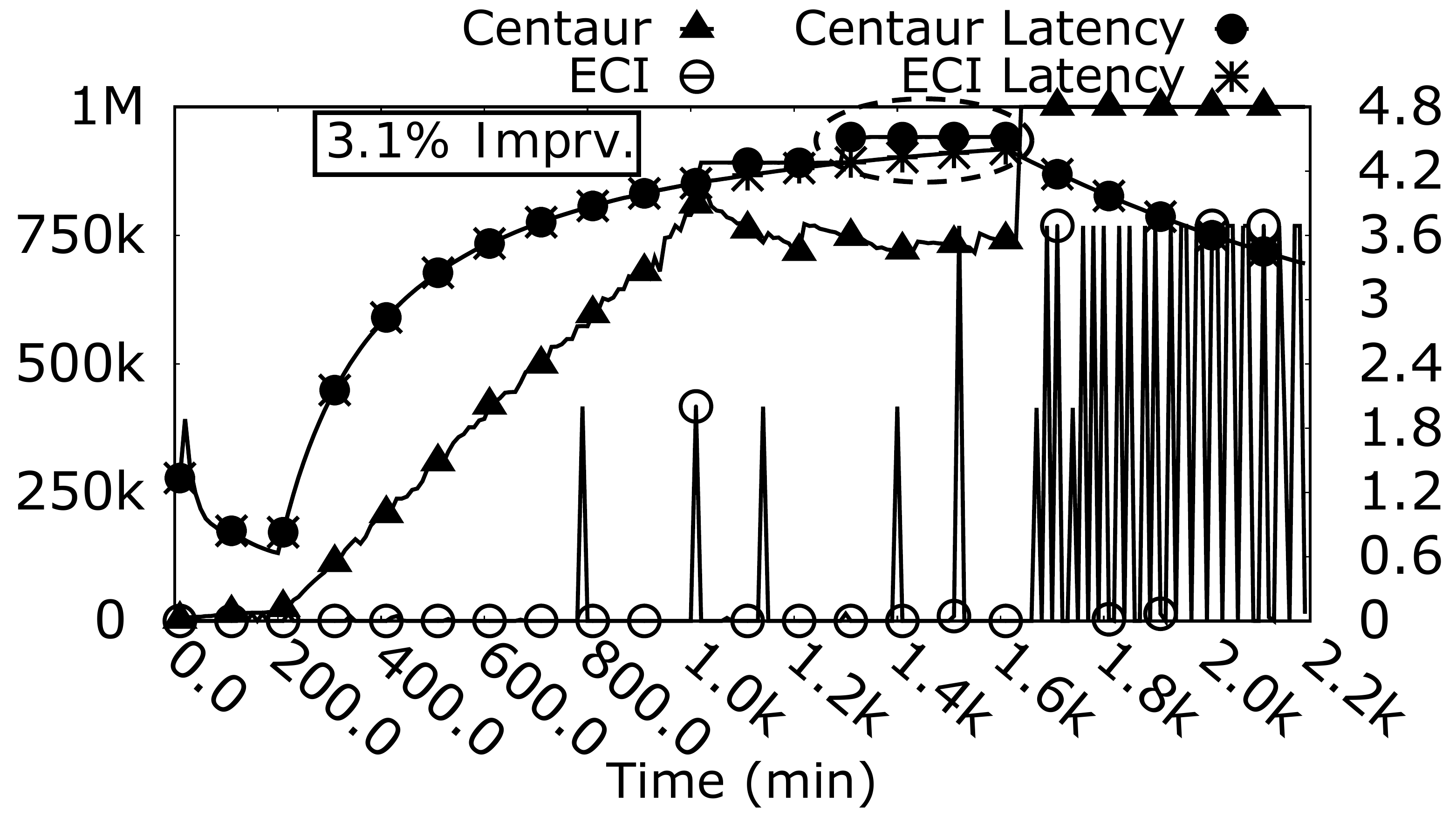}%
		\label{fig:stg1}}
	\hfil
	\subfloat[VM3: ts\underline{\hspace{.05in}}0]{\includegraphics[width=.25\textwidth]{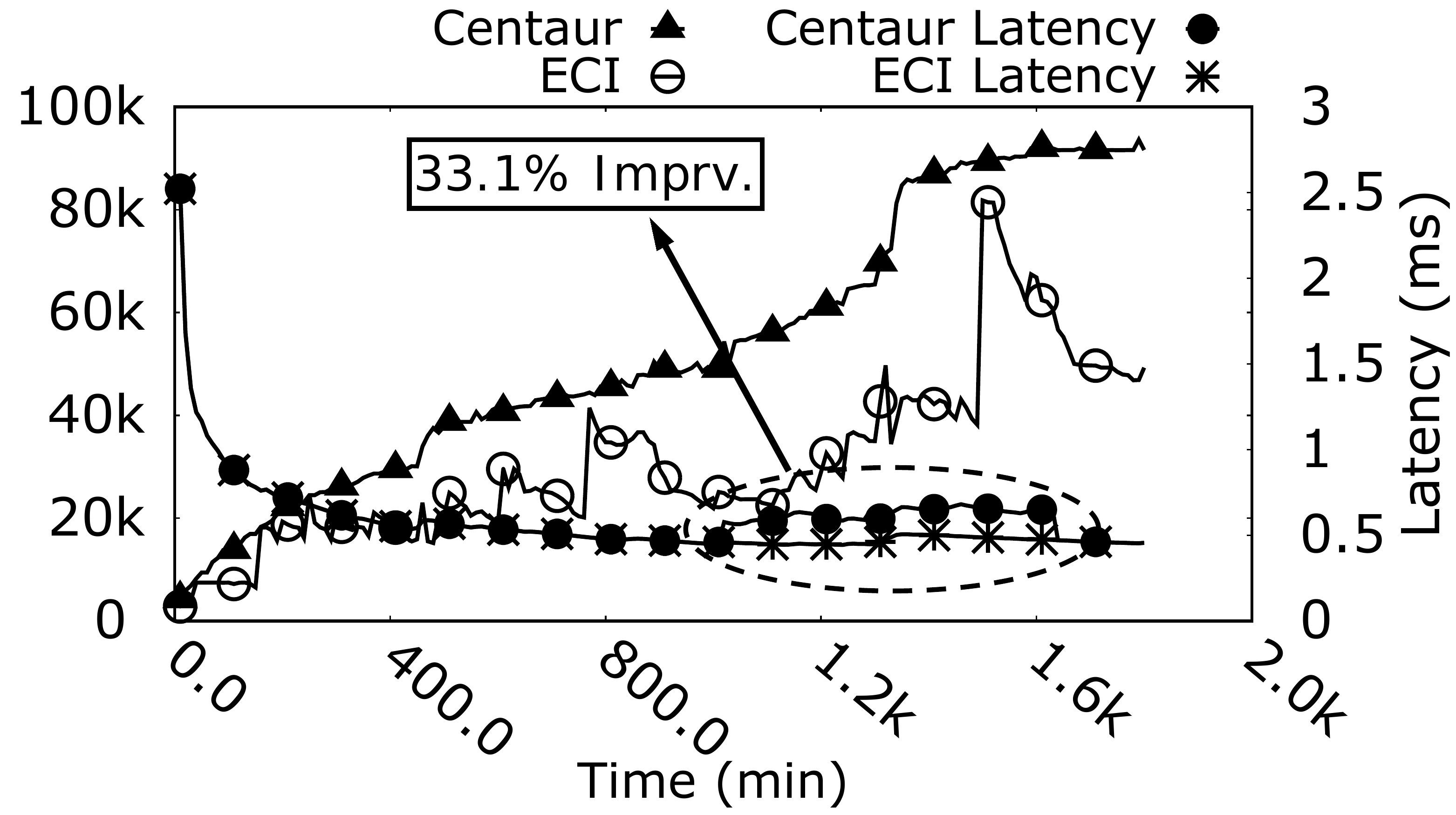}%
		\label{fig:ts0}}
	\hfil
	\subfloat[VM4: hm\underline{\hspace{.05in}}1]{\includegraphics[width=.25\textwidth]{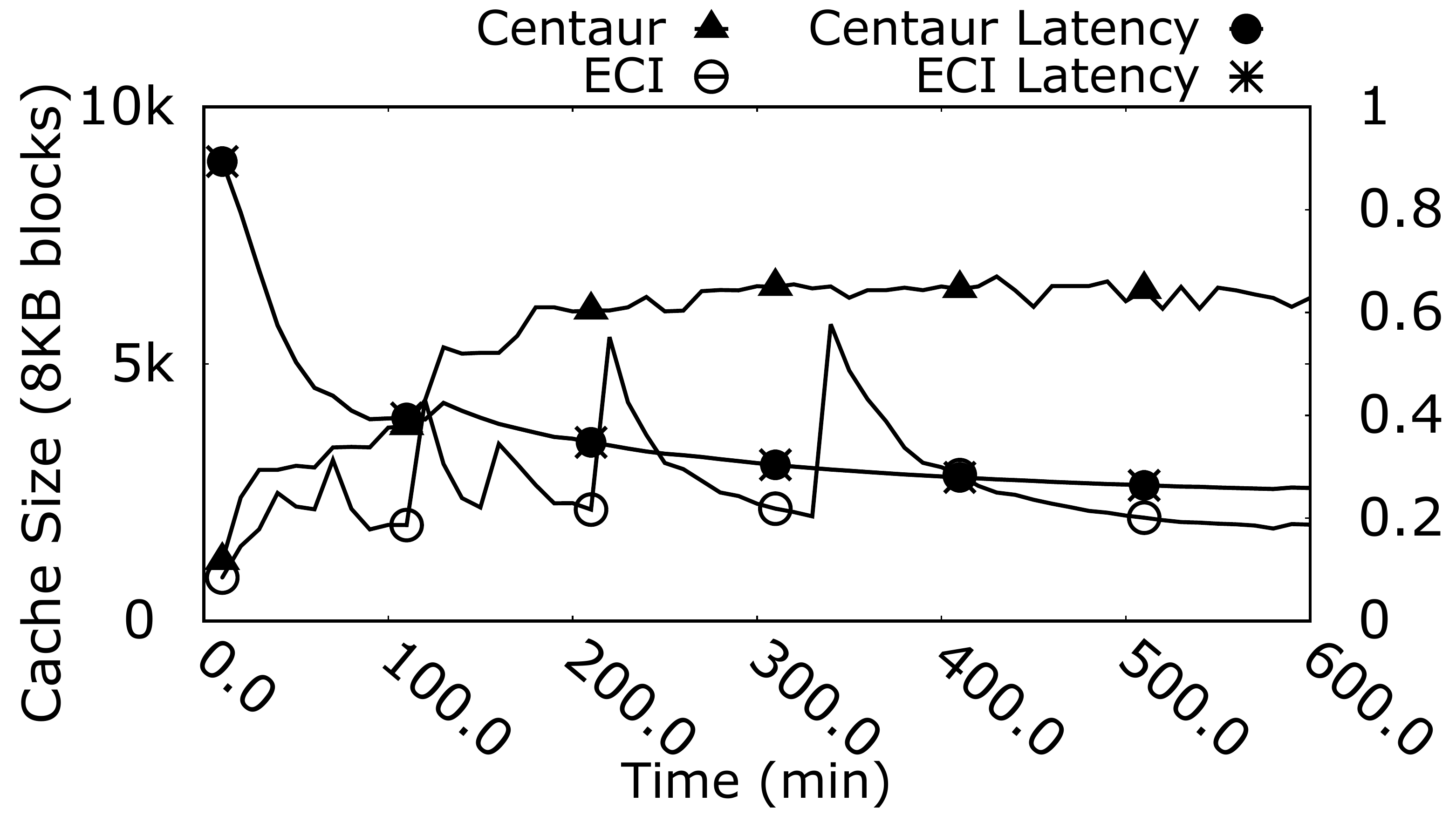}%
		\label{fig:hm1}}
	\hfil
	\subfloat[VM5: mds\underline{\hspace{.05in}}0]{\includegraphics[width=.25\textwidth]{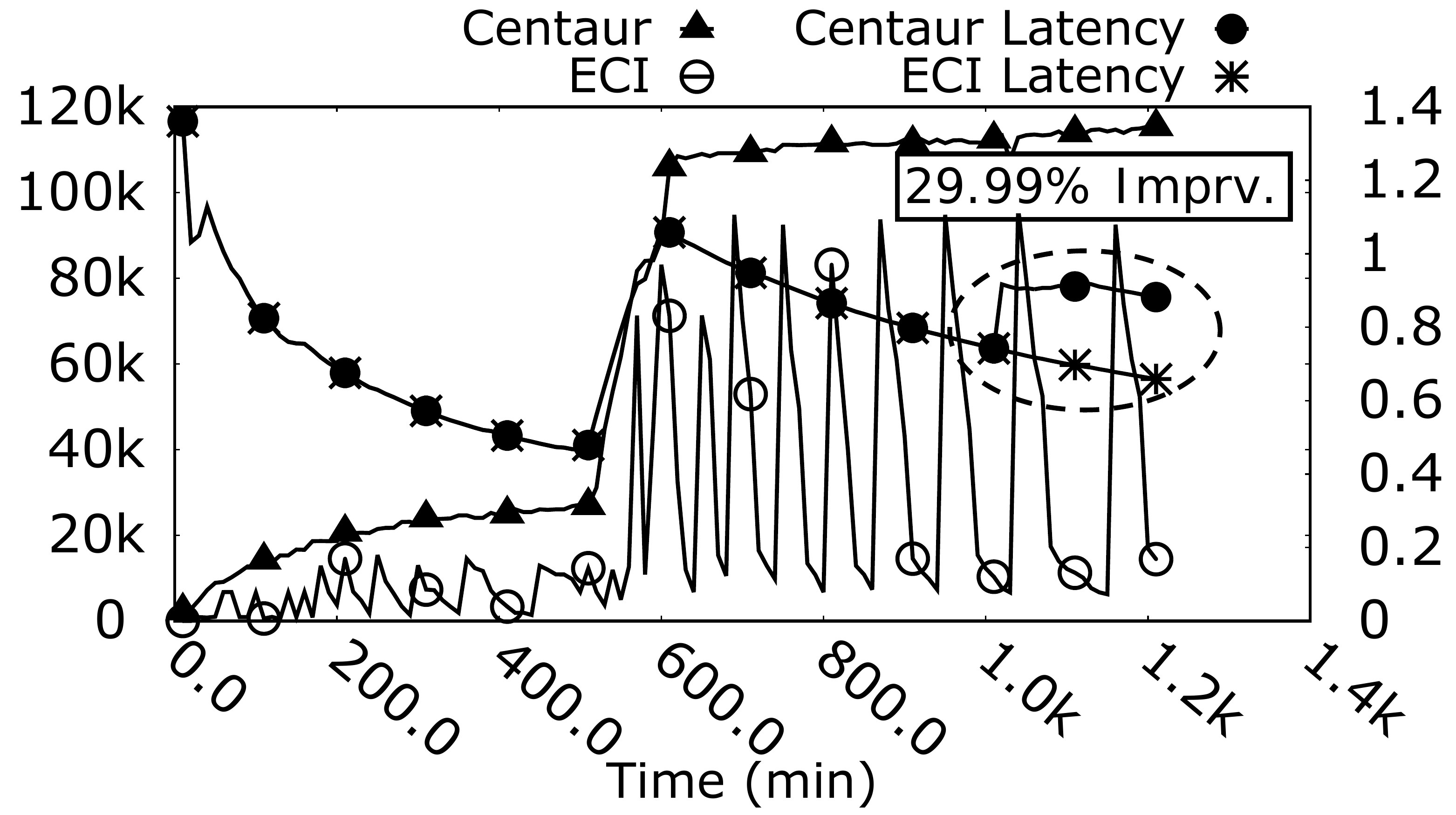}%
		\label{fig:mds_0}}
	\hfil
	\subfloat[VM6:‌ proj\underline{\hspace{.05in}}0]{\includegraphics[width=.25\textwidth]{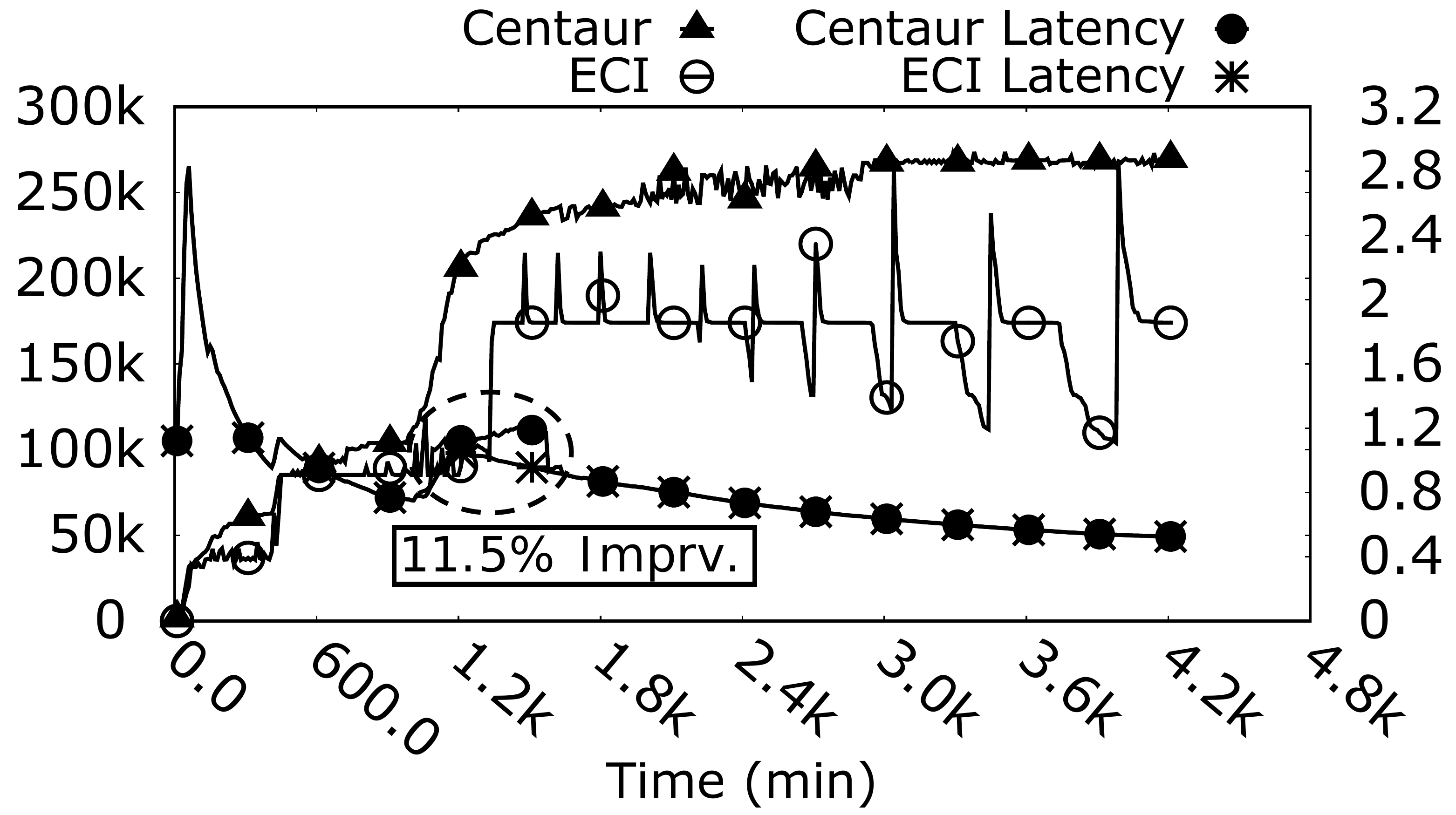}%
		\label{fig:proj_0}}
	\hfil
	\subfloat[VM7: prxy\underline{\hspace{.05in}}0]{\includegraphics[width=.25\textwidth]{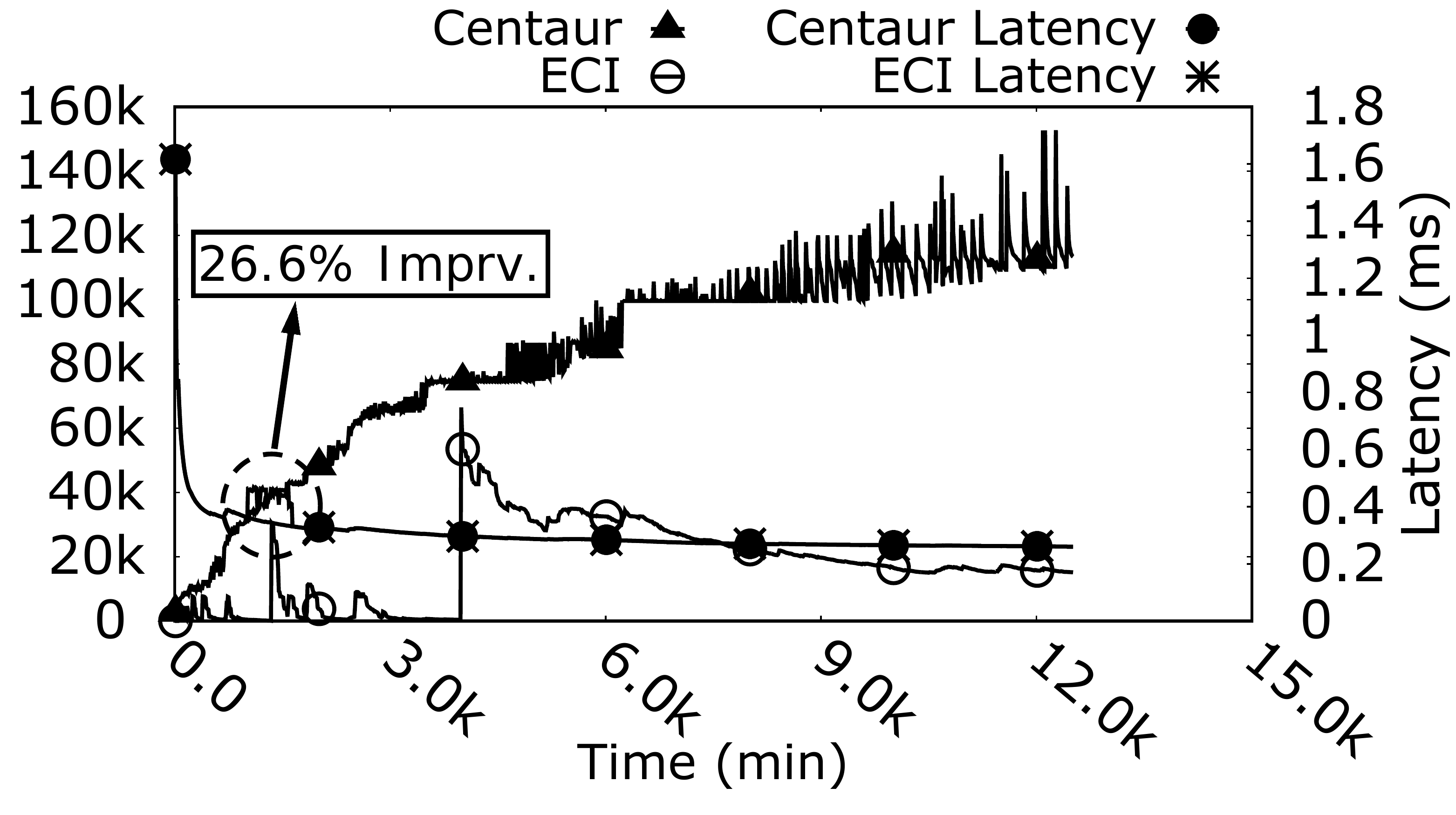}%
		\label{fig:prxy_0}}
	\hfil
	\subfloat[VM8: rsrch\underline{\hspace{.05in}}0]{\includegraphics[width=.25\textwidth]{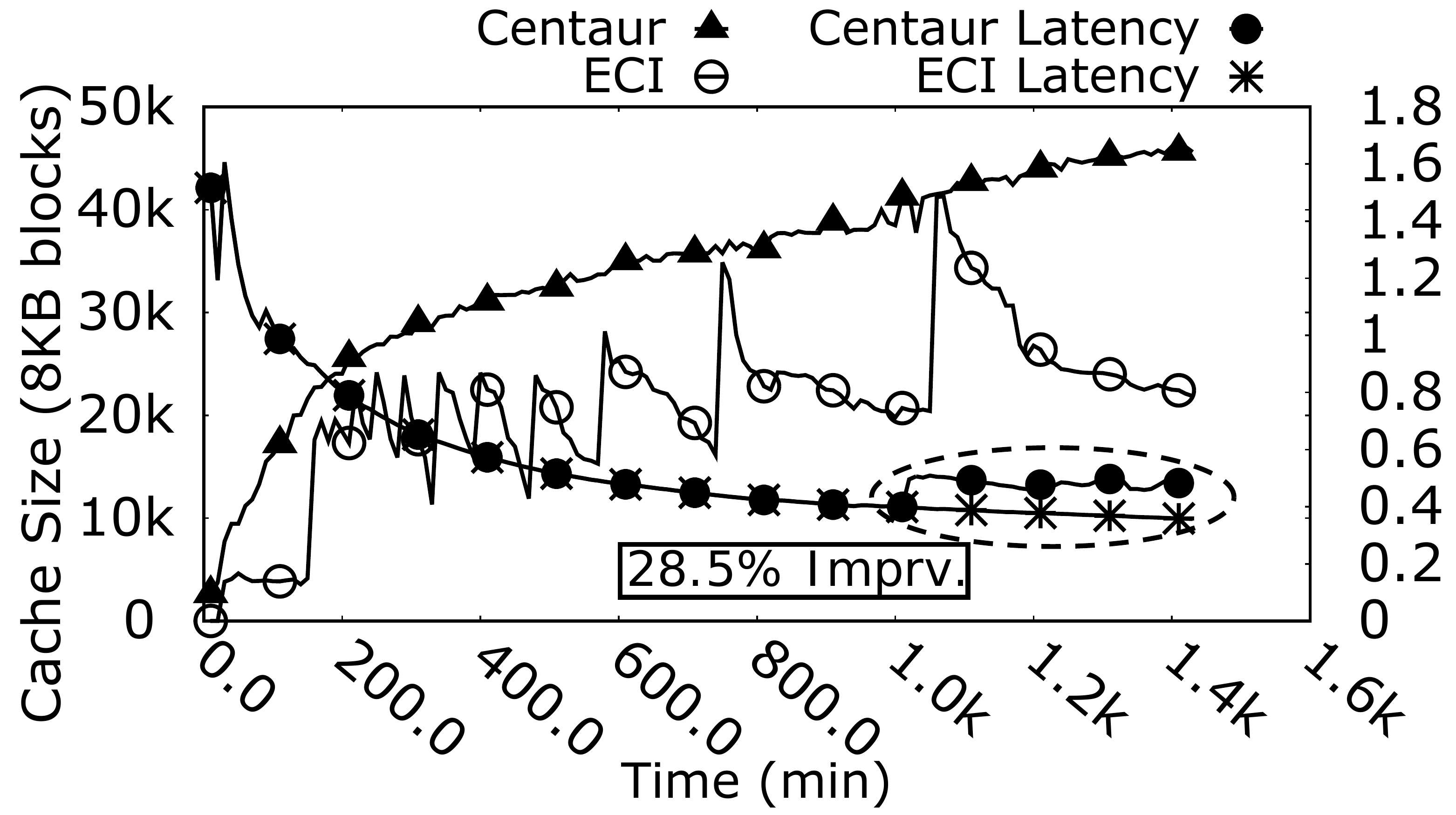}%
		\label{fig:rsrch_0}}
	\hfil
	\subfloat[VM9: src1\underline{\hspace{.05in}}2]{\includegraphics[width=.25\textwidth]{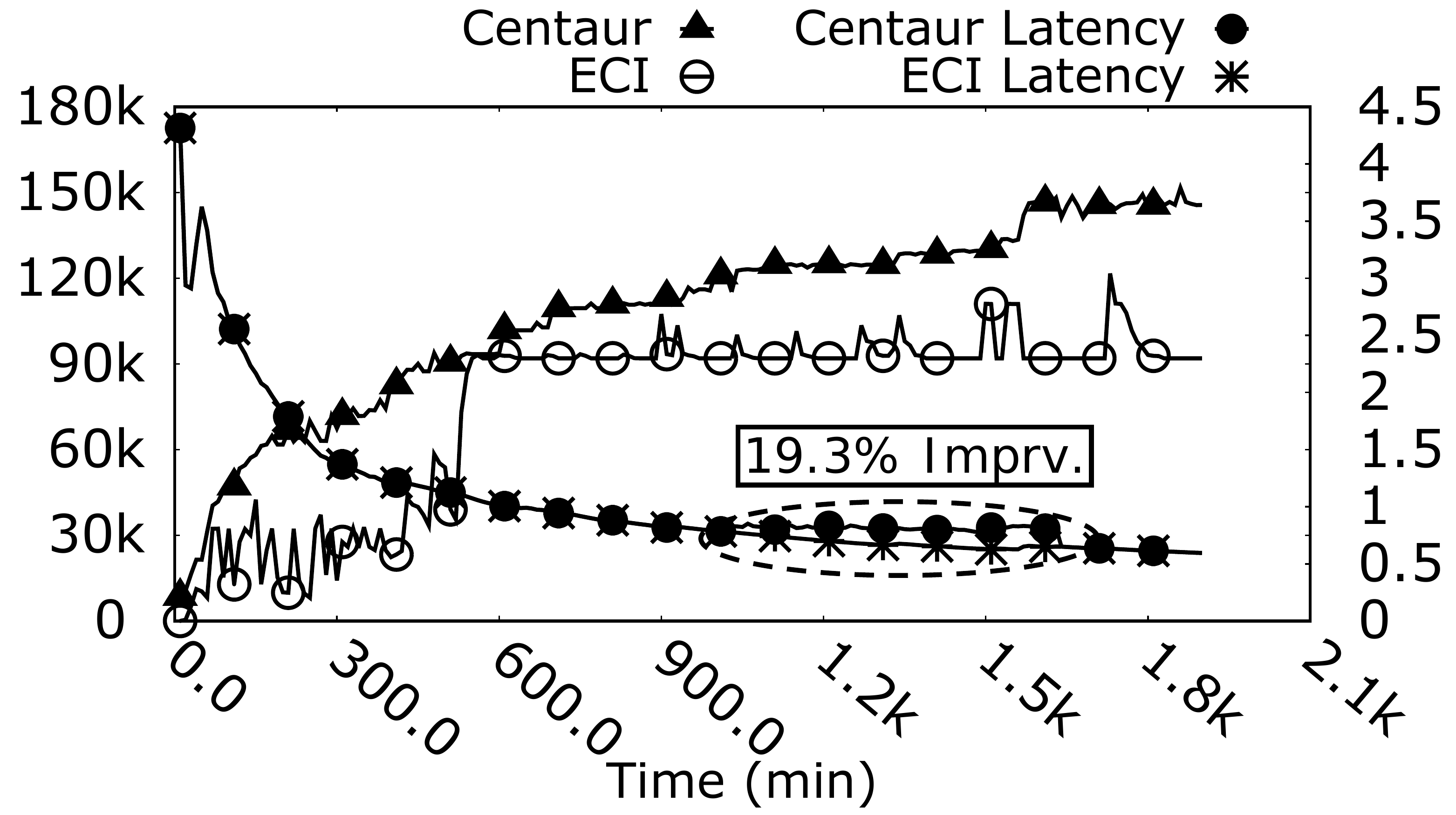}%
		\label{fig:src_1}}
	\hfil
	\subfloat[VM10: prn\underline{\hspace{.05in}}1]{\includegraphics[width=.25\textwidth]{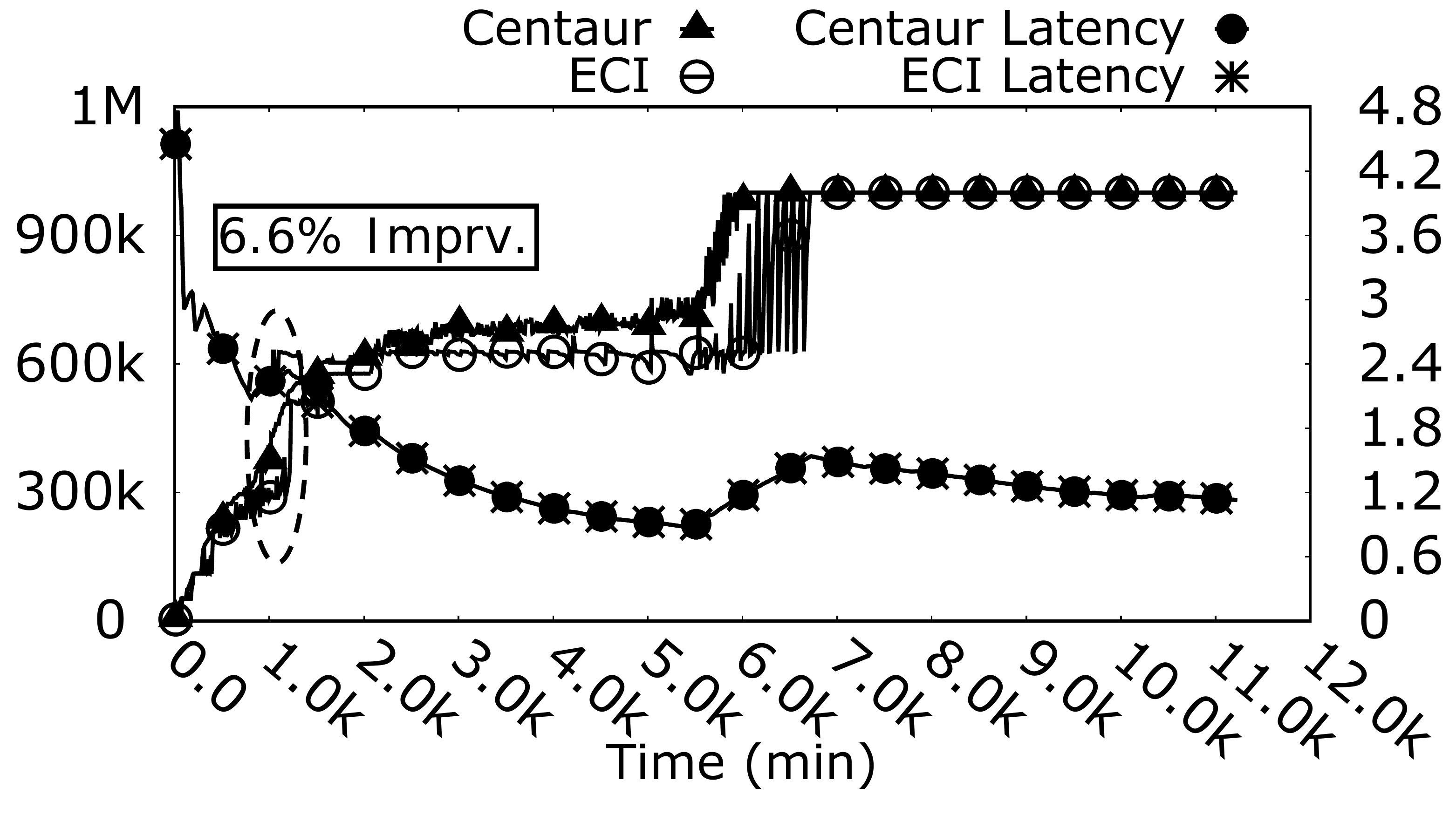}%
		\label{fig:prn_1}}
	\hfil
	\subfloat[VM11: src2\underline{\hspace{.05in}}0]{\includegraphics[width=.25\textwidth]{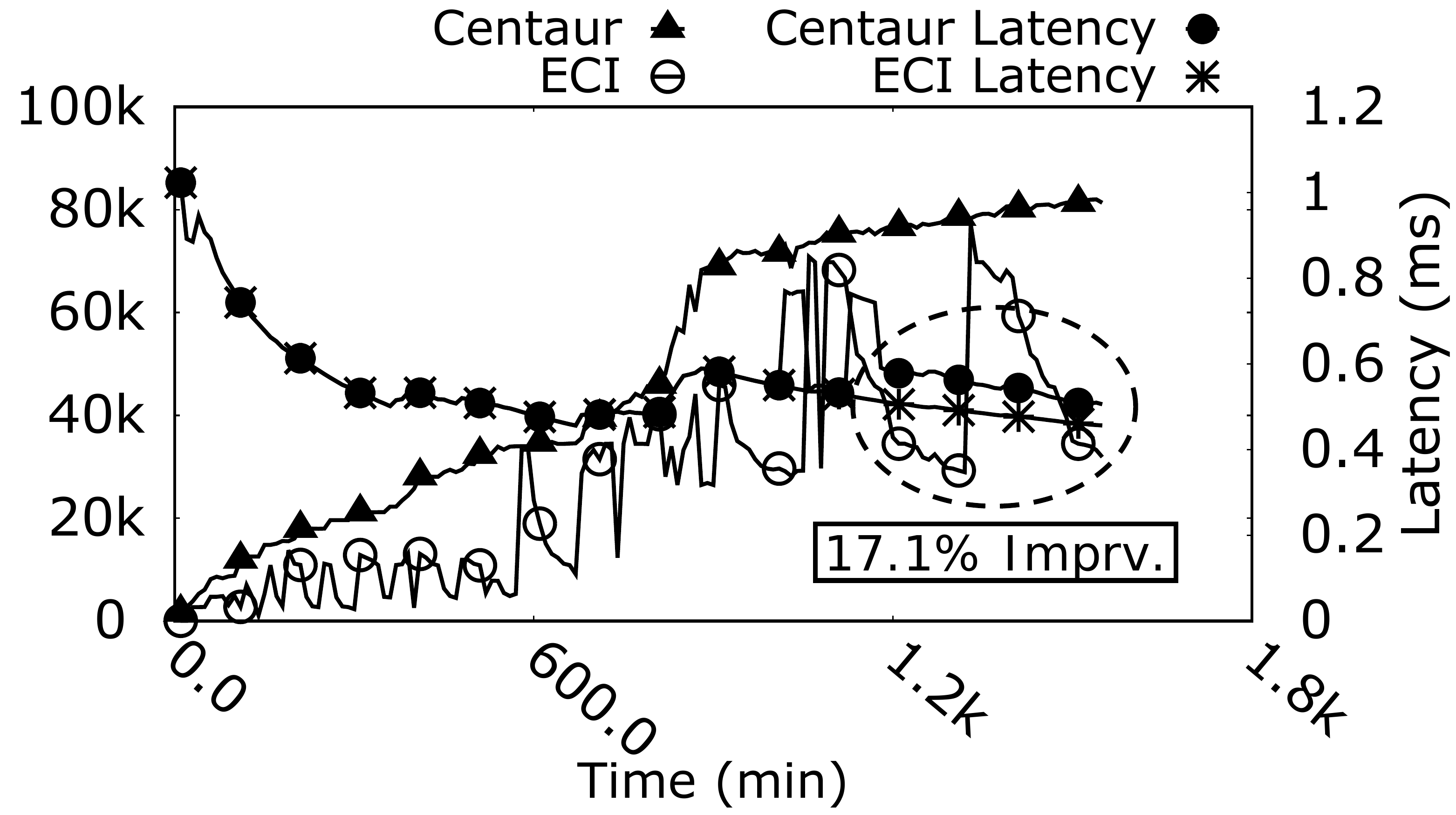}%
		\label{fig:src2_0}}
	\hfil
	\subfloat[VM12: web\underline{\hspace{.05in}}0]{\includegraphics[width=.25\textwidth]{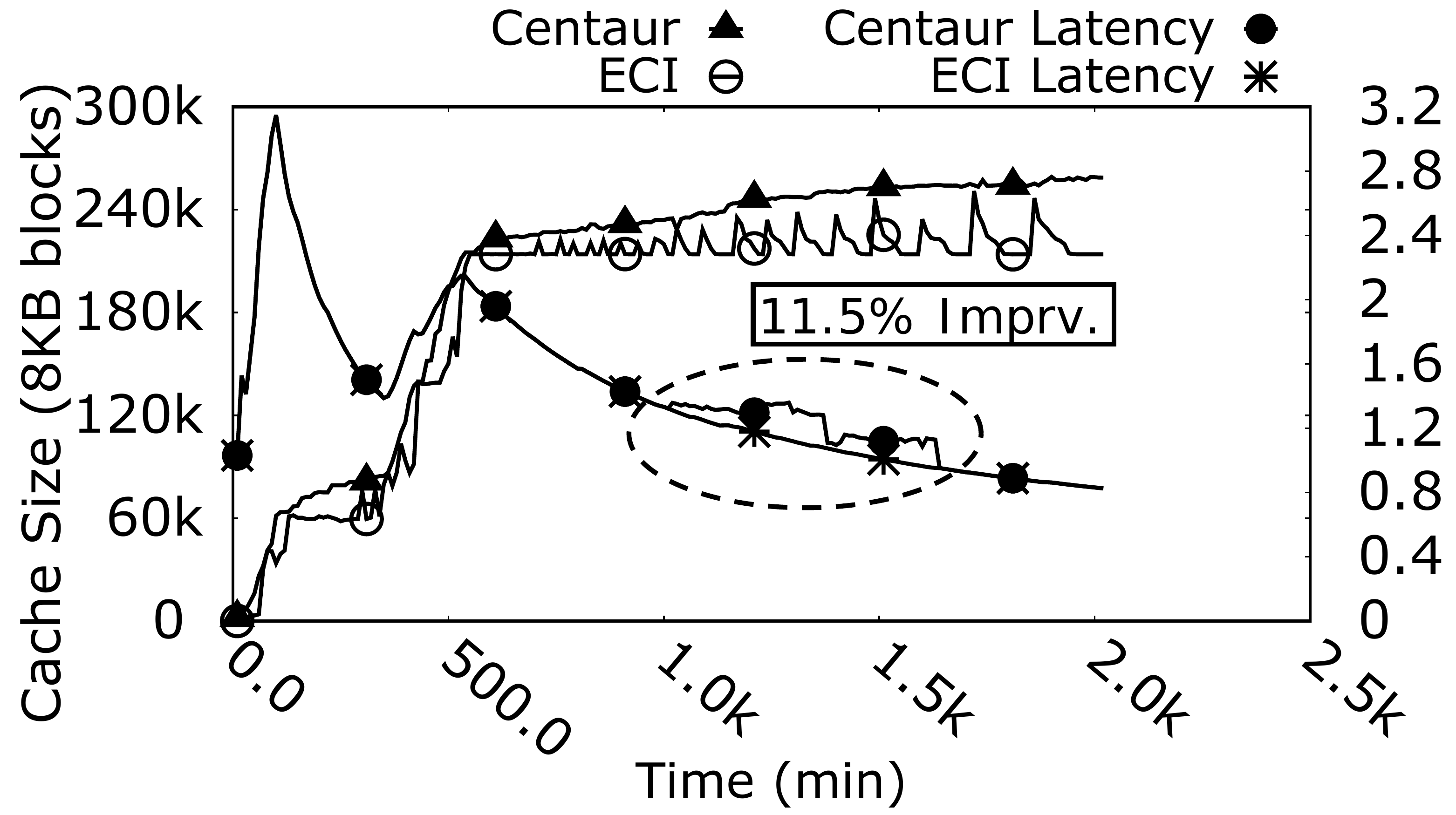}%
		\label{fig:web_0}}
	\hfil
	\subfloat[VM13: usr\underline{\hspace{.05in}}0]{\includegraphics[width=.25\textwidth]{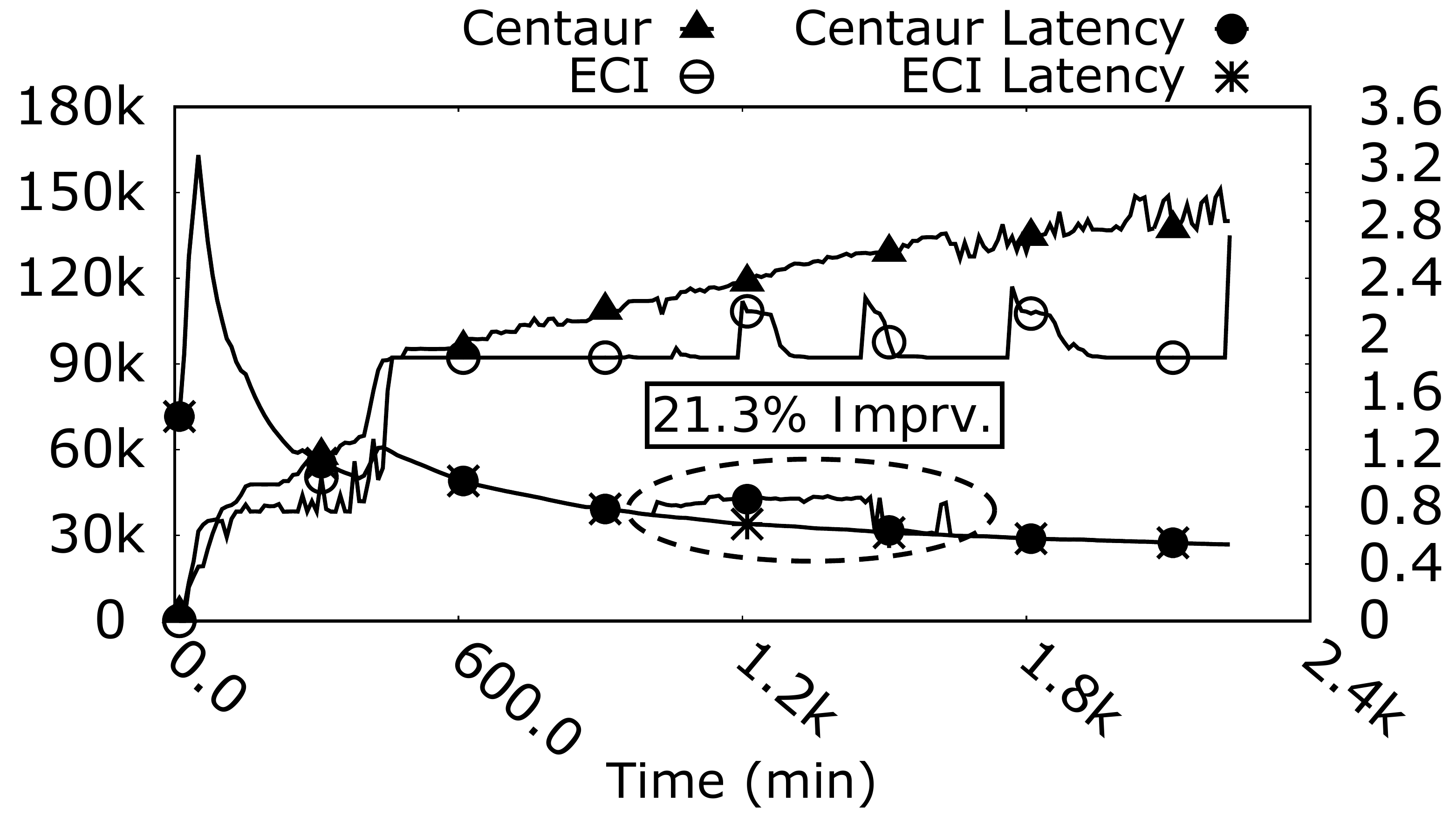}%
		\label{fig:usr_0}}
	\hfil
	\subfloat[VM14: rsrch\underline{\hspace{.05in}}2]{\includegraphics[width=.25\textwidth]{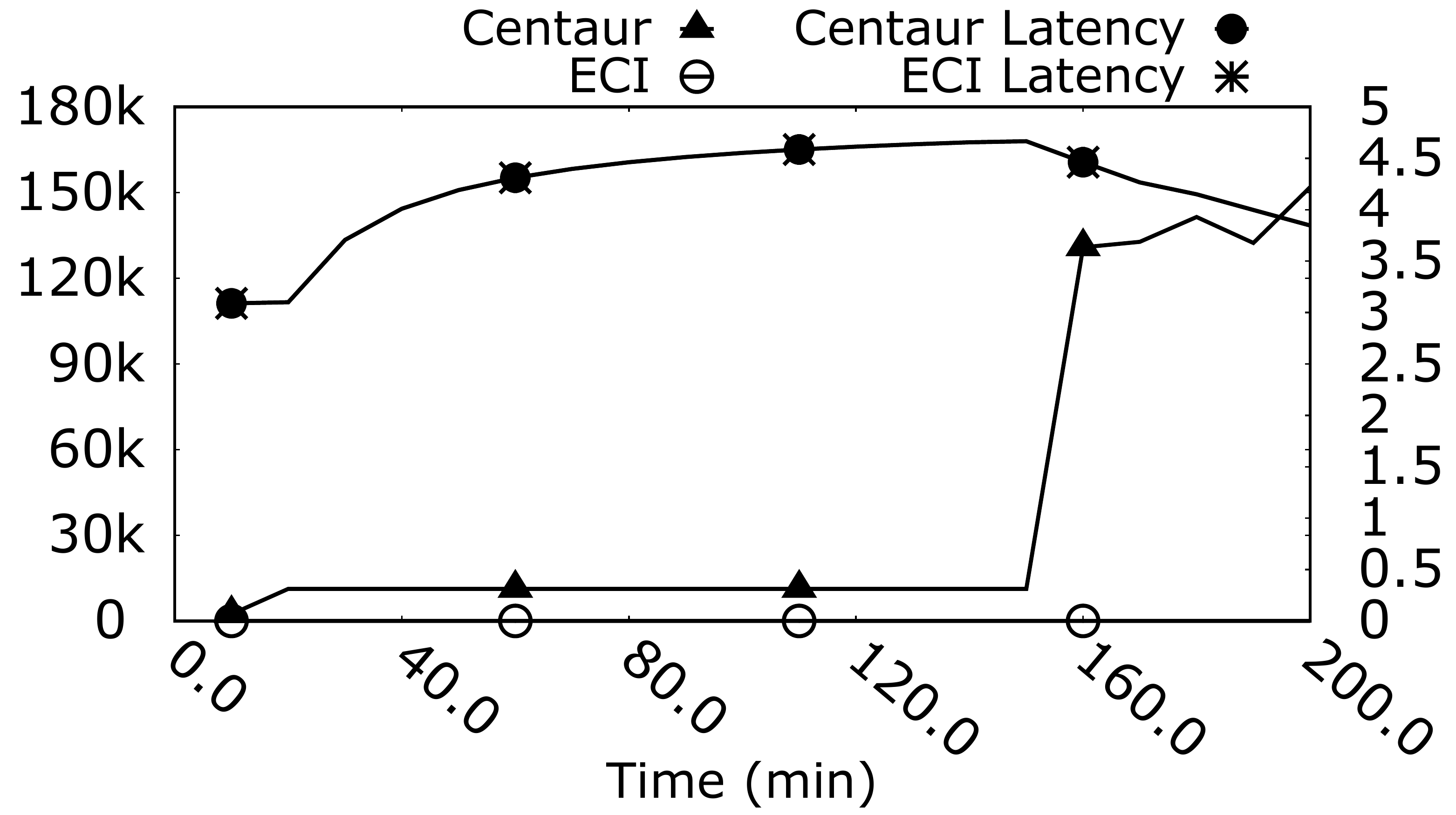}%
		\label{fig:rsrch_2}}
	\hfil
	\subfloat[VM15: mds\underline{\hspace{.05in}}1]{\includegraphics[width=.25\textwidth]{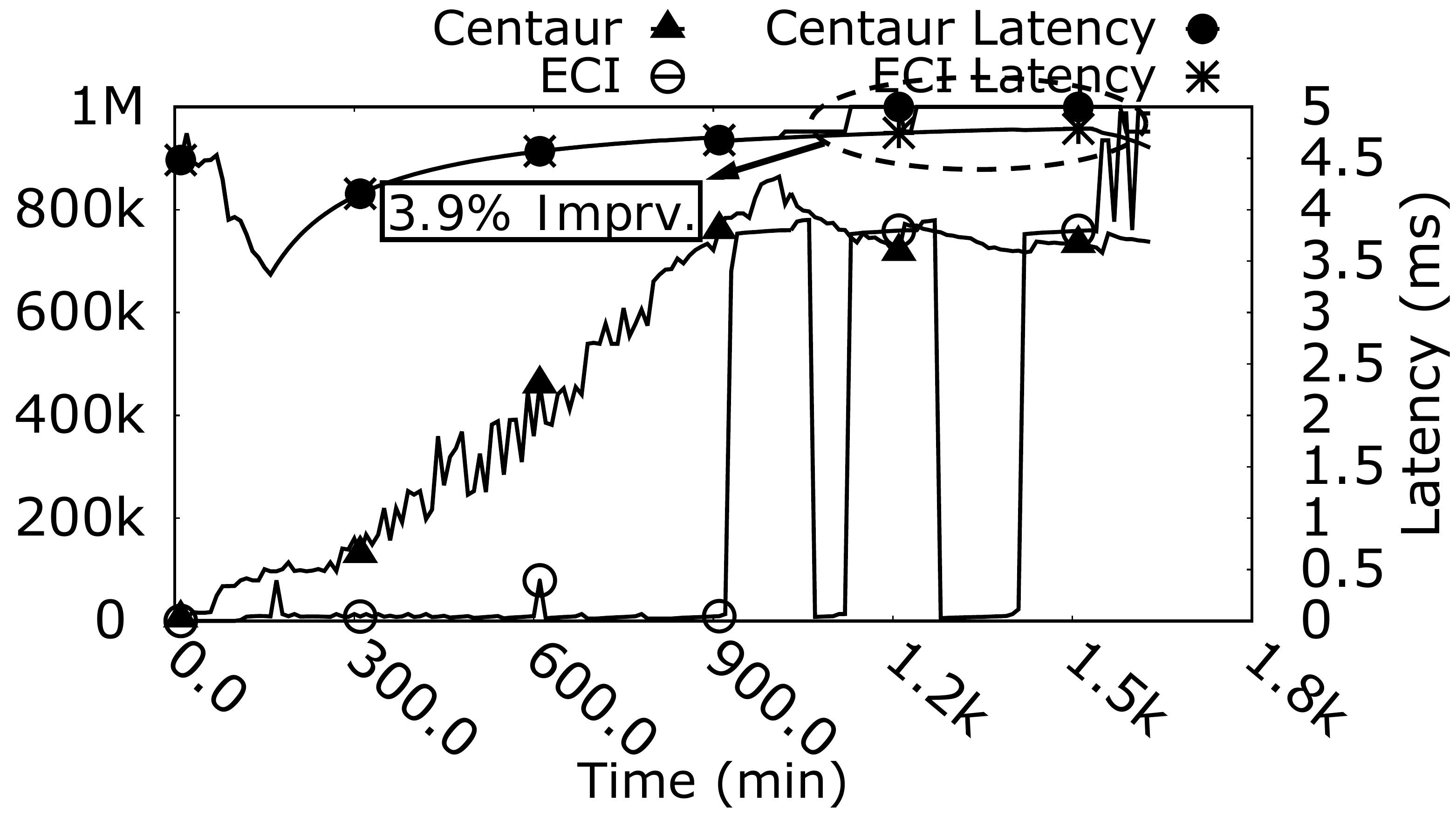}%
		\label{fig:mds_1}}
	\hfil
	
	\caption{{Allocated cache space of each VM and corresponding latency in infeasible state with limited SSD cache capacity (ECI-Cache vs. Centaur).}}
		\vspace{2em}
	\label{fig:TRD_URD_Hit}
\end{figure*}

{Fig. {\ref{fig:TRD_URD_Hit}} shows the details of allocated cache sizes for each individual VM by both Centaur and ECI-Cache. In addition, this figure shows the IO latency of the VMs for both schemes. During the experiment, the running workloads on the VMs finish one by one (because different workloads have different runtimes) and since VMs with the finished workload generates no more IO requests, ECI-Cache excludes those VMs from cache space partitioning. During the course of the experiment, we retrieve the allocated cache space of such VMs.
	In order to provide more detail, we present the results of each VM separately. The infeasible areas that lead to latency degradation in Centaur are shown by dashed circles around latency lines.
	We observe that in infeasible states, i.e., when  SSD cache capacity is limited, ECI-Cache improves performance and performance-per-cost by 17.08\% and 30\% compared to Centaur.}

\subsection{Write Policy Assignment}
\label{sec:write_pol_ass}

To show how ECI-Cache assigns an efficient write policy for each VM, we conduct experiments based on Algorithm \ref{alg:policy}. We characterize incoming requests from different VMs and calculate the ratio of \emph{WAW} and \emph{WAR} operations for the running VMs. Then, we assign an efficient write policy for the VM's cache. In the experiments, we set the value of $wThreshold$ to between 0.2 and 0.9 and achieve different results based on the value of $wThreshold$. Note that we assign the RO policy to a VM cache if the ratio of combined WAW and WAR requests over all requests is greater than or equal to $wThreshold$.

\begin{figure}[!h]
	\centering
	\includegraphics[scale=0.33]{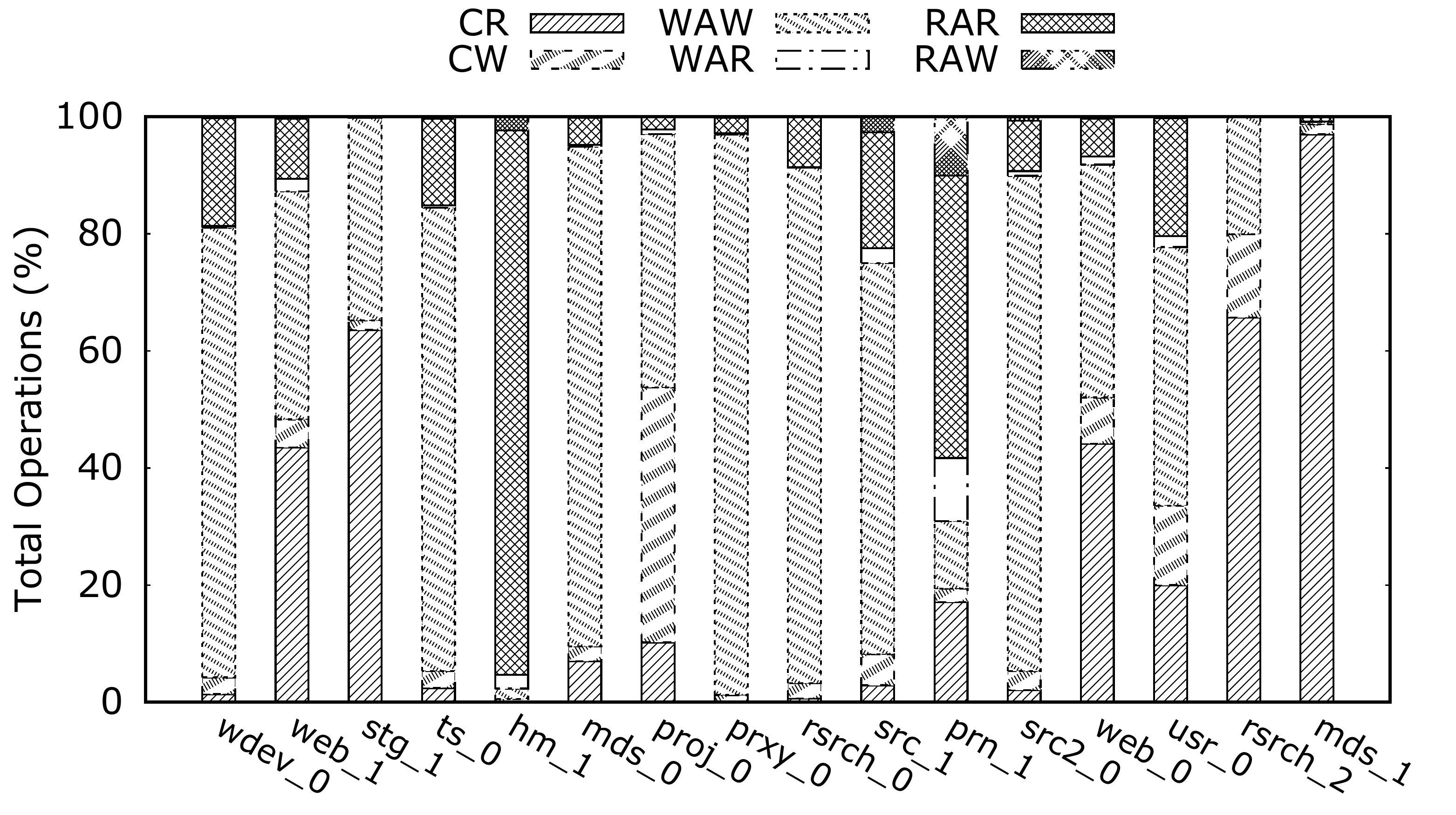}
	\caption{Ratio of different types of requests in the workloads.}
	\label{fig:all-vm-ratio}
\end{figure}

\begin{figure*}[!h]
	\centering
	\subfloat[VM0: wdev\underline{\hspace{.05in}}0]{\includegraphics[width=.25\textwidth]{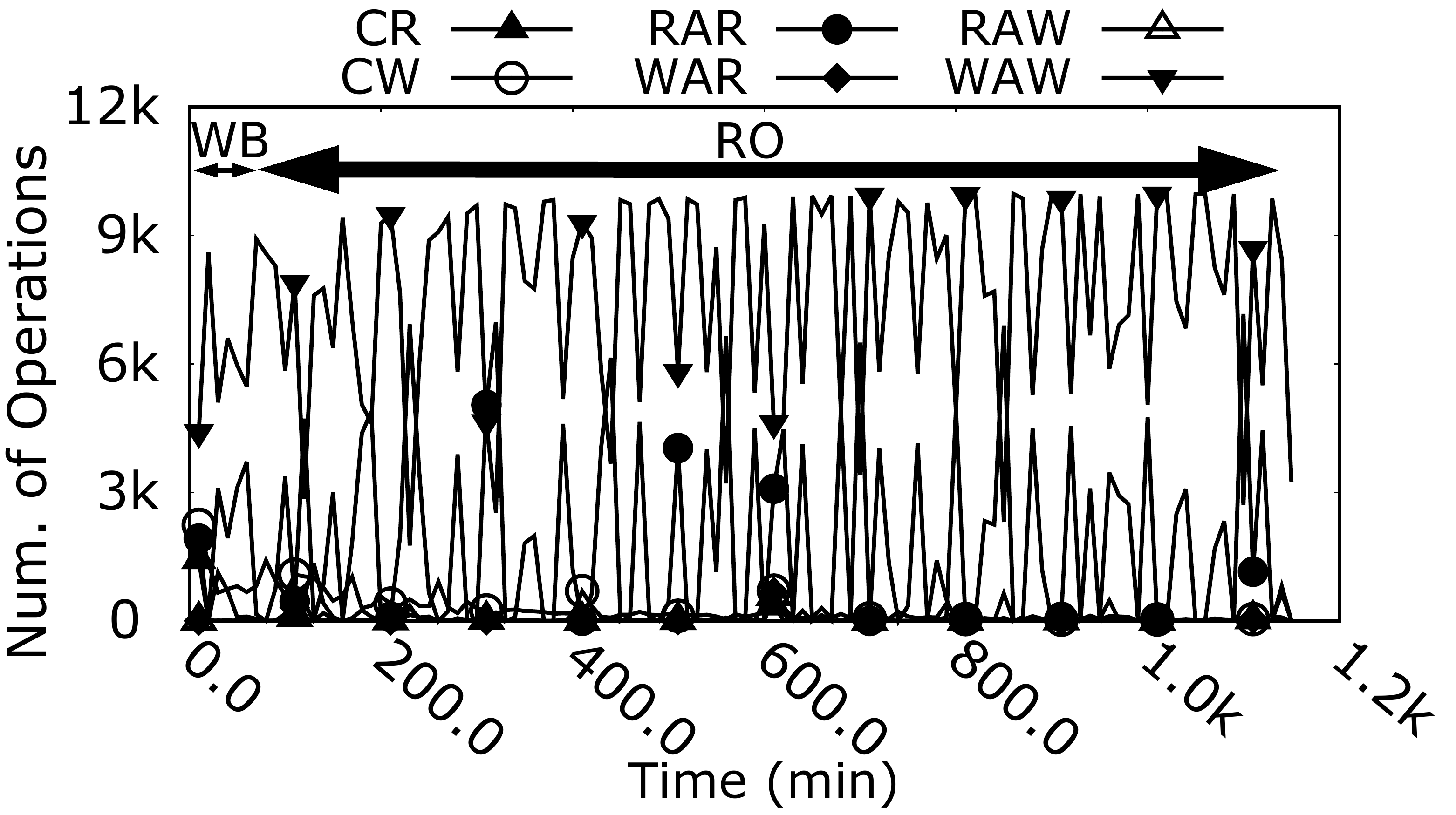}%
		\label{fig:wdev0-oper}}
	\hfil
	\subfloat[VM1: web\underline{\hspace{.05in}}1]{\includegraphics[width=.25\textwidth]{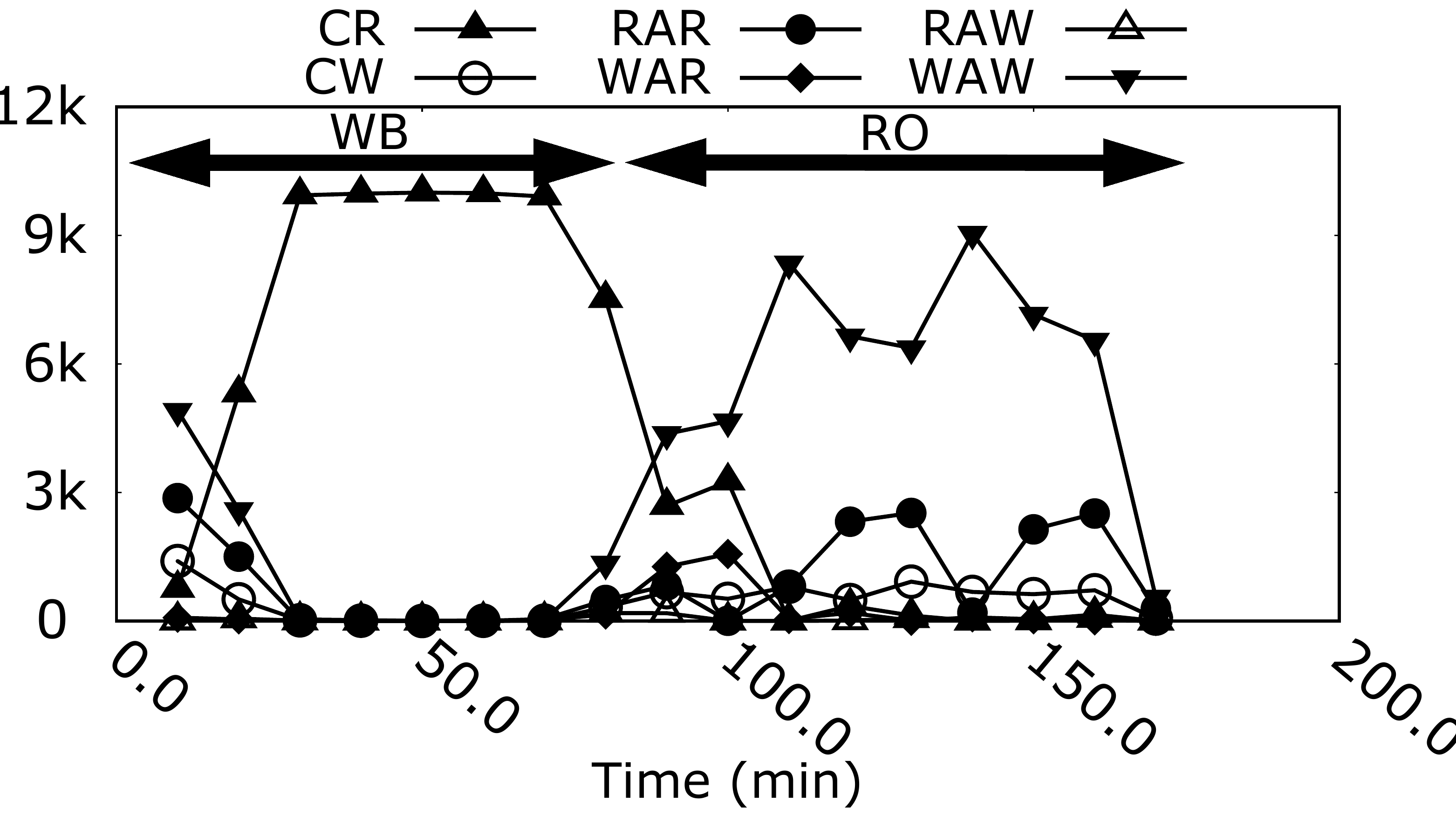}%
		\label{fig:web1-oper}}
	\hfil
	\subfloat[VM2: stg\underline{\hspace{.05in}}1]{\includegraphics[width=.25\textwidth]{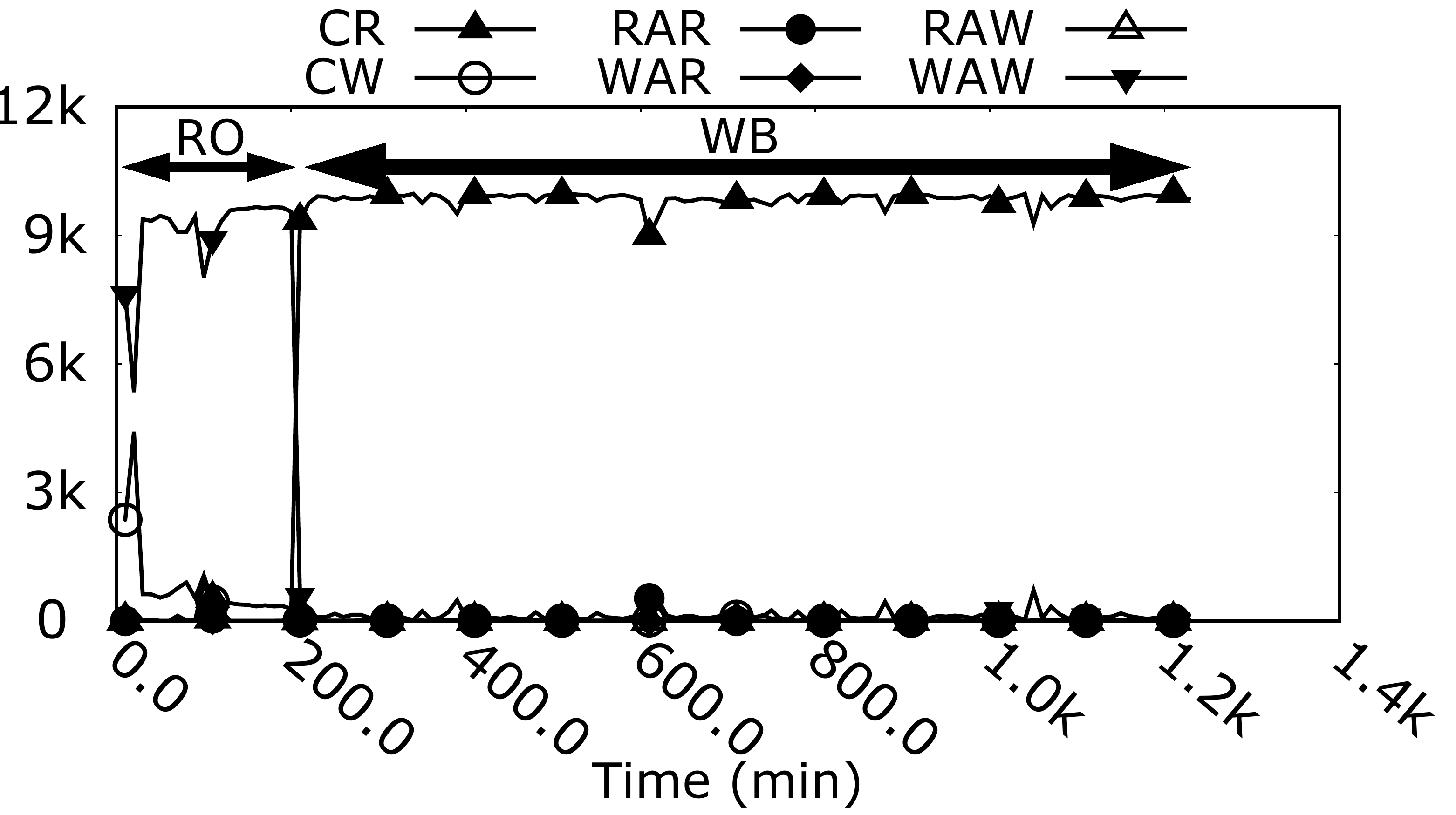}%
		\label{fig:stg1-oper}}
	\hfil
	\subfloat[VM3: ts\underline{\hspace{.05in}}0]{\includegraphics[width=.25\textwidth]{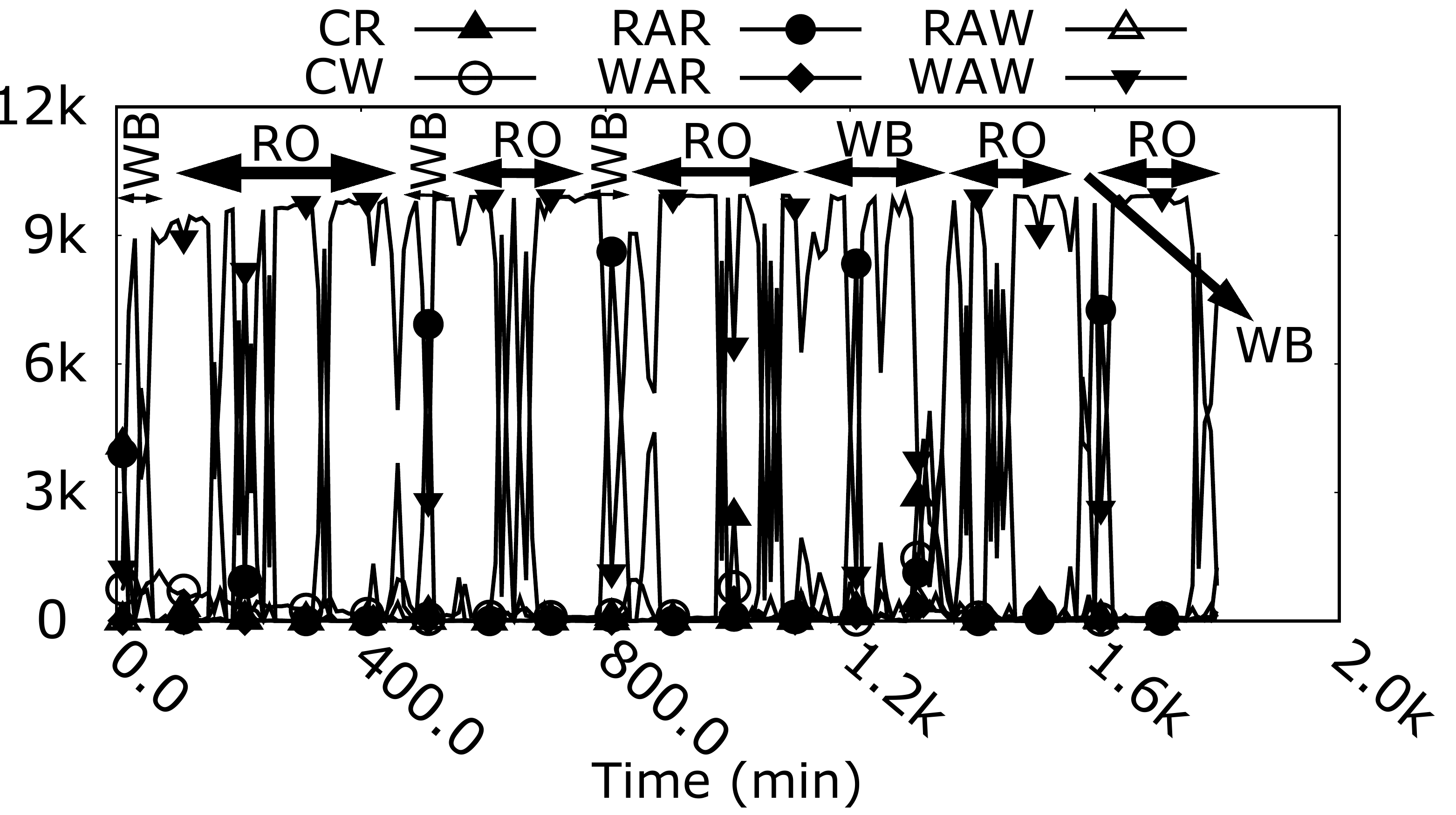}%
		\label{fig:ts0-oper}}
	\hfil
	\subfloat[VM4: hm\underline{\hspace{.05in}}1]{\includegraphics[width=.25\textwidth]{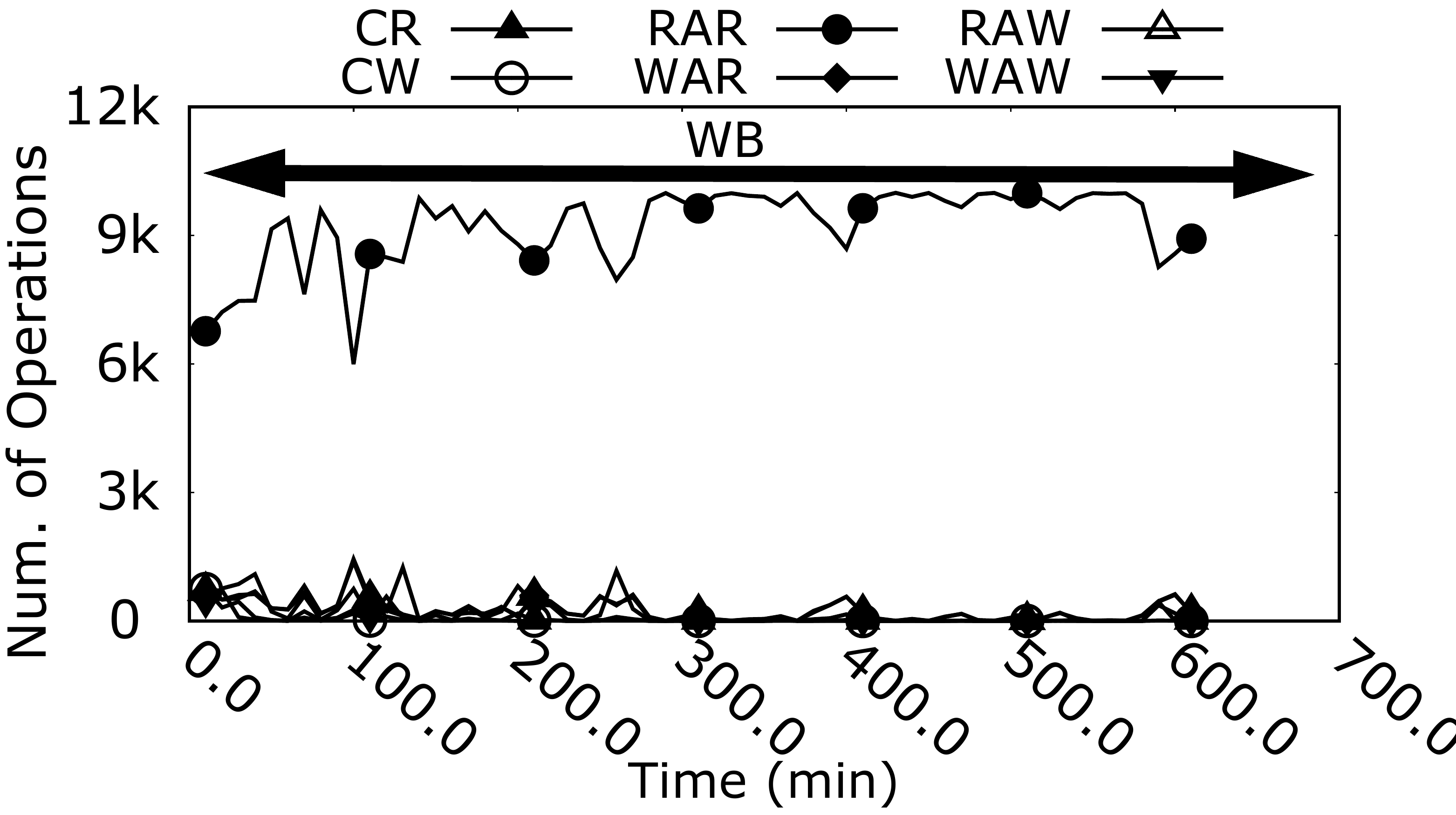}%
		\label{fig:hm1-oper}}
	\hfil
	\subfloat[VM5: mds\underline{\hspace{.05in}}0]{\includegraphics[width=.25\textwidth]{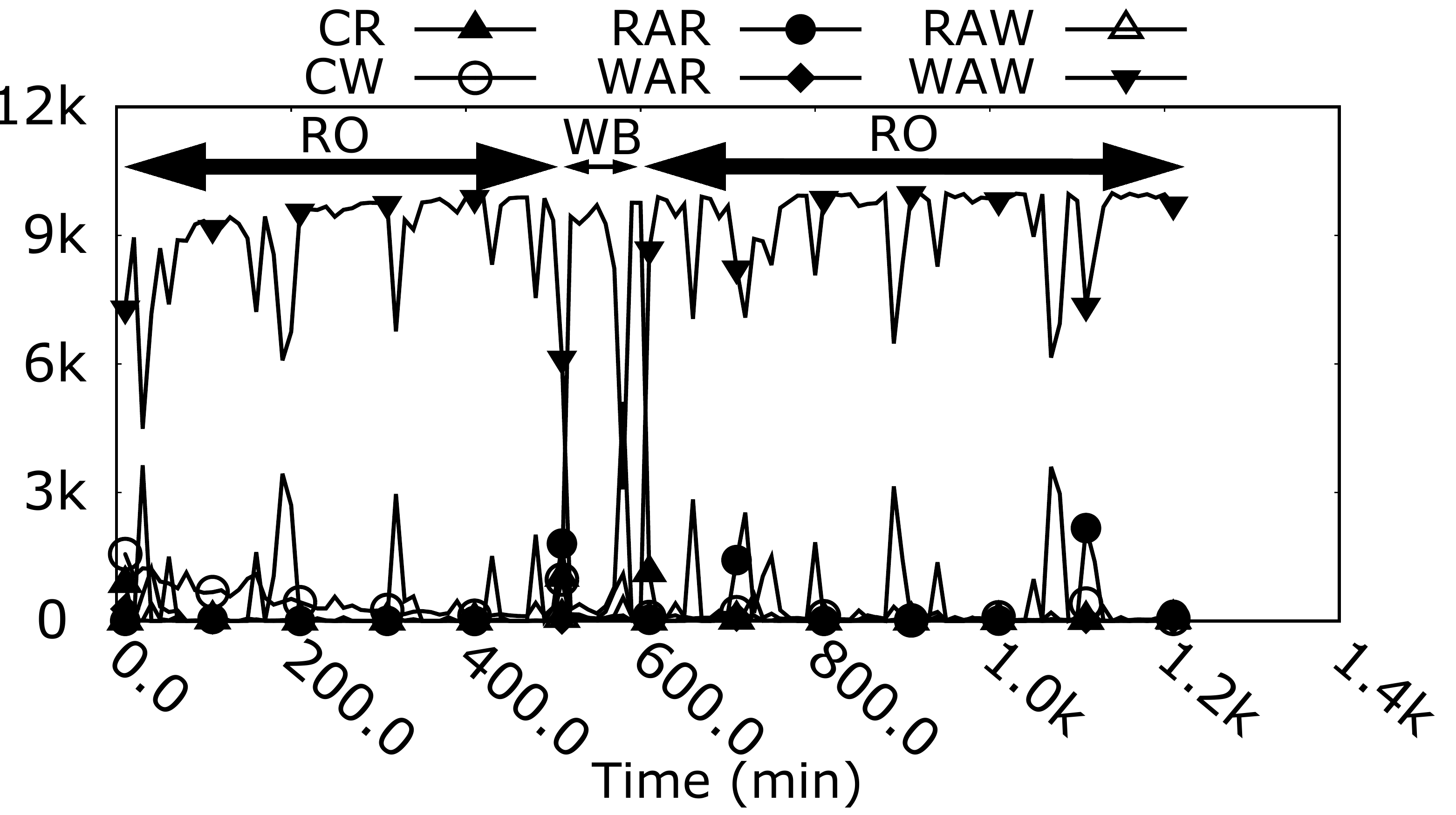}%
		\label{fig:mds_0-oper}}
	\hfil
	\subfloat[VM6:‌ proj\underline{\hspace{.05in}}0]{\includegraphics[width=.25\textwidth]{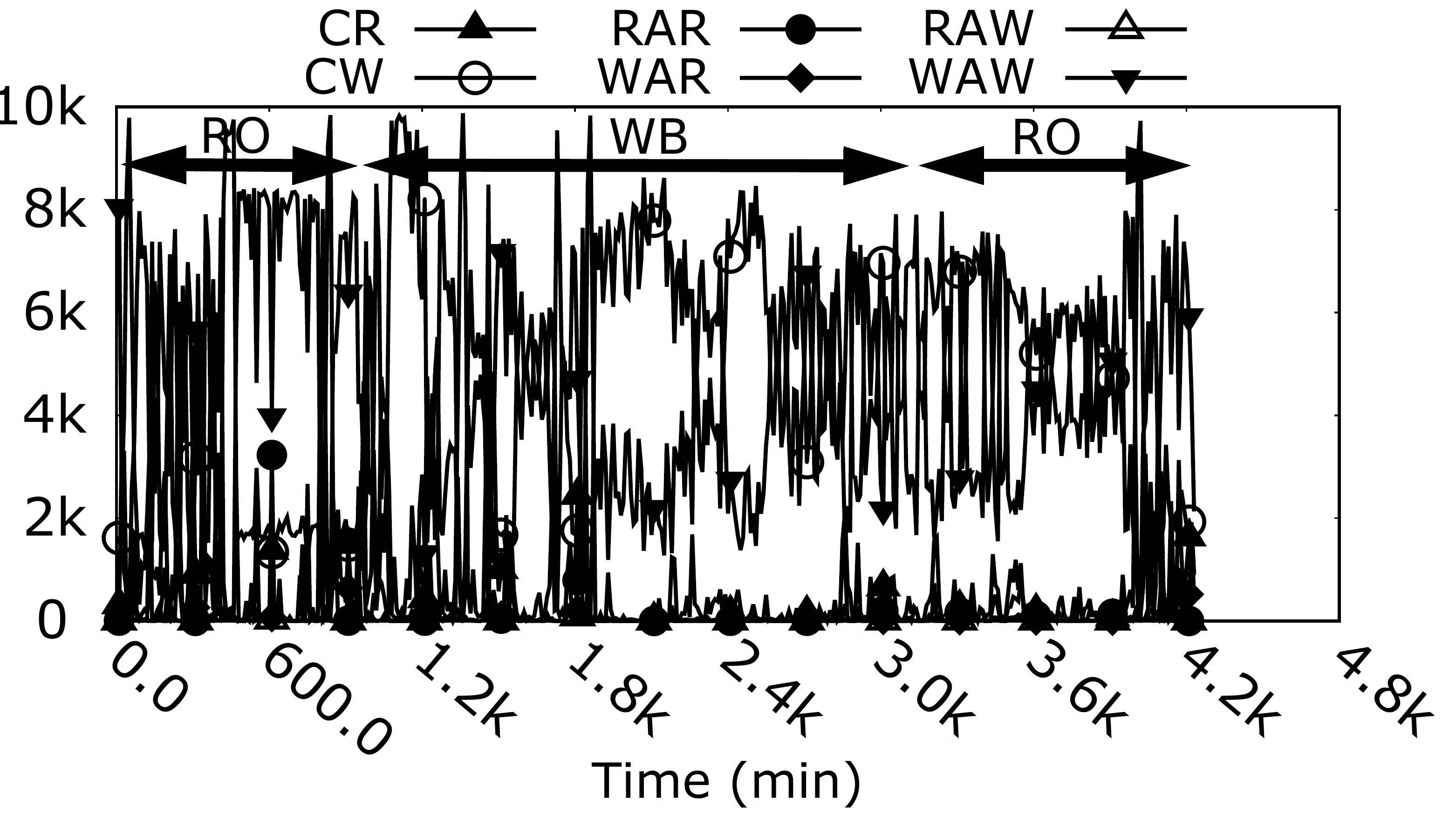}%
		\label{fig:proj_0-oper}}
	\hfil
	\subfloat[VM7: prxy\underline{\hspace{.05in}}0]{\includegraphics[width=.25\textwidth]{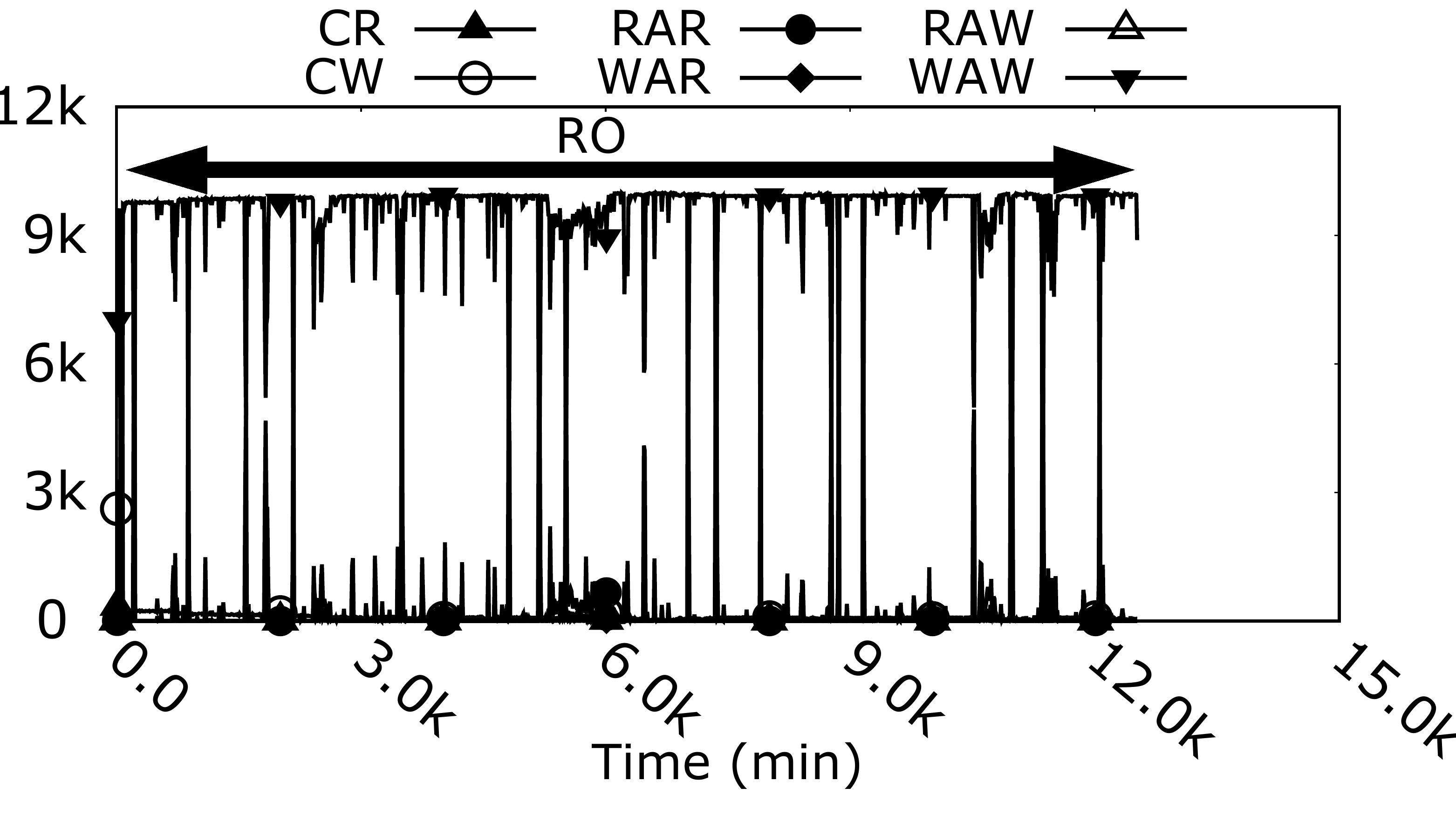}%
		\label{fig:prxy_0-oper}}
	\hfil
	\subfloat[VM8: rsrch\underline{\hspace{.05in}}0]{\includegraphics[width=.25\textwidth]{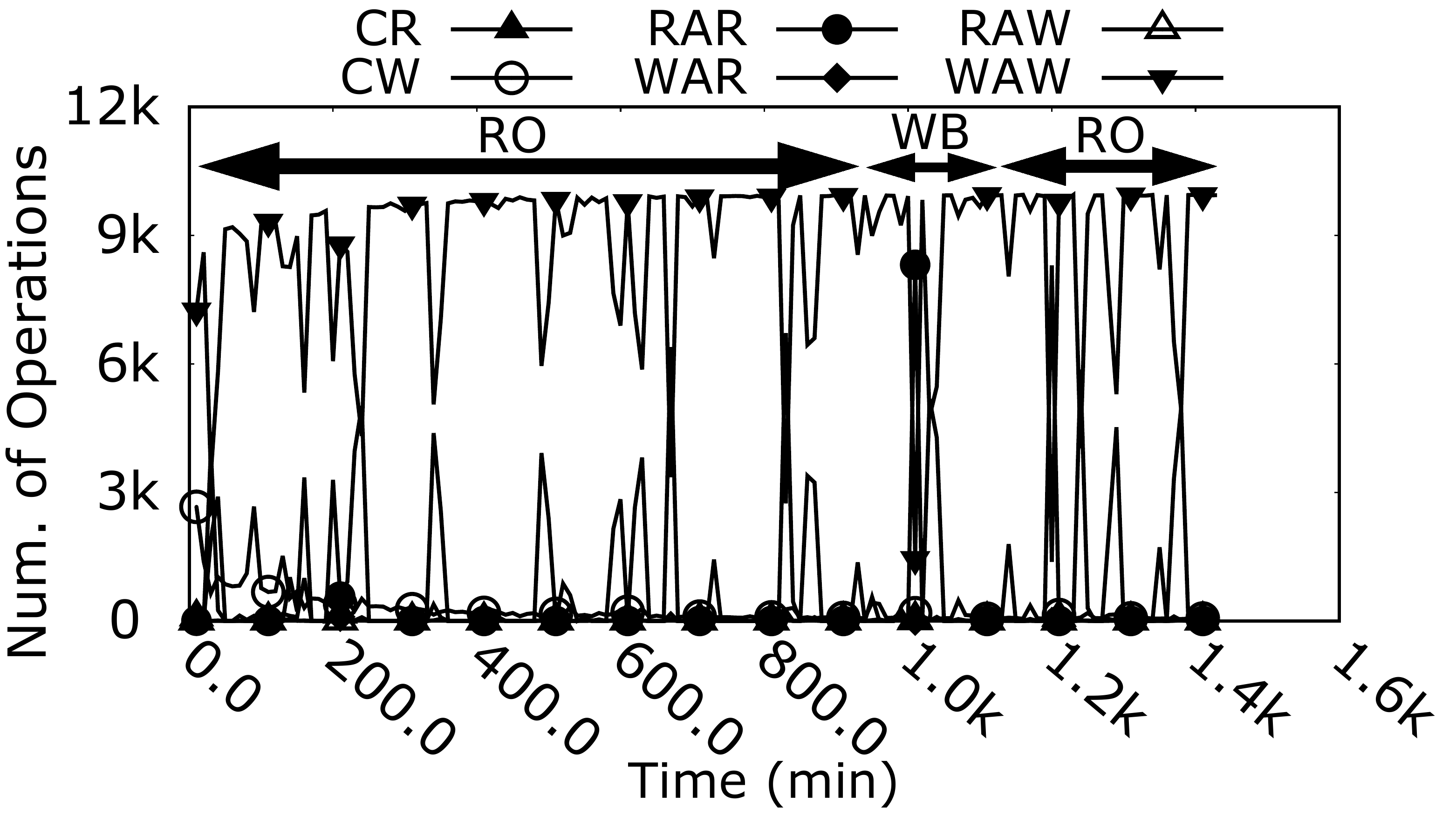}%
		\label{fig:rsrch_0-oper}}
	\hfil
	\subfloat[VM9: src1\underline{\hspace{.05in}}2]{\includegraphics[width=.25\textwidth]{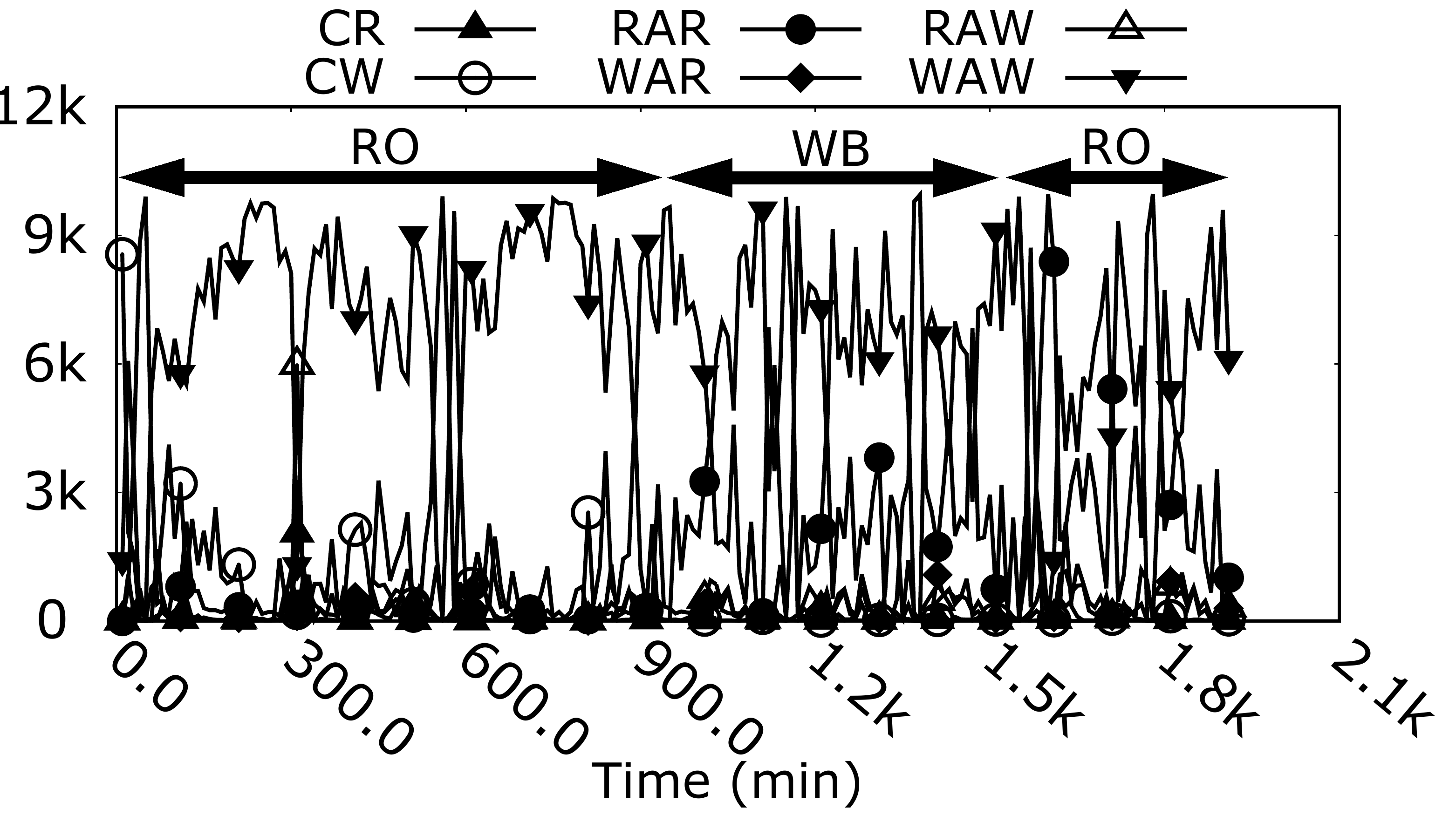}%
		\label{fig:src_1-oper}}
	\hfil
	\subfloat[VM10: prn\underline{\hspace{.05in}}1]{\includegraphics[width=.25\textwidth]{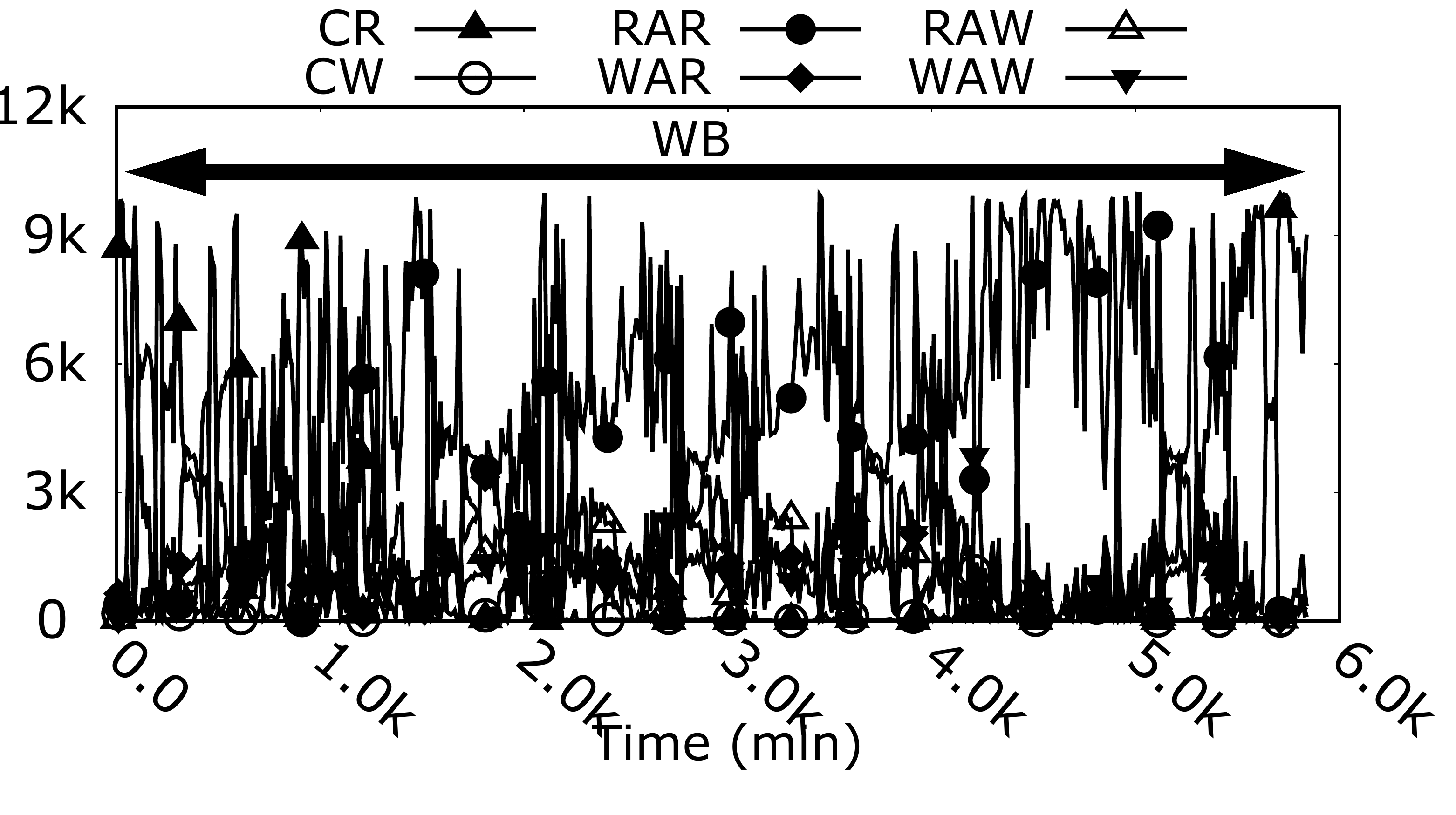}%
		\label{fig:prn_1-oper}}
	\hfil
	\subfloat[VM11: src2\underline{\hspace{.05in}}0]{\includegraphics[width=.25\textwidth]{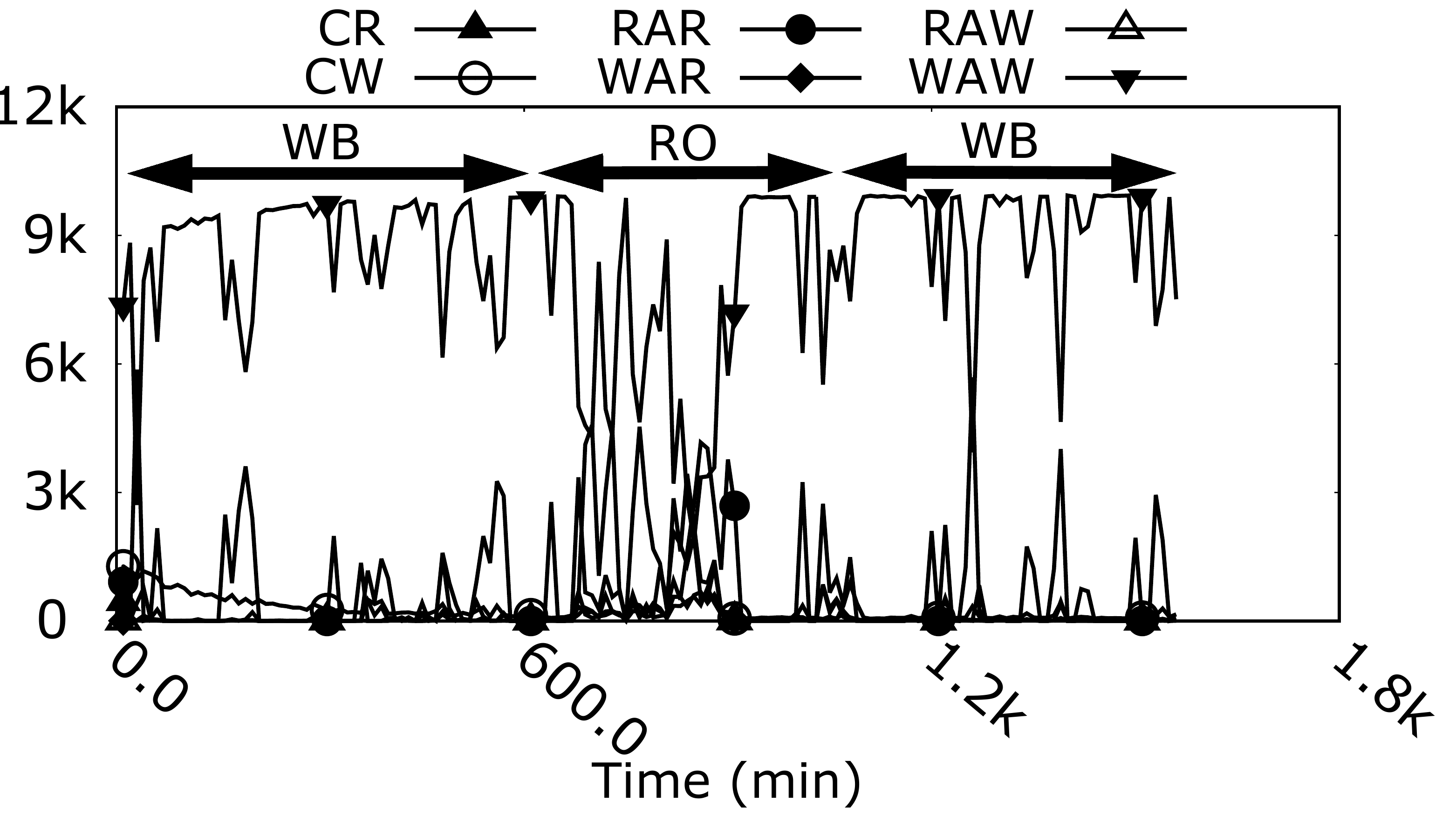}%
		\label{fig:src2_0-oper}}
	\hfil
	\subfloat[VM12: web\underline{\hspace{.05in}}0]{\includegraphics[width=.25\textwidth]{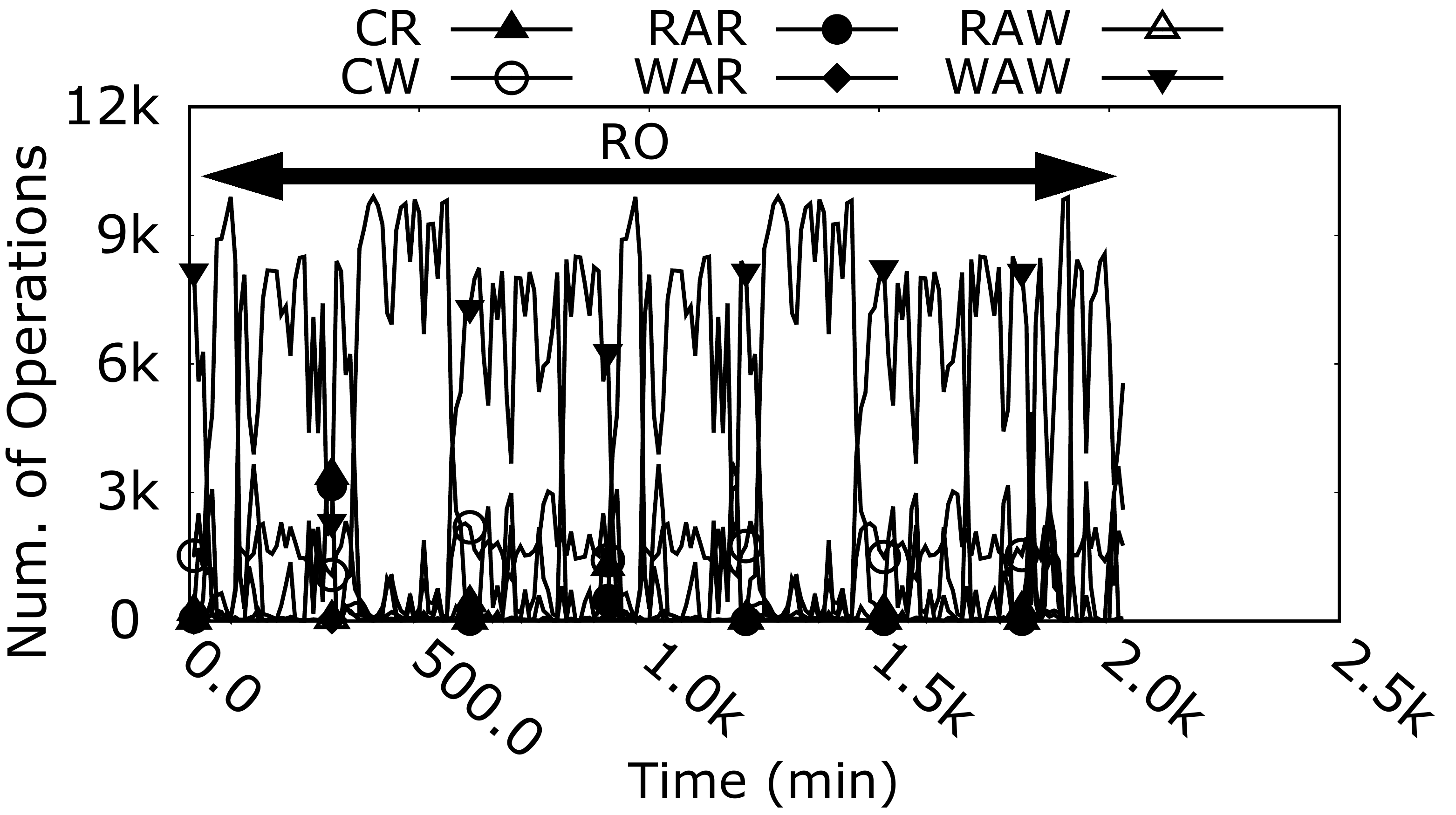}%
		\label{fig:web_0-oper}}
	\hfil
	\subfloat[VM13: usr\underline{\hspace{.05in}}0]{\includegraphics[width=.25\textwidth]{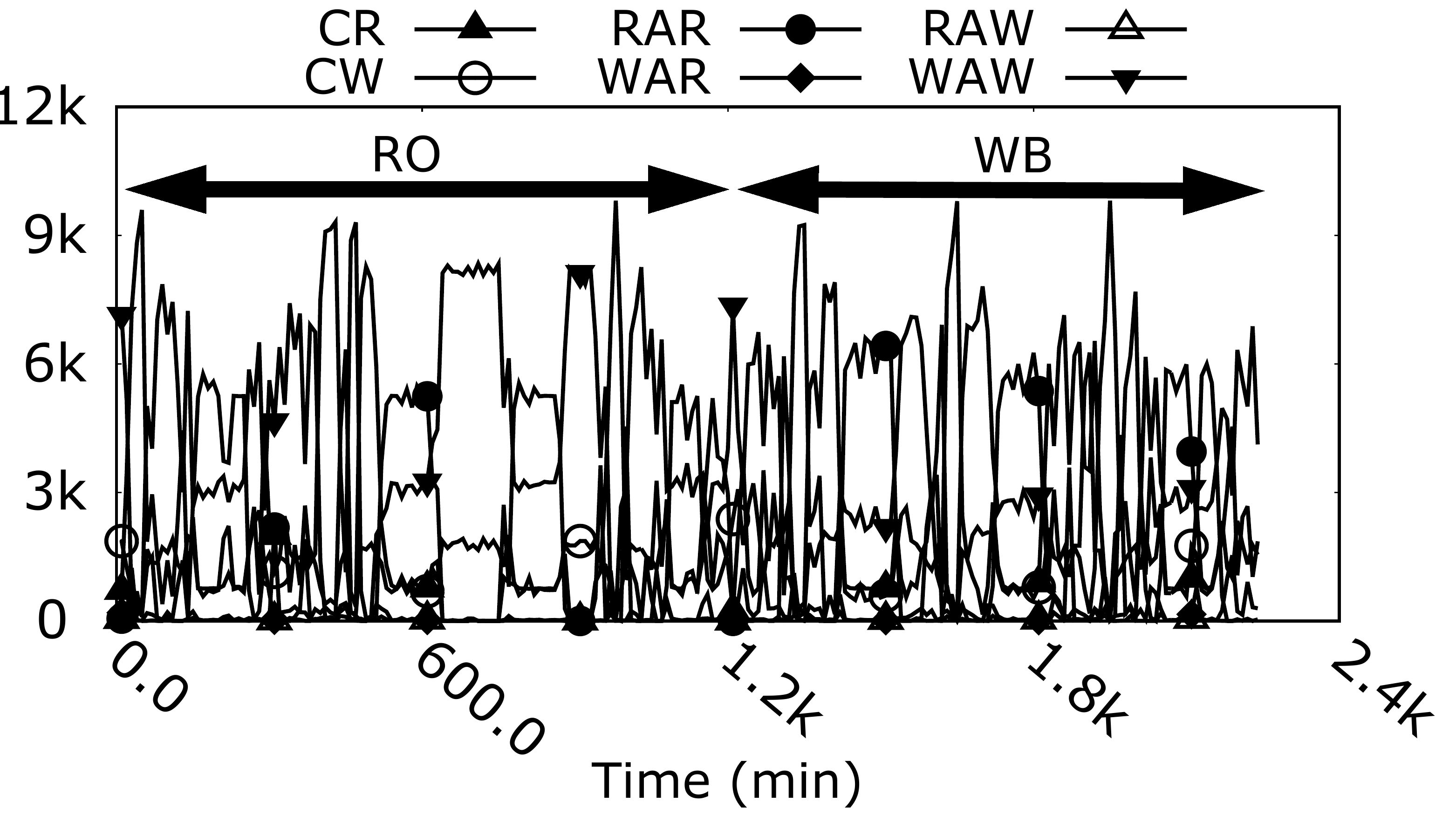}%
		\label{fig:usr_0-oper}}
	\hfil
	\subfloat[VM14: rsrch\underline{\hspace{.05in}}2]{\includegraphics[width=.25\textwidth]{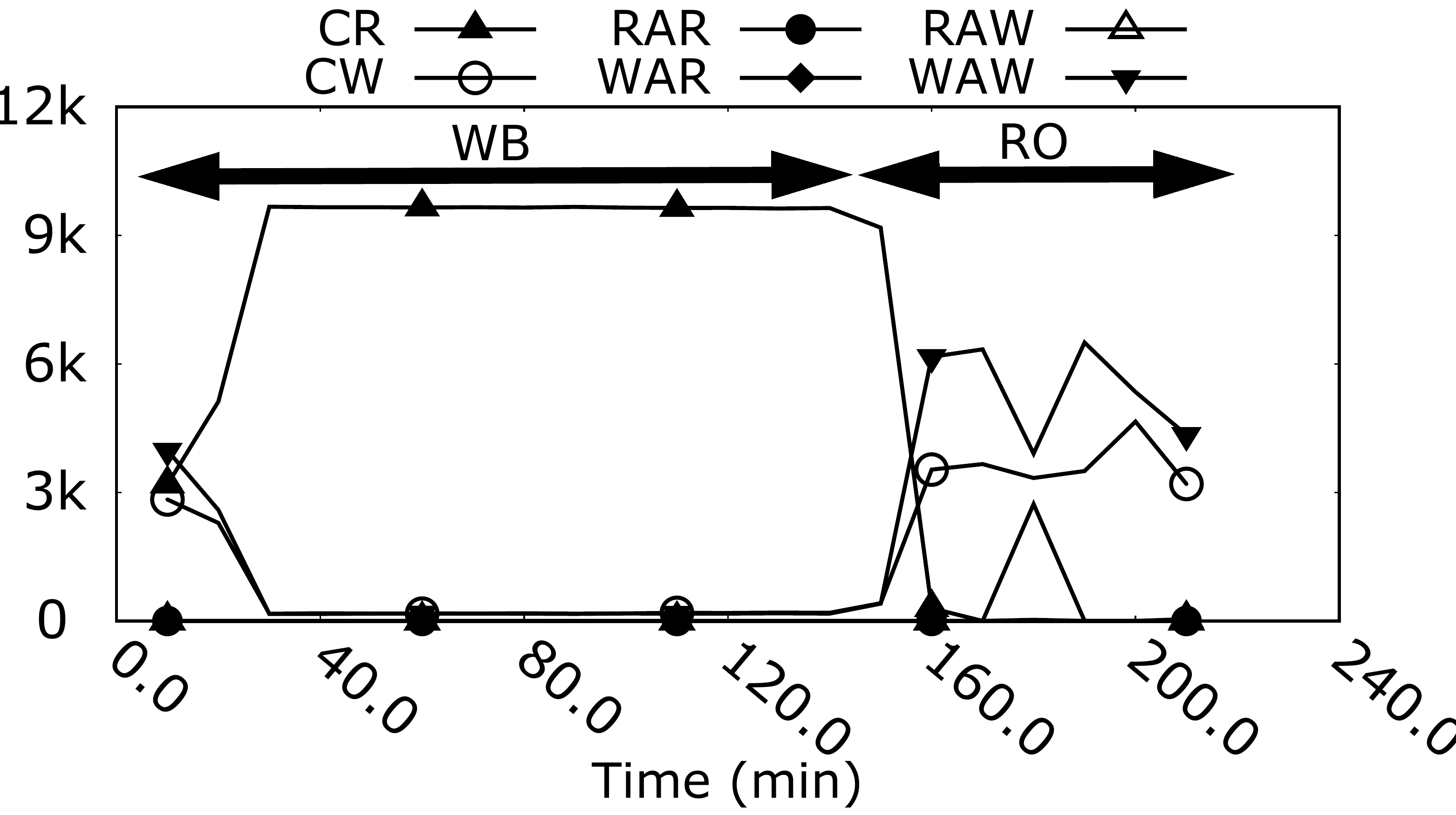}%
		\label{fig:rsrch_2-oper}}
	\hfil
	\subfloat[VM15: mds\underline{\hspace{.05in}}1]{\includegraphics[width=.25\textwidth]{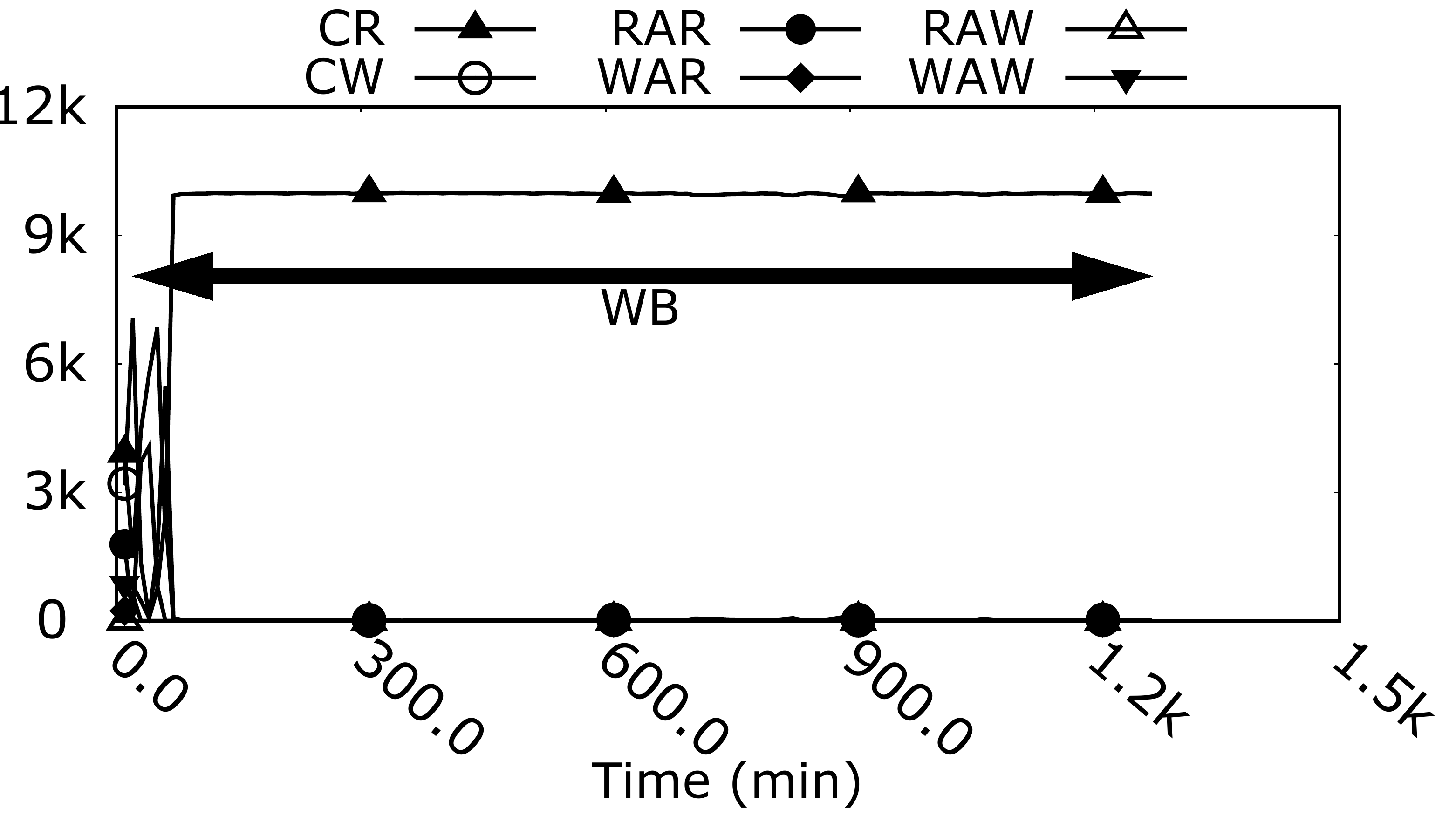}%
		\label{fig:mds_1-oper}}
	\hfil
	
	\caption{\techname{} write policy assignment to the VMs.}
	\label{fig:oper-per-time}
\end{figure*}

Fig. \ref{fig:all-vm-ratio} shows the ratio of different types of the requests in the running workloads. In addition, Fig. \ref{fig:oper-per-time} shows the number of WAW, WAR, RAR, and RAW accesses for the workloads of the running VMs, which are sampled in 10-minute intervals. 
Since the first access (either read or write request) to an arbitrary address within a sequence of I/O requests cannot be classified as either WAW, WAR, RAR, and RAW,
we denoted the first read and write access to an address as \emph{Cold Read} (CR) and \emph{Cold Write} (CW), respectively. 
Here we set $wThreshold=0.5$. 

{It can be seen that in the first 70-minute interval of web\underline{\hspace{.05in}}1 (running on VM1 as shown in Fig. \ref{fig:web1-oper} and Fig. \ref{fig:all-vm-ratio}), 
	neither WAW nor WAR accesses exist  and \techname{} assigns the WB policy for the allocated cache. Then at $t=100~min$, WAW operations become dominant and the RO policy is assigned to the cache. In VM0, with the running workload of wdev\underline{\hspace{.05in}}0, after about 50 minutes, we recognize that 77\% of the requests are WAW‌ and the remaining are mostly RAR (shown in Fig. \ref{fig:wdev0-oper} and Fig. \ref{fig:all-vm-ratio}) and thus at $t=50~min$, the RO policy is assigned to the cache of VM0 by \techname{}. As shown in Fig. \ref{fig:hm1-oper} and Fig. \ref{fig:all-vm-ratio}, hm\underline{\hspace{.05in}}1 running on VM4 consists of mostly RAR‌ operations (more than 92\%) without any WAR and WAW accesses and thus the RO policy is assigned to this VM.
	In time intervals between $t=0$ to $t=500~min$ and $t=610~min$ to $t=1200~min$, more than 86\% of the requests of proj\underline{\hspace{.05in}}0 running on VM6 are CW and WAW operations, and thus \techname{} assigns the RO policy to this VM. In the remaining interval ($t=500~min$ to $t=610~min$), the WB policy is assigned to this VM. \techname{} assigns the RO policy for VMs such as prxy\underline{\hspace{.05in}}0 and web\underline{\hspace{.05in}}0 that have a large number of WAW‌ and WAR operations. Doing so minimizes the number of unnecessary writes into the cache in these workloads.}

\subsection{Performance and Performance-Per-Cost Improvement}
\label{sec:perf_imp}
\begin{figure}[!htb]
	\centering
	\includegraphics[scale=0.42]{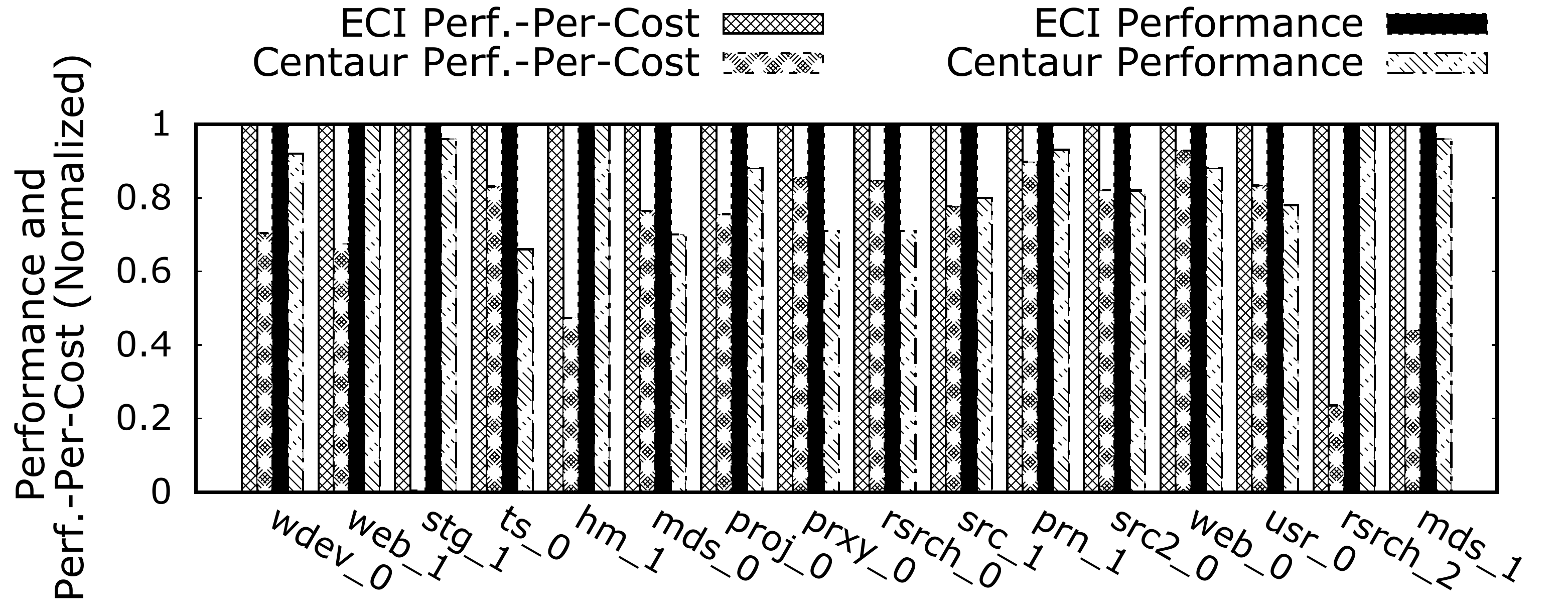}
	\caption{{Performance and performance-per-cost achieved by \techname{} and Centaur.}}
	\label{fig:perf-per-cost}
\end{figure}

The results of previous experiments on the proposed test platform indicate a significant performance and performance-per-cost {(i.e., performance per allocated cache space)} improvement for the running workloads on the VMs, as quantified in Fig. \ref{fig:perf-per-cost}. We observed that \techname{} is able to estimate a smaller cache space for each VM than Centaur, without any negative impact on performance. {In other words, ECI-Cache achieves similar hit ratio while allocating smaller cache space and hence improves performance-per-cost. In addition, \techname{} achieves higher performance compared to Centaur in infeasible cases for each VM.}
{Cache size estimation in the proposed scheme is based on the URD metric and thus ECI-Cache allocates much smaller cache space to the VMs compared to Centaur. Centaur estimates cache space based on TRD, which does \emph{not} consider the request type, leading to a higher cache size estimation for each VM.}

Fig. \ref{fig:perf-per-cost} shows the achieved performance and performance-per-cost of the VMs when we use the Centaur and \techname{} schemes. We observe that allocating cache space based on \techname{} for each VM improves the performance and performance-per-cost compared to Centaur in all workloads.

\begin{figure}[!h]
	\centering
	\includegraphics[scale=0.35]{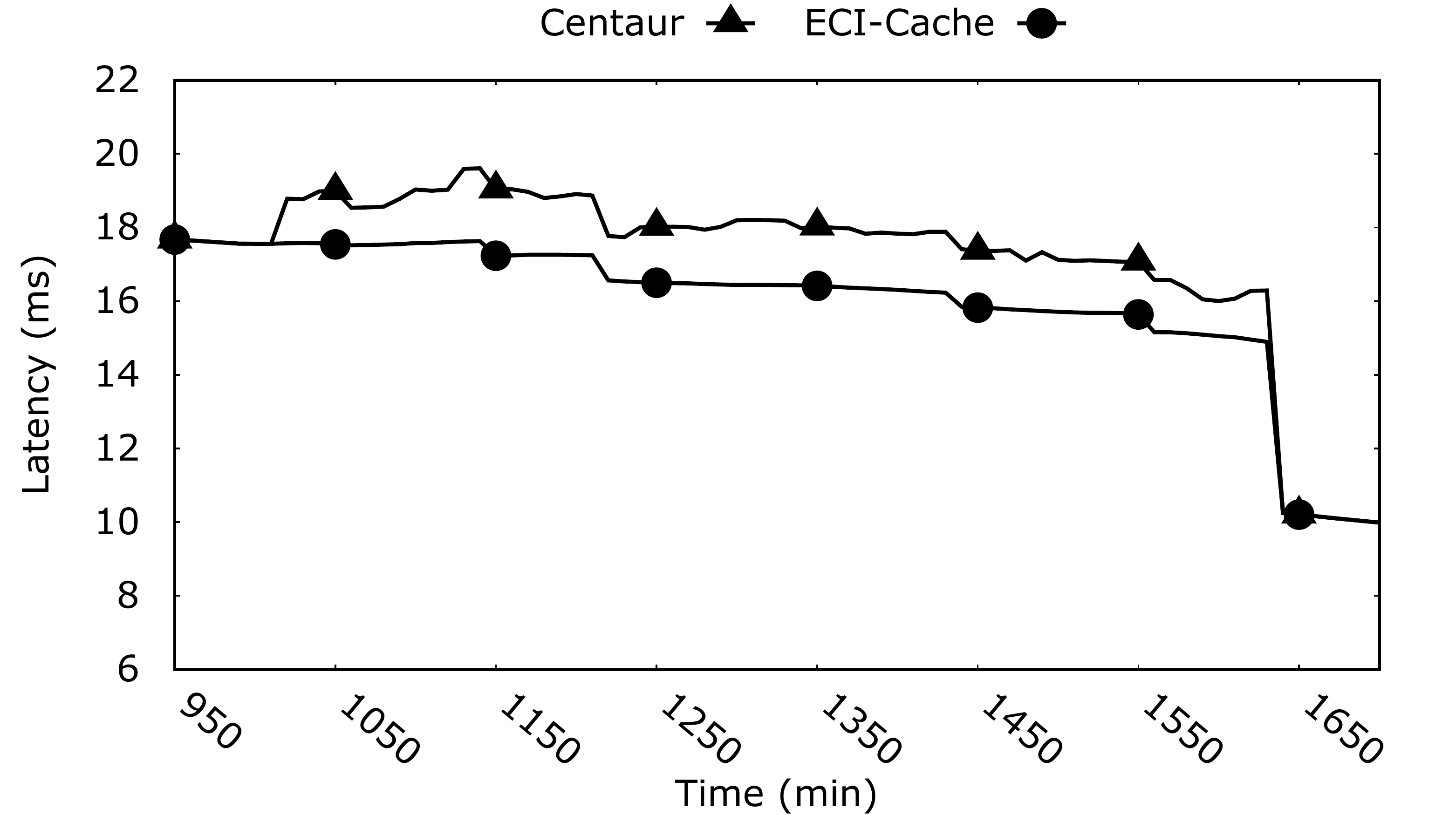}
	\caption{{Cumulative latency of VMs with ECI-Cache and Centaur in infeasible states.}}
	\label{fig:cum_lat}
\end{figure}

{Fig. \ref{fig:cum_lat} shows the cumulative latency of the running VMs with Centaur vs. ECI-Cache in \emph{infeasible} states (i.e., when the total SSD cache capacity is limited). In the time interval shown in Fig. \ref{fig:cum_lat}, 
	ECI-Cache achieves a higher hit ratio than Centaur. Therefore, in \emph{infeasible} states ECI-Cache reduces the latency of the workloads by 17\%, on average.} We conclude that \techname{} improves performance and performance-per-cost for the running VMs by 17.08\% and 30\%, by intelligently reducing the allocated cache space for each VM.

\subsection{Endurance Improvement}
\begin{figure}[!htb]
	\centering
	\includegraphics[scale=0.4]{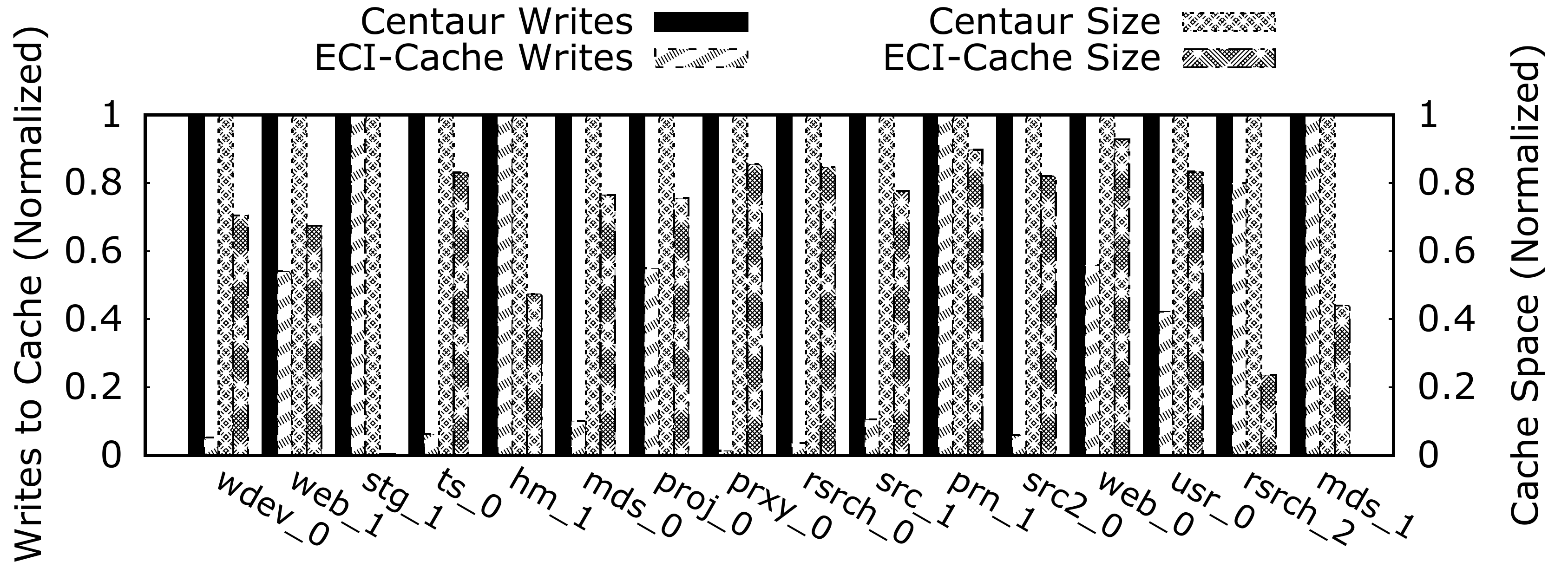}
	\caption{{Writes into the SSD cache and allocated cache space to the VMs by Centaur and \techname{}.}}
	\label{fig:writes_on_cache_with_RO}
\end{figure}

{To show the endurance improvement of \techname{}, we perform experiments by applying our write policy assignment algorithm and show the impact of the proposed scheme on the number of writes and also the performance of the VMs.
	Endurance of the SSD is affected by the number of writes committed to it.‌ Write operations on the SSD impose NAND flash memory writes and increase the P/E-cycle count {\cite{ssd-lifetime,tarihi2016hybrid,cai2017error,cai2012error,cai2013error,meza2015large}}. ECI-Cache has a positive impact on the endurance of the SSD because it \emph{reduces} the number of committed writes to the SSD. While ECI-Cache can effectively manage the committed writes to the SSD, it has no control on the writes initiated by the garbage collection and wear-leveling algorithms used within the SSD.
	Hence, we report endurance improvement of the SSD cache by using the reduced number of writes as a metric.
	A smaller number of writes is expected to lead to better endurance.
	Similar metrics are used in previous system-level studies, such as {\cite{anlj17,elasticQ,huang2016improving,liu2014plc}}.}
Note that the total number of writes for each workload (reported in the experiments) is calculated by Eq. \ref{equ:tot_writes} which includes writes from the disk subsystem to the SSD and also the writes from the CPU to the SSD:
\begin{equation} 
Total~Writes=\sum {(CR+CW+WAR+WAW)}
\label{equ:tot_writes}
\end{equation}
where CR (Cold Read) is the first read access to an address and CW (Cold Write) is the first write access to an address.
Fig. \ref{fig:writes_on_cache_with_RO} shows the number of writes into the SSD cache and the allocated cache space with Centaur and \techname{}. As this figure shows, \techname{} assigns a more efficient write policy for each VM as compared to Centaur. We re-conducted experiments of Sec. \ref{sec:write_pol_ass} and the results demonstrate that by using the RO policy, we can reduce the number of writes by 44\% compared to the caches using the WB policy. When ECI-Cache applies the RO policy on the VMs, only writes due to CRs (Cold Reads) will be written on the cache. Applying the RO policy on the VMs has a negative impact on the hit ratio of RAW operations (by 1.5\%).
As it can be seen in Fig. \ref{fig:writes_on_cache_with_RO}, applying \techname{} on the VM running hm\underline{\hspace{.05in}}1 has no impact on the number of writes into the cache. This workload mostly consists of RAR and RAW operations where it is necessary to cache referenced data by assigning the WB policy. It is important to note that as presented in Sec. \ref{sec:size_estimation} (shown in Fig. \ref{fig:perf-per-cost}), \techname{} allocates about 50\% smaller cache space to this VM, which is calculated based on the reuse distance of RAR and RAW operations by using URD. Similarly, for the VMs running stg\underline{\hspace{.05in}}1, mds\underline{\hspace{.05in}}1, and prn\underline{\hspace{.05in}}1, \techname{} allocates much less cache space based on the calculated URD for RAR and RAW operations and assigns the WB policy. Hence, ECI-Cache has no impact on the number of writes into the SSD‌ in such VMs.  For the remaining VMs running workloads such as mds\underline{\hspace{.05in}}0, \techname{} reduces the number of writes by 80\% and the allocated cache space by 25\%. \techname{} achieves 18\% cache size reduction by 90\% reduction in number of writes for ts\underline{\hspace{.05in}}0. For proj\underline{\hspace{.05in}}0, cache size and the number of writes is reduced by 22\% and 46\%, respectively. \techname{} reduces the number of writes for prxy\underline{\hspace{.05in}}0, wdev\underline{\hspace{.05in}}0, rsrch\underline{\hspace{.05in}}0, src2\underline{\hspace{.05in}}0, and src1\underline{\hspace{.05in}}2 by 90\%, 85\%, 87\%, 85\%, and 88\%, respectively. We conclude that \techname{} assigns an efficient write policy for each VM and thereby reduces the number of writes into the SSD‌ cache by 65\% on average across all VMs. 

%

\section{Conclusion}
\label{sec:conclusion}
In this paper, we presented the \techname{},  a new hypervisor-based I/O caching scheme for virtualized platforms. \techname{} maximizes the performance-per-cost of the I/O cache by 1) dynamically partitioning it between VMs, 2) allocating a small and efficient cache size to each VM, and 3) assigning a workload characteristic-aware cache write policy to each VM.
The proposed scheme also enhances the endurance of the SSD I/O cache by reducing the number of writes performed by each VM to the SSD.
\techname{} uses a new online partitioning algorithm to estimate an efficient cache size for each VM. To do so, \techname{} characterizes the running workloads on the VMs and makes two key decisions for each VM. First, \techname{} allocates cache space for each VM based on the \emph{Useful Reuse Distance} (URD) metric,‌ which considers \emph{only} the \emph{Read After Read} (RAR) and \emph{Read After Write} (RAW) operations in reuse distance calculation. This metric reduces the cost of the allocated cache space for each VM by assigning a smaller cache size because it does \emph{not} consider \emph{Write After Read} (WAR)‌ and \emph{Write After Write} (WAW) accesses to a block to be useful for caching purposes.
Second, \techname{} assigns an efficient write policy to each VM by considering the ratio of  WAR and  WAW operations of the VM's workload. \techname{} assigns the \emph{Write Back} (WB) policy for a VM with a large amount of re-referenced data due to RAW and RAR operations and the \emph{Read Only} (RO) write policy for a VM with a large amount of unreferenced (i.e., not re-referenced) writes due to WAR and WAW accesses. By allocating an efficient cache size and assigning an intelligent cache write policy for each VM, \techname{} 1) improves both performance and performance-per-cost (by 17\% and 30\%, respectively, compared to the state-of-the-art \cite{centaur}) and 2) enhances the endurance of the SSD by greatly reducing the number of writes (by 65\%).
We conclude that \techname{} is an effective method for managing the SSD cache in virtualized platforms.
\sloppypar

\begin{acks}
	
	We would like to thank the reviewers of SIGMETRICS 2017 and SIGMETRICS 2018 for their valuable comments and constructive suggestions. We especially thank Prof. Ramesh Sitaraman for serving as the shepherd for this paper. We acknowledge the support of Sharif ICT Innovation Center and HPDS Corp.\footnote{\href{http://hpdss.com/En/}{http://hpdss.com/En/}}

\end{acks}

\bibliographystyle{ACM-Reference-Format}
\bibliography{ref-2}

\newpage
\appendix
\begin{appendices}
	\section{Cache Allocation in Feasible State}
	\label{sec:unlimited_ssd}
\begin{figure*}[!t]
	\centering
	\subfloat[VM0: wdev\underline{\hspace{.05in}}0]{\includegraphics[width=.25\textwidth]{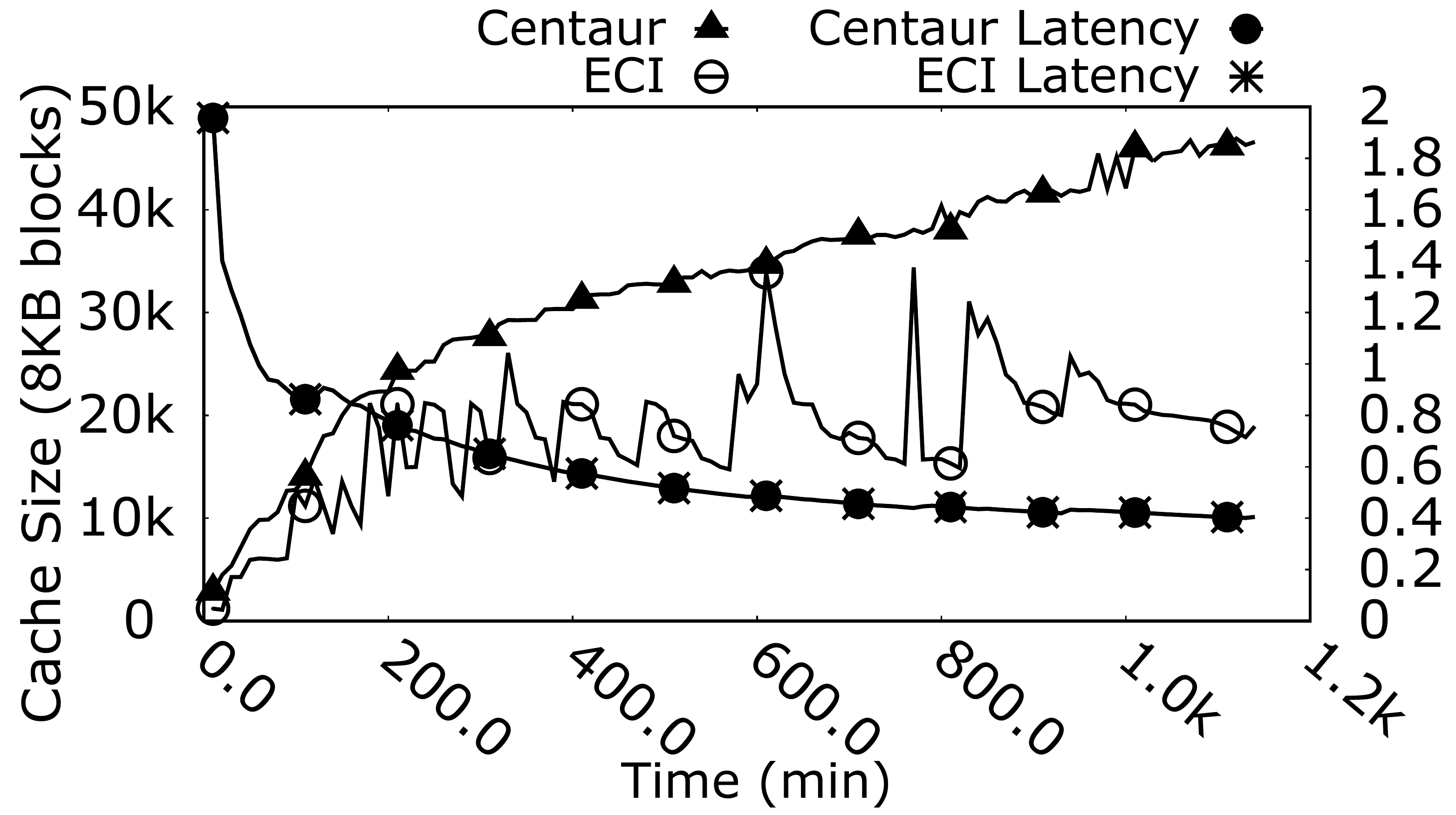}%
		\label{fig:wdev0_nl}}
	\hfil
	\subfloat[VM1: web\underline{\hspace{.05in}}1]{\includegraphics[width=.25\textwidth]{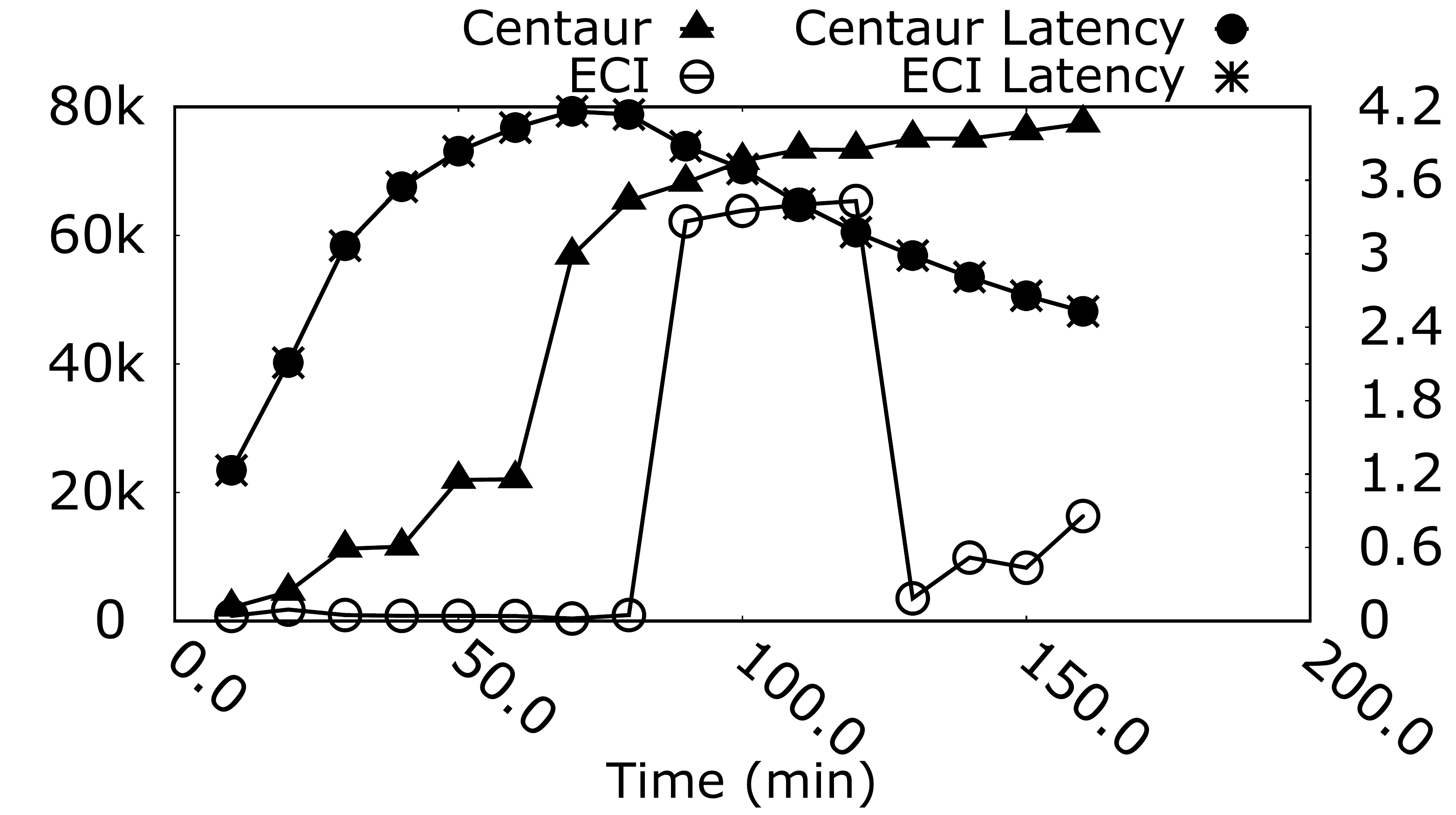}%
		\label{fig:web1_nl}}
	\hfil
	\subfloat[VM2: stg\underline{\hspace{.05in}}1]{\includegraphics[width=.25\textwidth]{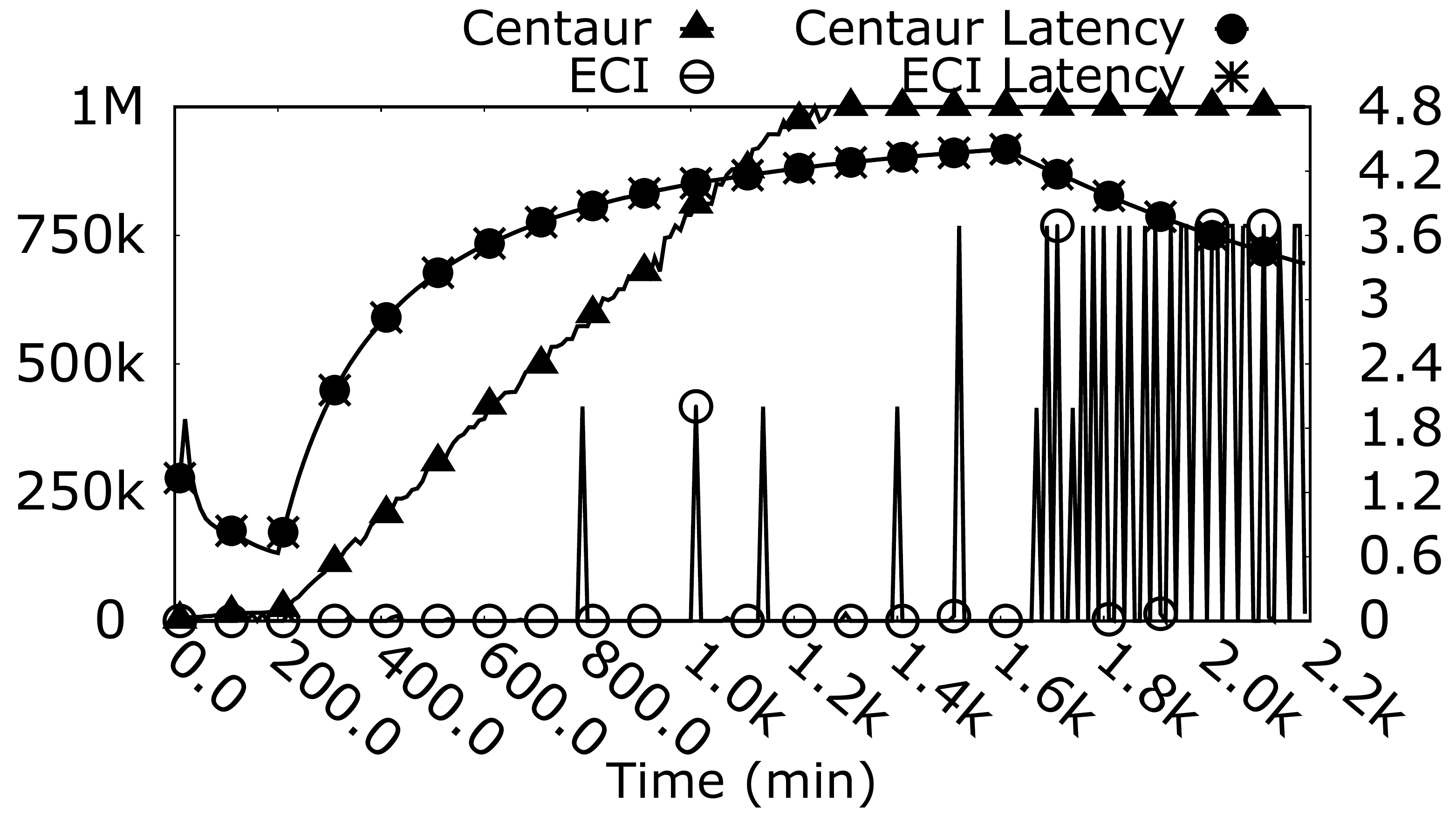}%
		\label{fig:stg1_nl}}
	\hfil
	\subfloat[VM3: ts\underline{\hspace{.05in}}0]{\includegraphics[width=.25\textwidth]{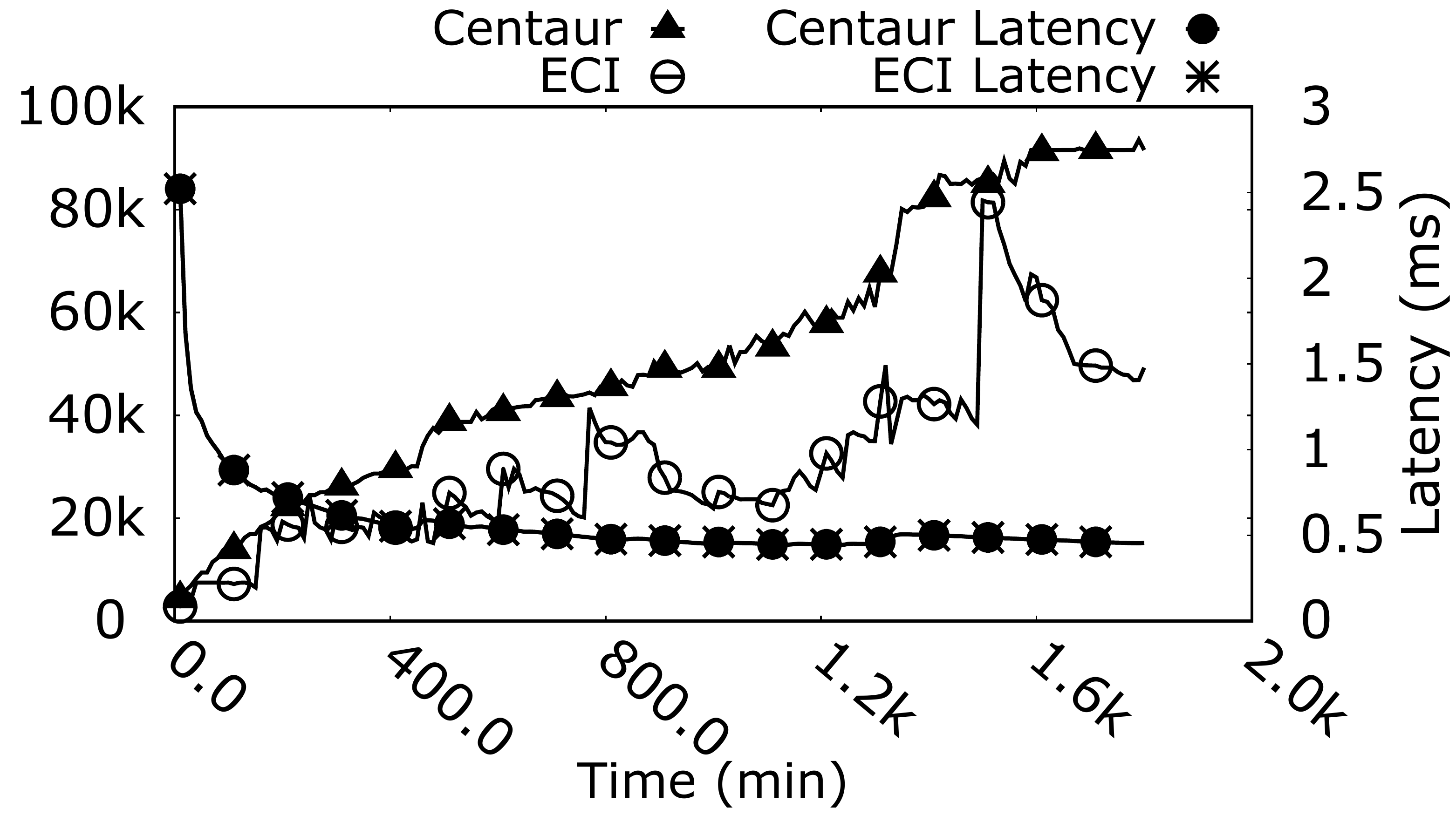}%
		\label{fig:ts0_nl}}
	\hfil
	\subfloat[VM4: hm\underline{\hspace{.05in}}1]{\includegraphics[width=.25\textwidth]{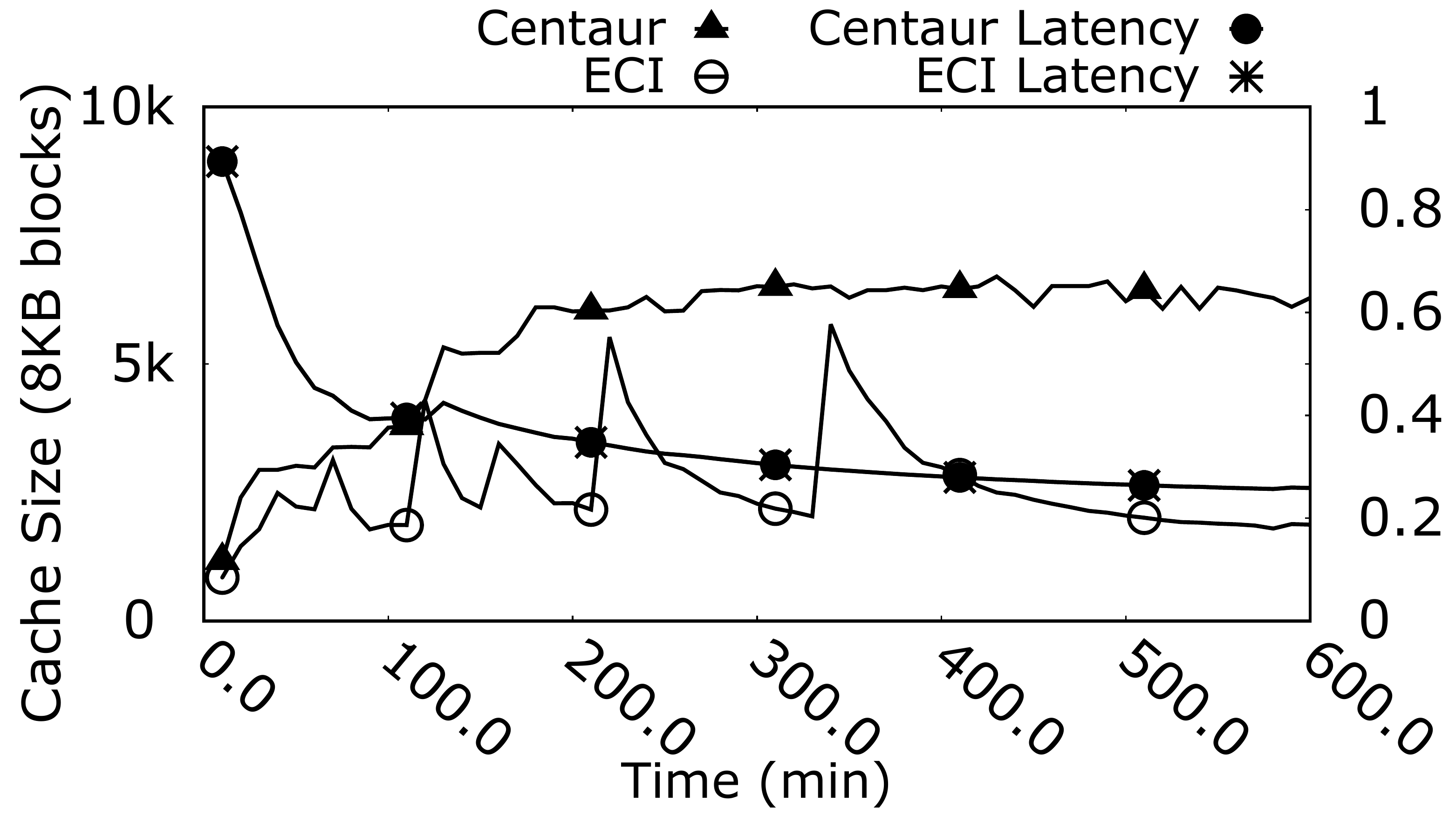}%
		\label{fig:hm1_nl}}
	\hfil
	\subfloat[VM5: mds\underline{\hspace{.05in}}0]{\includegraphics[width=.25\textwidth]{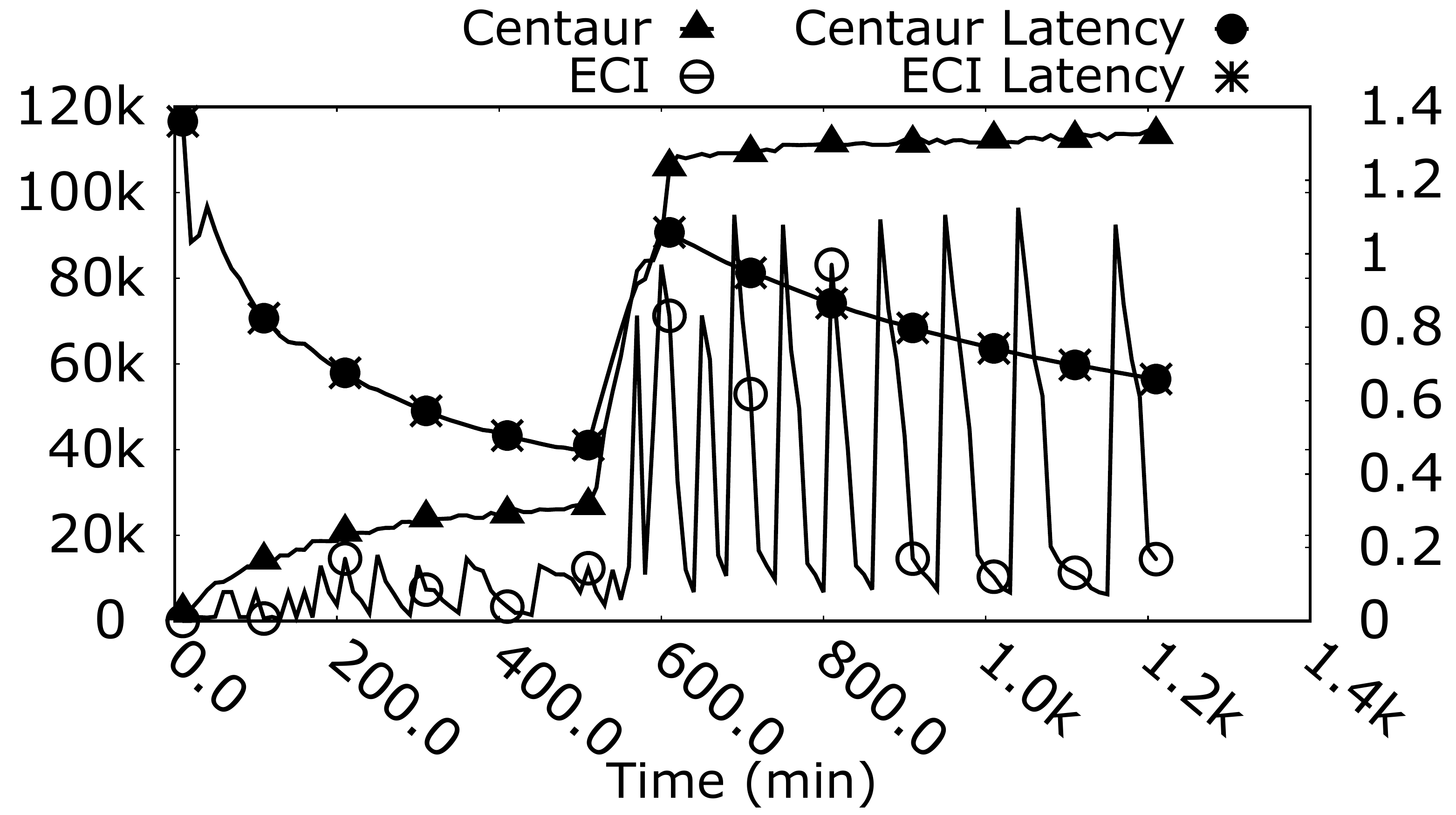}%
		\label{fig:mds_0_nl}}
	\hfil
	\subfloat[VM6:‌ proj\underline{\hspace{.05in}}0]{\includegraphics[width=.25\textwidth]{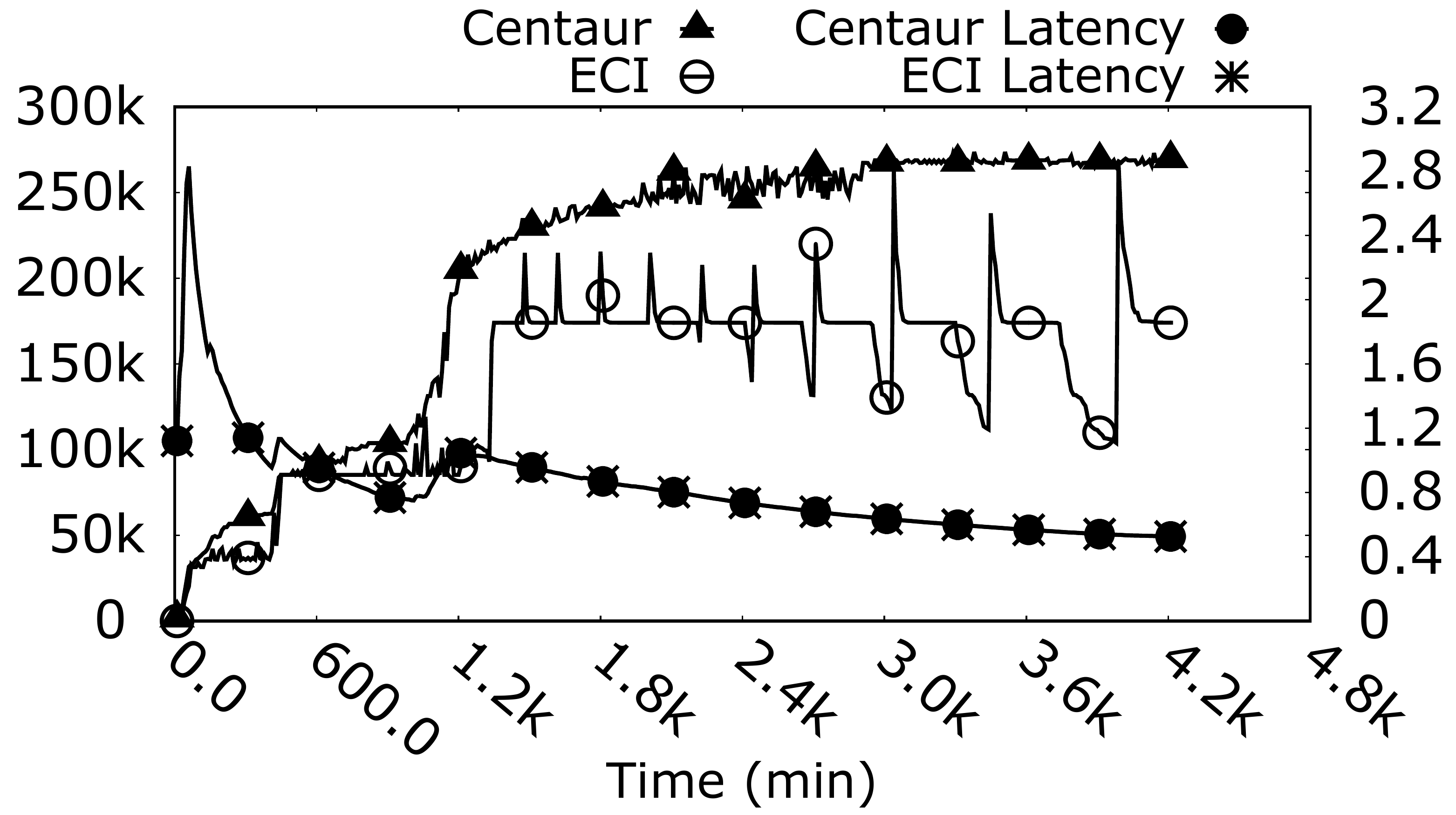}%
		\label{fig:proj_0_nl}}
	\hfil
	\subfloat[VM7: prxy\underline{\hspace{.05in}}0]{\includegraphics[width=.25\textwidth]{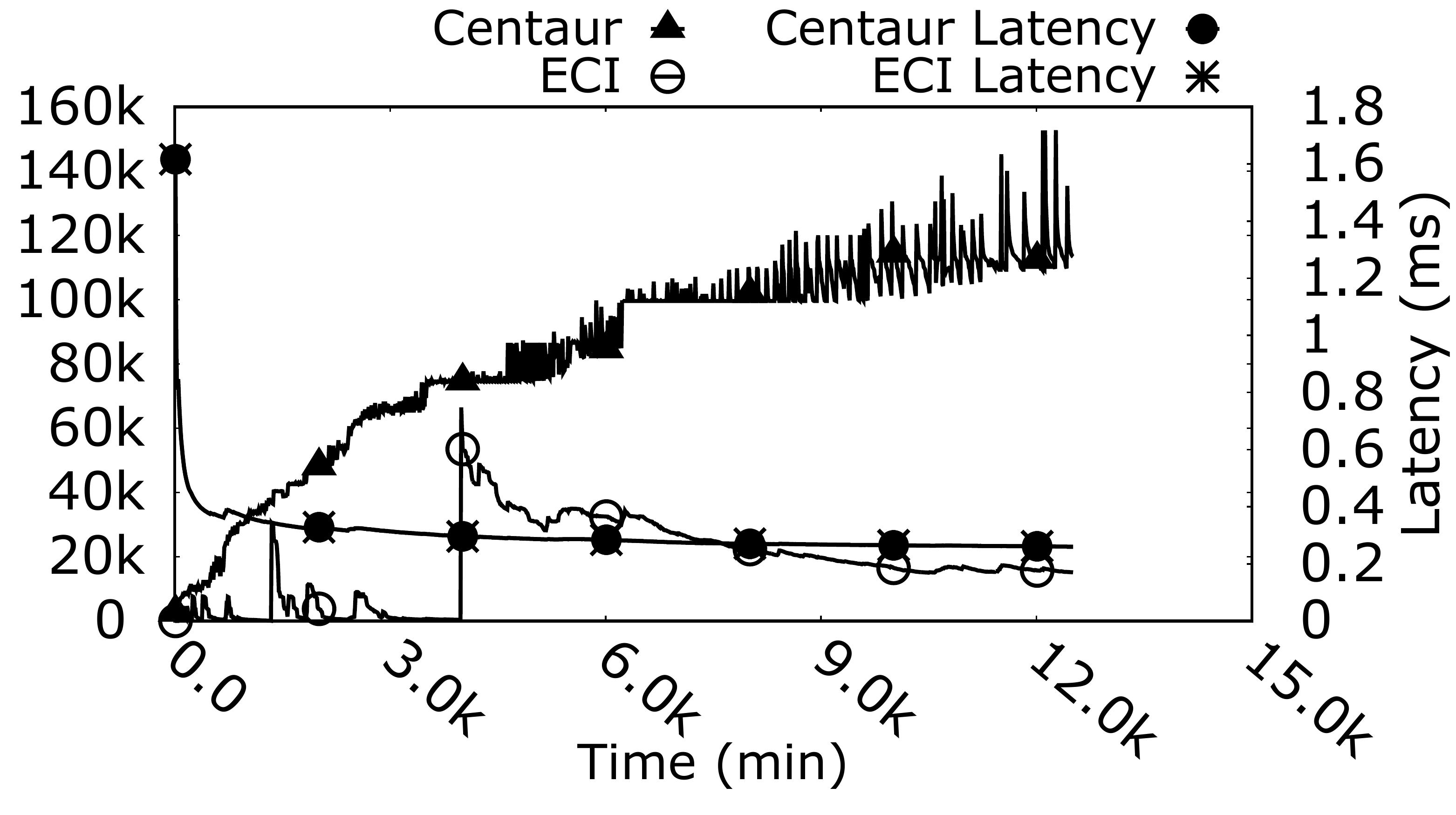}%
		\label{fig:prxy_0_nl}}
	\hfil
	\subfloat[VM8: rsrch\underline{\hspace{.05in}}0]{\includegraphics[width=.25\textwidth]{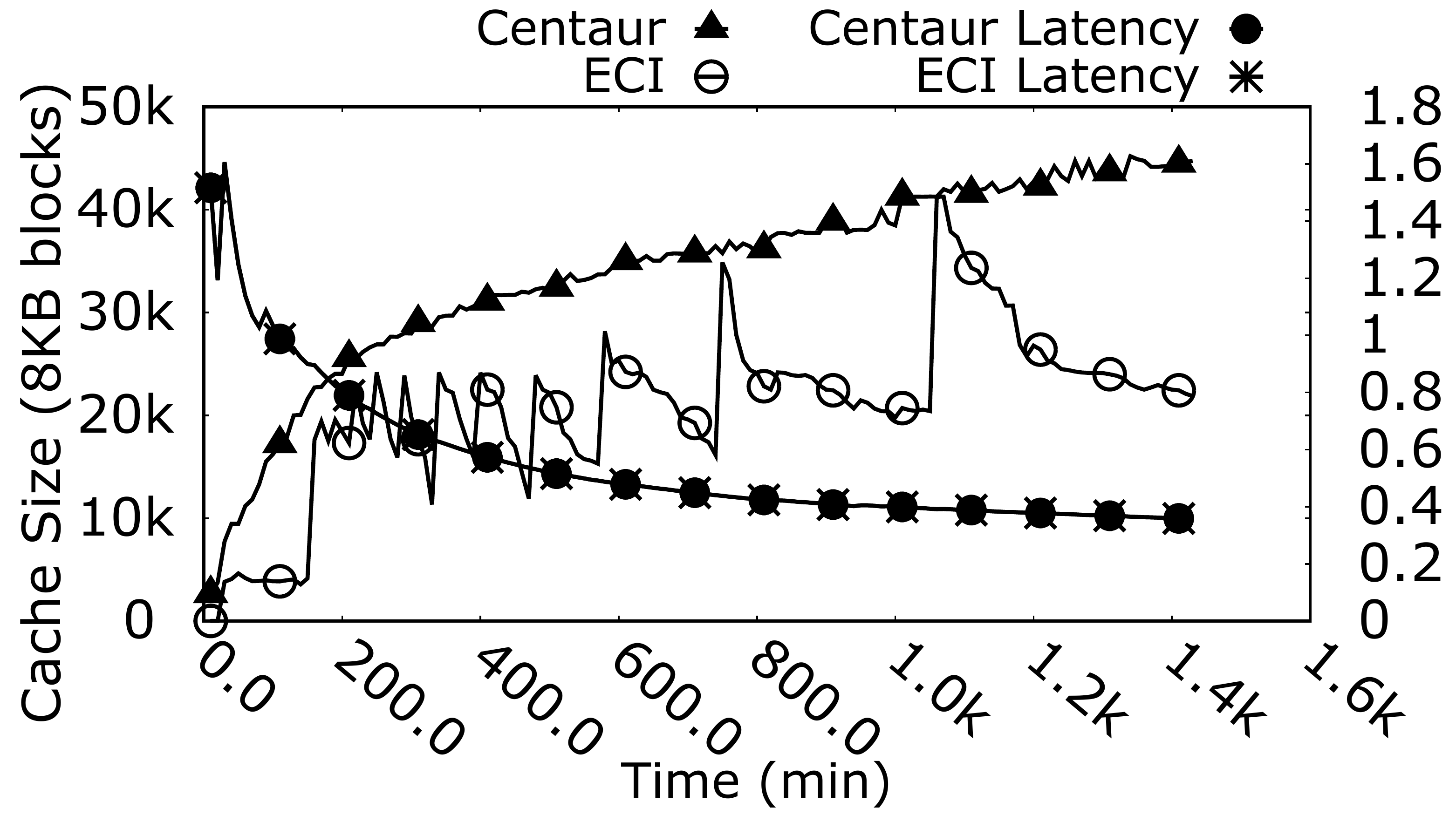}%
		\label{fig:rsrch_0_nl}}
	\hfil
	\subfloat[VM9: src1\underline{\hspace{.05in}}2]{\includegraphics[width=.25\textwidth]{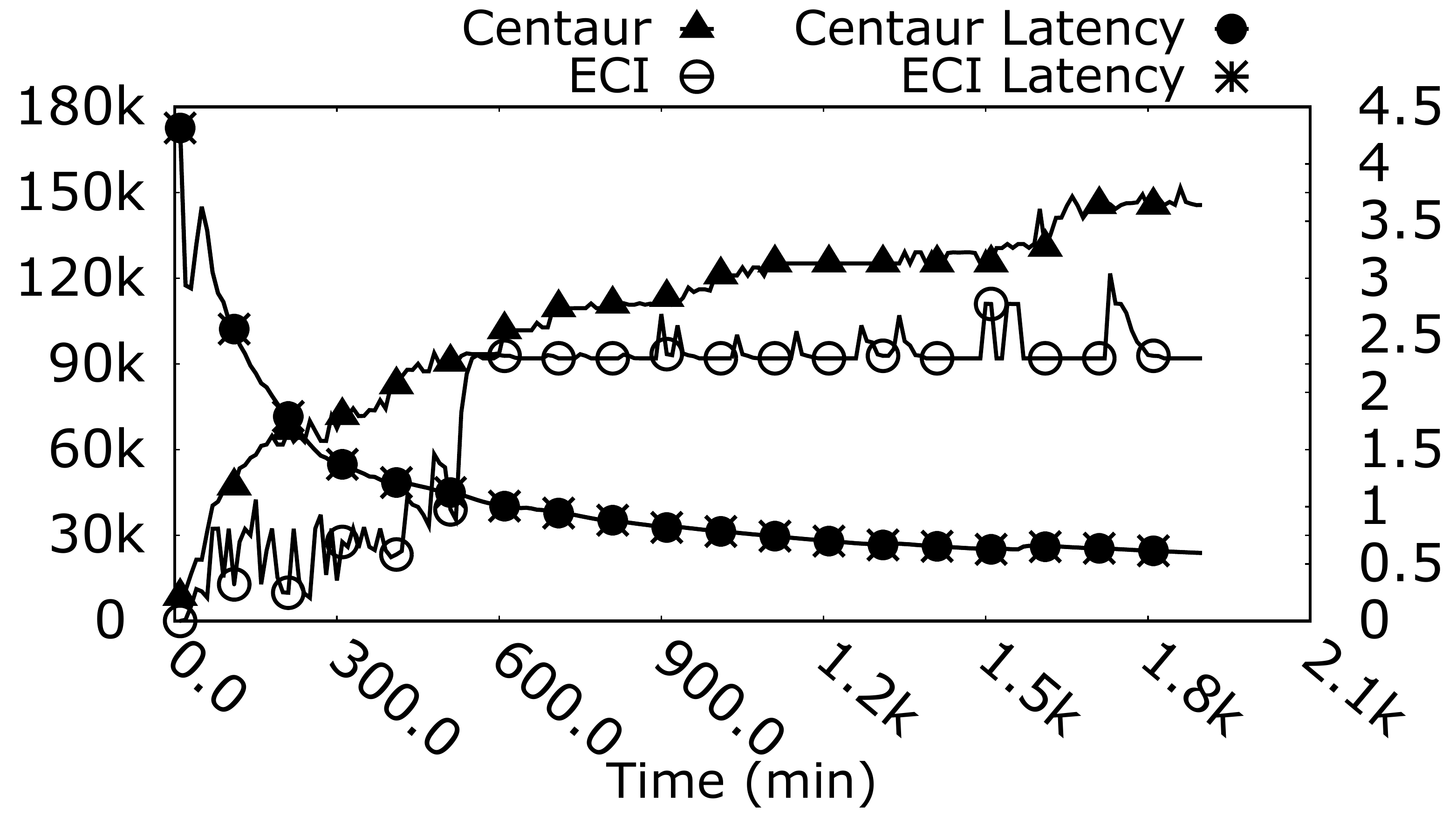}%
		\label{fig:src_1_nl}}
	\hfil
	\subfloat[VM10: prn\underline{\hspace{.05in}}1]{\includegraphics[width=.25\textwidth]{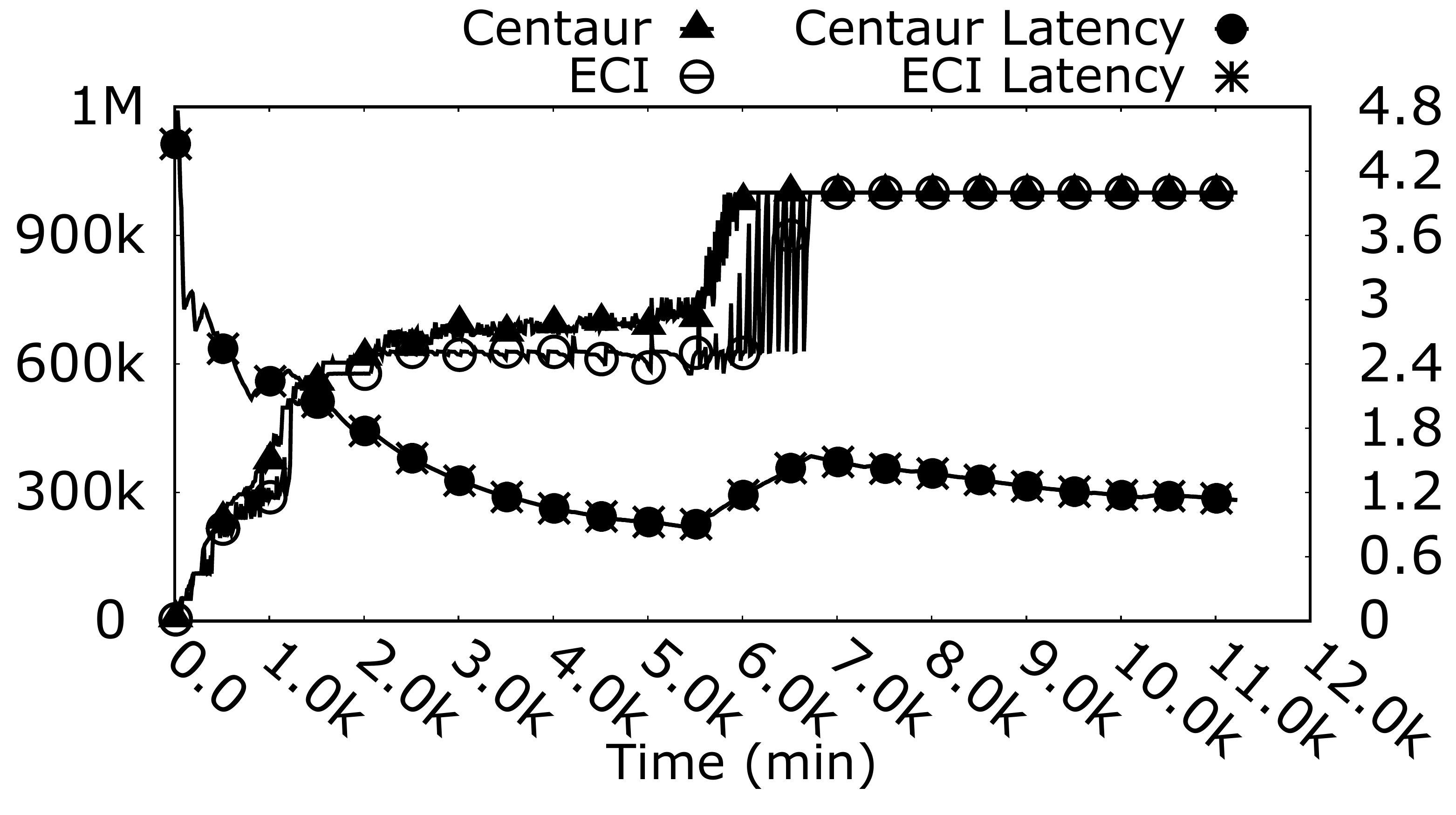}%
		\label{fig:prn_1_nl}}
	\hfil
	\subfloat[VM11: src2\underline{\hspace{.05in}}0]{\includegraphics[width=.25\textwidth]{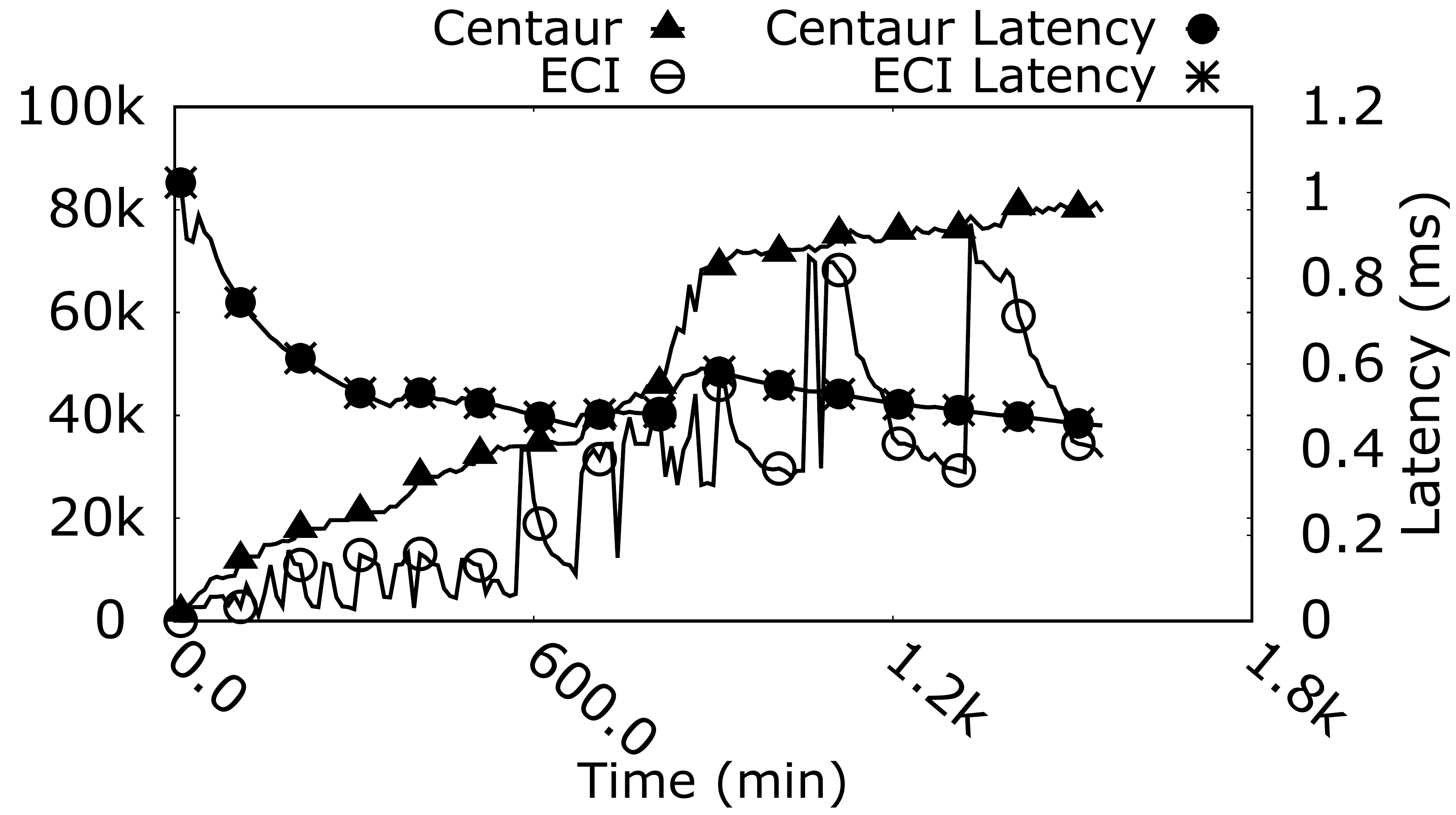}%
		\label{fig:src2_0_nl}}
	\hfil
	\subfloat[VM12: web\underline{\hspace{.05in}}0]{\includegraphics[width=.25\textwidth]{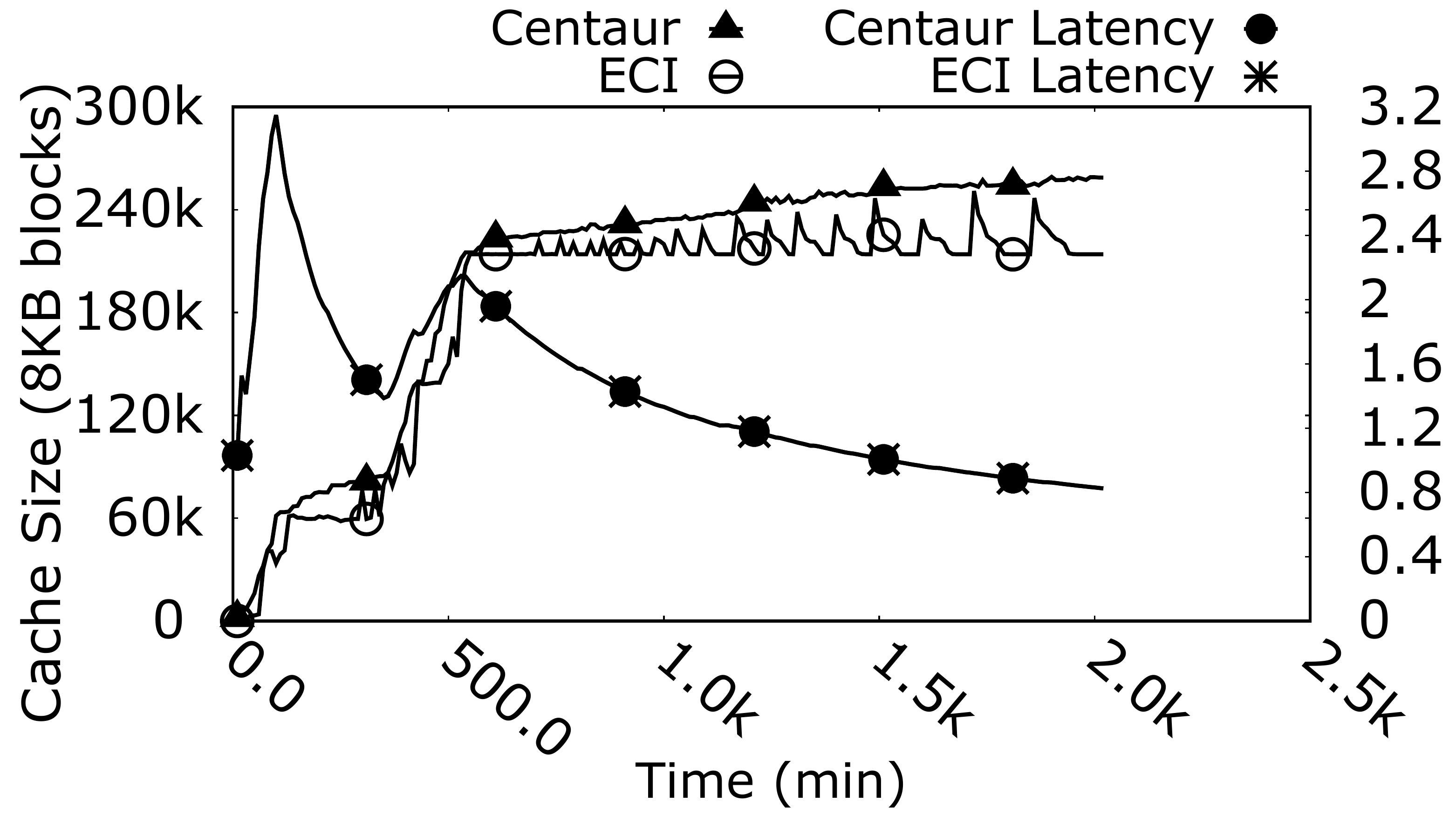}%
		\label{fig:web_0_nl}}
	\hfil
	\subfloat[VM13: usr\underline{\hspace{.05in}}0]{\includegraphics[width=.25\textwidth]{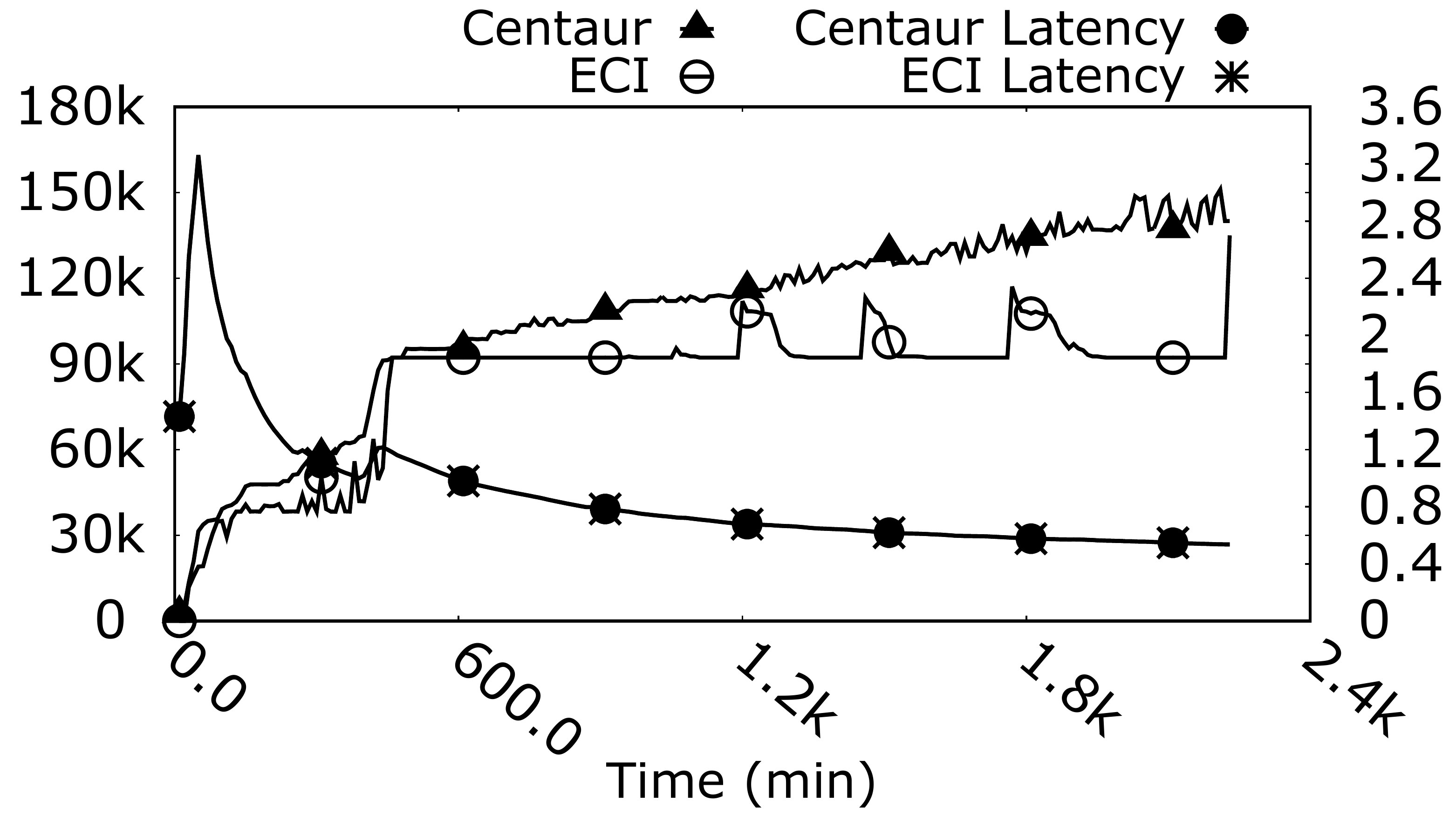}%
		\label{fig:usr_0_nl}}
	\hfil
	\subfloat[VM14: rsrch\underline{\hspace{.05in}}2]{\includegraphics[width=.25\textwidth]{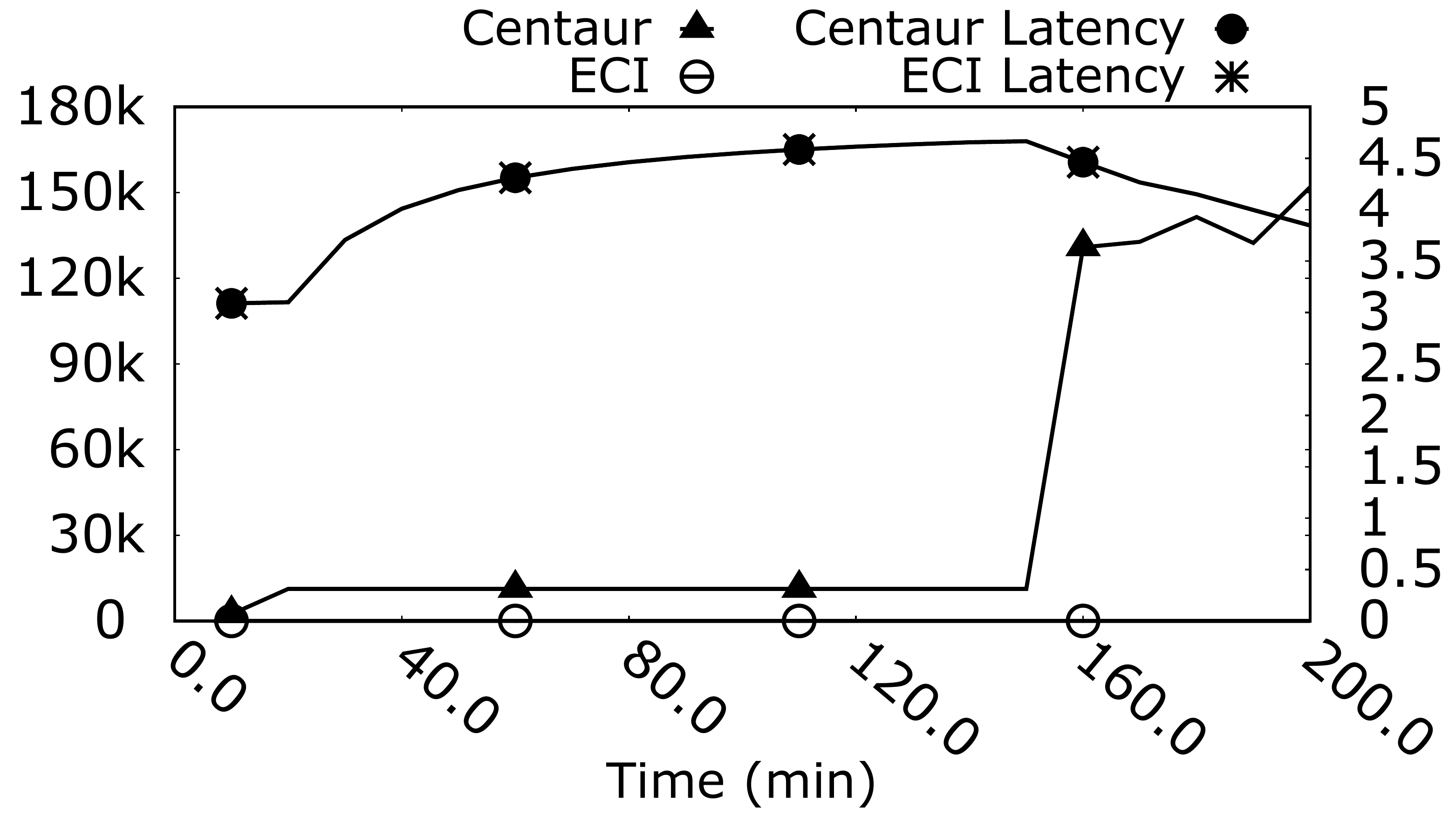}%
		\label{fig:rsrch_2_nl}}
	\hfil
	\subfloat[VM15: mds\underline{\hspace{.05in}}1]{\includegraphics[width=.25\textwidth]{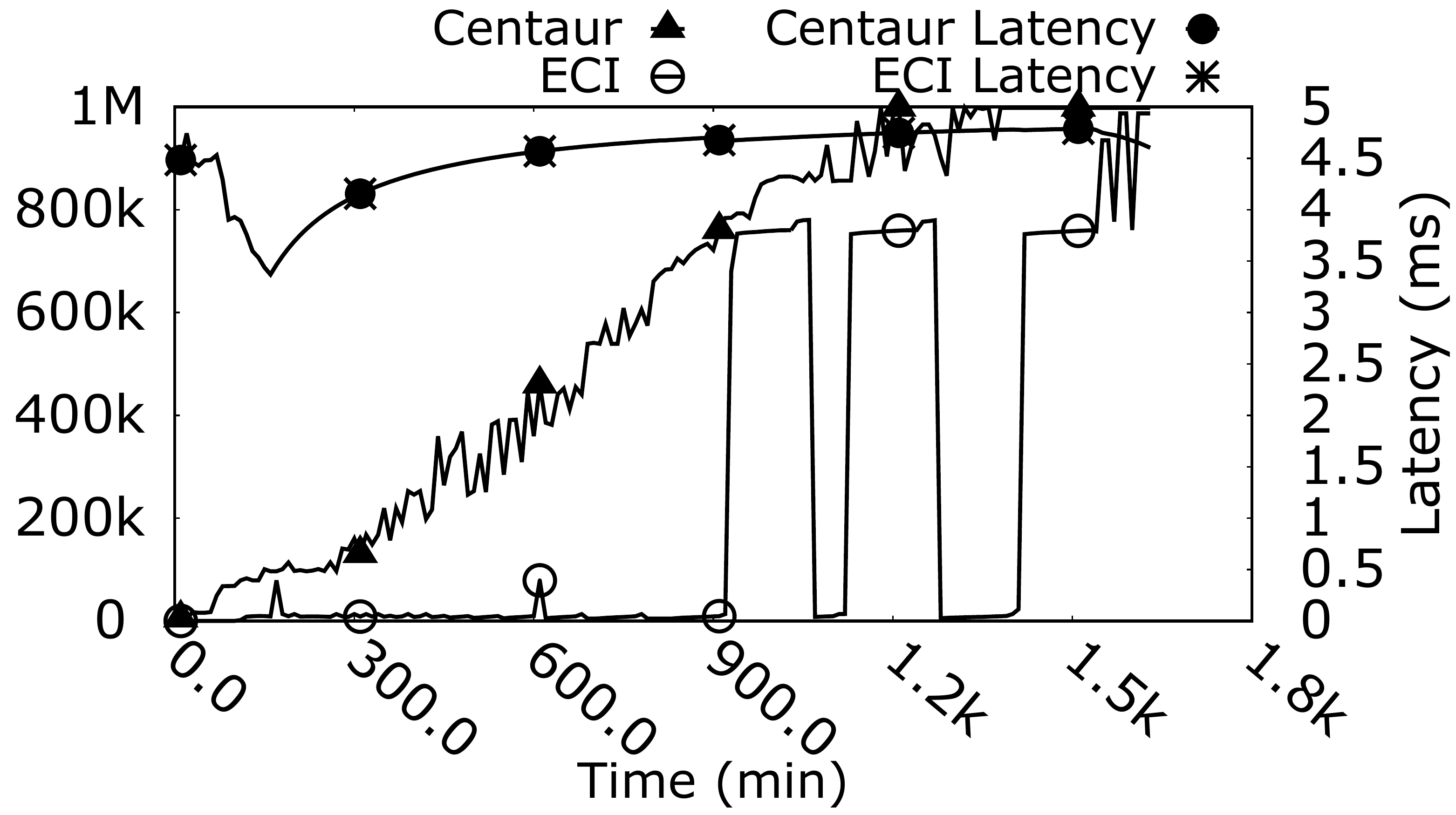}%
		\label{fig:mds_1_nl}}
	\hfil
	
	\caption{Allocated cache space of each VM and corresponding latency in feasible state with unlimited SSD cache capacity (ECI-Cache vs. Centaur).}
	\label{fig:TRD_URD_Hit_nl}
\end{figure*}

	In this section, we show how ECI-Cache and Centaur allocate cache space for the VMs in feasible state. We conduct experiments by applying both schemes in the hypervisor when the SSD cache capacity is unlimited. The experiments are performed on 16 running VMs with an initial cache space equal to $10,000$ cache blocks for each VM (block size is equal to 8KB) and using the WB policy for each VM's cache space. 
	Fig. {\ref{fig:TRD_URD_Hit_nl}} shows the allocated cache space by ECI-Cache and Centaur scheme for the VMs separately. In addition, this figure shows the latency of each VM.

	{We observed that for write-intensive workloads with a large amount of unreferenced (i.e., not re-referenced) data, such as stg\underline{\hspace{.05in}}1 in $VM2$, \techname{} allocates significantly smaller cache space (about $14,000$ cache blocks) for caching referenced data while Centaur allocates about 1000X larger cache space to that VM.
		The allocated cache space by \techname{} for $VM5$, $VM6$, $VM7$, and $VM12$ in some cases becomes equal to the allocated cache space by Centaur (as shown in Fig. \ref{fig:mds_0}, Fig. \ref{fig:proj_0}, Fig. \ref{fig:prxy_0}, and Fig. \ref{fig:web_0}, respectively). This is because the maximum reuse distance of the workloads is mainly affected by RAR‌ and RAW requests.
		In Fig. \ref{fig:mds_0}, at $t=600~min$, there is a hit ratio drop which is recovered by increasing the allocated cache space by both \techname{} and Centaur. It can be seen that the allocated cache space by \techname{} is much smaller than the allocated space by Centaur. In addition, there is a hit ratio drop in $VM1$ at the first 50-minute interval where increasing the cache space using the Centaur scheme does not have any positive impact on the hit ratio. 
		This is due to the lack of  locality in references of the requests, which mostly include WAR and WAW operations.
		It can be seen that \techname{} does not increase the cache space at this time. At $t=60~min$, \techname{} increases the allocated cache space, which results in improving the hit ratio.
		\techname{} allocates the minimum cache space for rsrch\underline{\hspace{.05in}}2 which is running on $VM14$ while Centaur allocates a much larger cache space (more than $50,000$X) than \techname{}. This is because this workload mostly consists of WAR and WAW operations with poor locality of reference. Hence, \techname{} achieves the same hit ratio by allocating much smaller cache space to this workload.
		We conclude that in feasible state, both ECI-Cache and Centaur achieve the same performance while ECI-Cache allocates \emph{much smaller} cache space for the VMs compared to Centaur.}

	\begin{figure}[!h]
	\centering
	
	\subfloat[]{\includegraphics[scale=0.2]{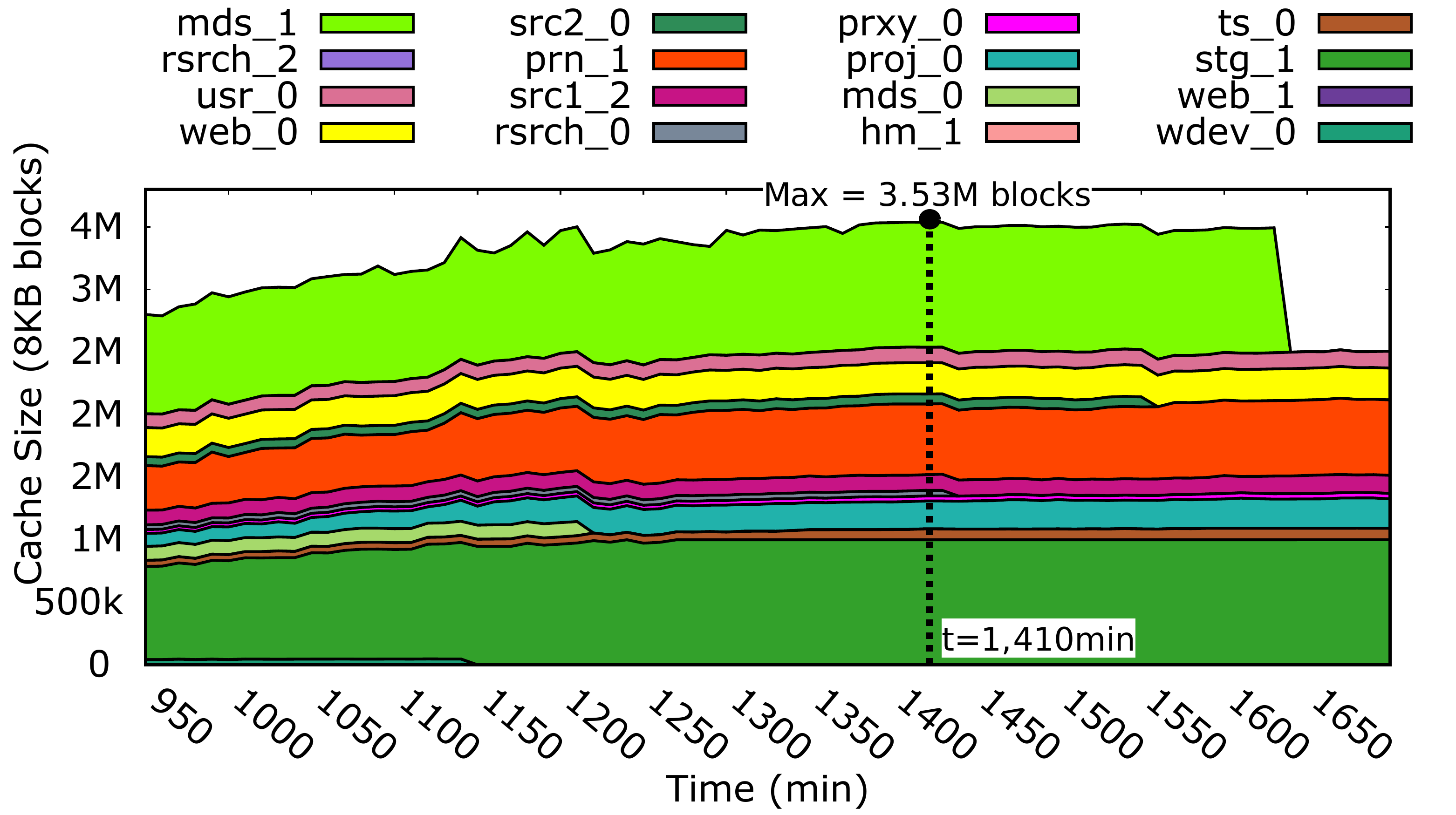}
		\label{fig:centaur_non_limited_cs}}
	\hfil
	\subfloat[]{\includegraphics[scale=0.2]{Figures/Size_allocation_stacked/URD_stacked_size_first_2000min_sum_edited.pdf}
		\label{fig:eci_non_limited_cs_both}}
	\hfil
	\caption{{Cache allocation for the VMs in feasible state by (a) Centaur and (b) ECI-Cache.}}
	
\end{figure}
	
	{Fig. {\ref{fig:centaur_non_limited_cs}} and Fig. {\ref{fig:eci_non_limited_cs_both}} show cache space allocation for the VMs in the time interval from $t=950~min$ to $t=1,700~min$ by Centaur and ECI-Cache, respectively. It can be seen that in all time intervals, ECI-Cache allocates much smaller cache space to the VMs compared to Centaur. We observe that in feasible state (i.e., unlimited SSD cache), ECI-Cache allocates much smaller cache space (29.45\%, on average) than Centaur and hence reduces performance-per-cost significantly.}

	\section{URD Overhead and Time Interval Trade-off  }
	\label{sec:urd-overhead}
	
	{In this section, we report the average delay of calculating URD for the workloads running on the VMs in our experiments. 
		We select the intervals of URD calculation in the experiments in order to limit the time overhead of URD calculation (the time that we should wait until URD‌ is calculated).
		Table {\ref{table:time_overhead}} reports the average time it takes to calculate URD for each workload. As reported in the table, the maximum time it takes to calculate URD is about 22.67 seconds, which is reported for the prn{\underline{\hspace{.05in}}}1 workload. In addition, the average time overhead of the URD calculation for all workloads is about 4.82 seconds. We selected 10 min time intervals which is calculated based on maximum URD calculation delay to reduce the time overhead of URD calculation to less than 5\% in our experiments.}
	\begin{table*}[!tbh]
	\begin{center}
		\caption{{Time overhead of URD‌ calculation in the running VMs.}}\label{table:time_overhead}
		\tiny
		\begin{tabular}{|m{0.1\textwidth}||m{0.035\textwidth}m{0.028\textwidth}m{0.028\textwidth}m{0.028\textwidth}m{0.028\textwidth}m{0.028\textwidth}m{0.028\textwidth}m{0.028\textwidth}m{0.032\textwidth}m{0.028\textwidth}m{0.028\textwidth}m{0.028\textwidth}m{0.028\textwidth}m{0.028\textwidth}m{0.031\textwidth}m{0.031\textwidth}|}
			\hline
			 $VM_{ID}$ & \begin{turn}{90} VM0 \end{turn} & \begin{turn}{90} VM1 \end{turn} & \begin{turn}{90} VM2 \end{turn} & \begin{turn}{90} VM3 \end{turn} & \begin{turn}{90} VM4 \end{turn} & \begin{turn}{90} VM5 \end{turn} & \begin{turn}{90} VM6 \end{turn} & \begin{turn}{90} VM7 \end{turn} & \begin{turn}{90} VM8 \end{turn} & \begin{turn}{90} VM9 \end{turn} & \begin{turn}{90} VM10 \end{turn} & \begin{turn}{90} VM11 \end{turn} & \begin{turn}{90} VM12 \end{turn} & \begin{turn}{90} VM13 \end{turn} & \begin{turn}{90} VM14 \end{turn} & \begin{turn}{90} VM15\end{turn}\\ \hline
			Workload & \begin{turn}{90} wdev\underline{\hspace{.05in}}0 \end{turn} & \begin{turn}{90}web\underline{\hspace{.05in}}1 \end{turn} & \begin{turn}{90} stg\underline{\hspace{.05in}}1 \end{turn} & \begin{turn}{90} ts\underline{\hspace{.05in}}0 \end{turn} & \begin{turn}{90} hm\underline{\hspace{.05in}}1 \end{turn} & \begin{turn}{90} mds\underline{\hspace{.05in}}0 \end{turn} & \begin{turn}{90} proj\underline{\hspace{.05in}}0 \end{turn} & \begin{turn}{90} prxy\underline{\hspace{.05in}}0 \end{turn} & \begin{turn}{90} rsrch\underline{\hspace{.05in}}0 \end{turn} & \begin{turn}{90} src1\underline{\hspace{.05in}}2 \end{turn} & \begin{turn}{90} prn\underline{\hspace{.05in}}1 \end{turn} & \begin{turn}{90} src2\underline{\hspace{.05in}}0 \end{turn} & \begin{turn}{90} web\underline{\hspace{.05in}}0 \end{turn} & \begin{turn}{90} usr\underline{\hspace{.05in}}0 \end{turn} & \begin{turn}{90} rsrch\underline{\hspace{.05in}}2 \end{turn} & \begin{turn}{90} mds\underline{\hspace{.05in}}1\end{turn}\\ \hline
				URD Calculation Time (s) & \begin{turn}{90} 1.57 \end{turn} & \begin{turn}{90} 0.387 \end{turn} & \begin{turn}{90} 5.531 \end{turn} & \begin{turn}{90} 2.557 \end{turn} & \begin{turn}{90} 0.719 \end{turn} & \begin{turn}{90} 1.777 \end{turn} & \begin{turn}{90} 6.857 \end{turn} & \begin{turn}{90} 15.401 \end{turn} & \begin{turn}{90} 1.919 \end{turn} & \begin{turn}{90} 3.035 \end{turn} & \begin{turn}{90} 22.674 \end{turn} & \begin{turn}{90} 2.187 \end{turn} & \begin{turn}{90} 3.499 \end{turn} & \begin{turn}{90} 3.417 \end{turn} & \begin{turn}{90} 0.376 \end{turn} & \begin{turn}{90} 5.233 \end{turn}\\
			\hline
		\end{tabular}
	\end{center}
\end{table*}
	
	{
		\section{Worst-Case Analysis}
		In this section, we provide three examples that illustrate corner cases where ECI-Cache fails in cache size estimation, and as a result, does \emph{not} have a positive impact on performance improvement. We find that these cases are uncommon in the workloads we examine.}\newline
	
\begin{figure}[!h]
		\centering
		\subfloat[{Sequential-Random workload: ECI-Cache underestimates cache size}]{\includegraphics[width=.3\textwidth]{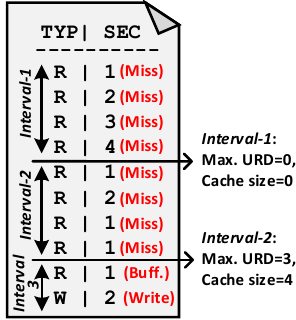}%
			\label{fig:seq-rand}}
		\hfil
		\subfloat[{Random-Sequential workload: ECI-Cache overestimates cache size}]{\includegraphics[width=.3\textwidth]{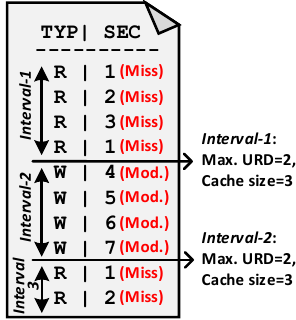}%
			\label{fig:rand-seq}}
		\hfil
		\subfloat[{Semi-Sequential workload: ECI-Cache wastes cache space}]{\includegraphics[width=.3\textwidth]{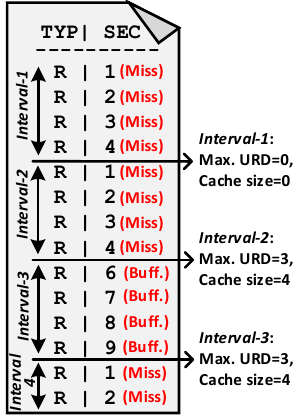}%
			\label{fig:case3}}	
		\caption{{Corner cases where ECI-Cache fails in size estimation (TYP: Type, SEC: Sector, W: Write, R: Read, Buff.: Buffer, and Mod.: Modify).}}
		\label{fig:corner}
	\end{figure}

	\noindent
	{\uline{Case 1. \emph{Sequential-Random workload}:}
		The workload has two intervals: 1) sequential accesses followed by 2) random and repetitive requests (i.e., requests to the previously-accessed addresses). In the first interval, ECI-Cache does not allocate cache space for the workload while the requests in the second interval are random accesses to the previously (not buffered) accesses.
		Thus, in the first interval, ECI-Cache underestimates the allocated cache size for the workload.
		Fig. {\ref{fig:seq-rand}} shows an example of such workloads. We elaborate on how ECI-Cache works in each interval for the example workload:}
	\begin{enumerate}
		\item {At the end of the first interval (i.e., when ECI-Cache recalculates URD and cache size), the URD of the workload is equal to 0. Hence, no cache space is allocated to this VM.}
		\item {In the second interval, the workload accesses become random with repetitive addresses where all requests are provided by the HDD and none of them are buffered in the cache (since no cache space is allocated to this VM).}
		\item {At the end of the second interval, the maximum URD of the workload is equal to three and hence ECI-Cache allocates cache space equal to four  blocks for this VM.}
		\item {In the last interval, the workload issues two accesses to the storage subsystem but neither of them can be supplied by the cache (since the cache has no valid data) and the requests are  buffered in the cache without any performance improvement.}
	\end{enumerate}
	\noindent
	{We find that allocating cache space in such a manner (i.e., only at the end of the second interval) \emph{cannot} improve the performance of requests of the last interval in this workload. In this case, the Centaur scheme works similar to ECI-Cache.}\newline
	
	\noindent
	{\uline{Case 2. \emph{Random-Sequential workload}:} In this case, in the first interval, the workload issues random accesses to the storage subsystem and ECI-Cache allocates cache space based on the URD of the requests. In the second interval, the accesses of the workload become sequential \emph{without} any access to previously-buffered requests. 
		ECI-Cache overestimates the allocated cache space for the workload.
		Fig. {\ref{fig:rand-seq}} shows an example workload for this case. We show how ECI-Cache overestimates the cache space for this example workload:}
	\begin{enumerate}
		\item {In the first interval, the workload is random with locality of accesses and at the end of interval. The maximum URD is equal to two. Hence, at the end of the interval, three cache blocks are allocated to this VM.}
		\item {In the second interval, requests become sequential writes and are buffered in the allocated cache space.}
		\item {At the end of the second interval, the maximum URD of the workload is equal to two and hence ECI-Cache allocates three cache blocks for this workload.}
		\item {In the last interval,  repetitive requests (that are the same as requests of the first interval) cannot be supplied by the cache since the entire cache space is used up by the accesses of the second interval (sequential writes).}
	\end{enumerate}
	\noindent
	{We observe that although enough cache space is allocated for the VM, ECI-Cache cannot improve the performance of the workload compared to the HDD-based system due to the access behavior. In this case, Centaur works similar to ECI-Cache.}\newline
	
	
	\noindent
	{\uline{Case 3. \emph{Semi-Sequential workload}:} Such a workload includes similar sequential accesses in different intervals, which creates a large maximum URD without any locality of reference. In this case, ECI-Cache allocates a large cache space for the VM. Further requests use up the entire cache space without any read hit from allocated cache. Fig. {\ref{fig:case3}} shows an example workload. We elaborate on how ECI-Cache allocates cache size for this example workload:}
	\begin{enumerate}
		\item {At the end of the first interval, the maximum URD‌ of the workload is 0 and no cache space is allocated to this VM.}
		\item {In the second interval, all accesses of the first interval are repeated and are provided by the HDD. Since we have no cache space, none of them are buffered in the cache.}
		\item {At the end of the second interval, the maximum URD of the workload is equal to three and ECI-Cache allocates four blocks to this VM.}
		\item {In the third interval, the workload issues read accesses to the storage subsystem and all of them are provided by the HDD (since the allocated cache has no valid data). In this case, these requests are buffered in the cache (all cache blocks are used).}
		\item {In the last interval, since cache space is used up by accesses of the third interval, the future requests that are identical to requests of the \emph{second} interval miss in the cache and are supplied by the HDD.}
	\end{enumerate}
	\noindent
	{In this example, we observe that ECI-Cache allocates large cache space to the VM and buffers unnecessary data \emph{without} any improvement in performance. We can resolve this problem by changing the length of the intervals.}
	
	{
		\section{Convex Optimization}
		In the following, we show why the objective function of the proposed algorithm (Eq. \ref{equ:convex})
		is convex and then show how we solve the function using the MATLAB optimization toolbox. To do so, we first show that the objective function is convex. Then we show that the objective function and its constraints are in canonical form of convex.}
	\begin{equation} 
	\begin{cases}
	LatencyVM_i=h_i(c_i) \times T_{ssd} + (1-h_i(c_i)) \times T_{hdd} \\
	\text{Objective: }Minimize[\sum_{i=1}^{N}LatencyVM_i]\\
	\text{Constraint1: }\sum_{i=1}^{N}c_{i} \leq C \\
	\text{Constraint2: }0 \leq c_i \leq c_{{urd}_i}
	\end{cases}
	\label{equ:convex}
	\end{equation}
	{$LatencyVM_i$ is a linear function and is convex. The objective function is the sum of $LatencyVM_i$ functions because the sum of convex functions is convex. We express the first constraint of the objective function (constraint 1) as follows ($\mathcal{I}$ is a unity matrix of N by N):}
	\begin{equation} 
	\text{Constraint1: }\sum_{i=1}^{N}c_{i} \leq C \\ \longrightarrow
	\mathcal{I}
	\times
	\begin{bmatrix}
	c_0 \\ .. \\ .. \\ c_{N-1}
	\end{bmatrix}
	\leq
	C
	\label{equ:cons1}
	\end{equation}
	{We express the second constraint (constraint 2) as:}
	\begin{equation} 
	\text{Constraint2: }0 \leq c_i \leq c_{{urd}_i} \\ \longrightarrow
	\mathcal{I}
	\times
	\begin{bmatrix}
	c_0 \\ .. \\ .. \\ c_{N-1}
	\end{bmatrix}
	\leq
	\begin{bmatrix}
	c_{{urd}_0} \\ .. \\ .. \\ c_{{urd}_{N-1}}
	\end{bmatrix}\\
	\text{and~~~} \\
	-
	\mathcal{I}
	\times
	\begin{bmatrix}
	c_0 \\ .. \\ .. \\ c_{N-1}
	\end{bmatrix}
	\leq
	\begin{bmatrix}
	0 \\ .. \\ .. \\ 0
	\end{bmatrix}\\
	\label{equ:cons2}
	\end{equation}
	{Then, we have:}
	\begin{equation}
	\begin{bmatrix}
	\begin{bmatrix}
	1 & ... & 1
	\end{bmatrix} \\
	\mathcal{I} \\
	-\mathcal{I}
	\end{bmatrix} 
	\times
	\begin{bmatrix}
	c_0 \\ .. \\ .. \\ c_{N-1}
	\end{bmatrix}
	\preceq
	\begin{bmatrix}
	C\\
	\begin{bmatrix}
	c_0 \\ .. \\ .. \\ c_{N-1}
	\end{bmatrix}\\
	\begin{bmatrix}
	0 \\ .. \\ .. \\ 0
	\end{bmatrix}
	\end{bmatrix}
	\label{equ:canonical}
	\end{equation}
	{Eq. \ref{equ:canonical} is the canonical form of convex and hence the objective function with constraints is convex. We employ MATLAB optimization toolboxes to minimize the objective function.}
	
	\section{Details of ECI-Cache Size Estimation Algorithm}
	\label{sec:size_detail_alg}
	In this section, we present Algorithm \ref{alg:size_detail}, which provides the
	details of Algorithm \ref{alg:size}. {Table \ref{notation} summarizes the key notational  elements used in Algorithm \ref{alg:size} and Algorithm \ref{alg:size_detail}.}
	
	The main objective of Algorithm \ref{alg:size_detail} (similar to Algorithm \ref{alg:size}) is to find the most appropriate cache sizes for the running VMs such that the system is able to meet the following conditions: 1) Sum of allocated cache spaces for all VMs  is less than or equal to the total SSD cache capacity. 2) Aggregate latency of all VMs is minimum. 3) Allocated cache space for each VM is less than or equal to the estimated cache space by URD.
	This objective is obtained by the $calculateEffSize$ function via the use of the  $ObjectiveFunction$. $ObjectiveFunction$ minimizes $diff$ which is the difference between sum of allocated cache spaces for the VMs ($csum$) and the SSD cache capacity ($C$) (i.e., assigns the maximum cache space for each VM). In addition, $ObjectiveFunction$ minimizes the aggregate latency of the running VMs.
\begin{table}[!h]
	\centering
	\caption{{Description of notation used in Algorithm \ref{alg:size} and Algorithm \ref{alg:size_detail}.}}
	\label{notation}
	\scriptsize
	
	\begin{tabular}{|l|l|}
		\hline
\multicolumn{2}{|c|}{\textbf{Variables}}                                                                                                                                                                                                                                          \\ \hline

		Notation     & Description                                                                                                                                                                                               \\ \hline\hline
		$N$            & Number of running VMs.                                                                                                                                                                                    \\ \hline
		$C$            & SSD cache capacity.                                                                                                                                                                                       \\ \hline
		$T_{HDD}$         & HDD delay (i.e., HDD service time).                                                                                                                                                                       \\ \hline
		$T_{SSD}$         & SSD delay (i.e., SSD service time).                                                                                                                                                                       \\ \hline
		$C_{eff}[i]$  & Efficient cache size allocated for each VMi.                                                                                                                                                              \\ \hline
		$URD_i$         & Useful Reuse Distance (URD) for $VM_i$, which is the output of $calculateURD()$ function.                                                                                                                             \\ \hline
		$Size_{URD}[i]$     & Initial efficient cache size suggested by URD for $VM_i$.                                                                                                                                                    \\ \hline
		$csum$         & Sum of initial efficient cache sizes (i.e., sum of $Size_{URD}[i]$ of all running VMs).                                                                                                                          \\ \hline
		$cacheBlkSize$ & Size of cache blocks (equal to 8KB in our experiments).                                                                                                                                                   \\ \hline
		$initialSize$  & Array of initial cache sizes allocated to each VM.                                                                                                                                                       \\ \hline
		$lowerBound$   & Array of minimum cache sizes that can be assigned to the VMs.                                                                                                                                             \\ \hline
		$upperBound$   & \begin{tabular}[c]{@{}l@{}}Array of efficient cache sizes suggested by URD. \\ The efficient cache sizes by ECI-Cache for each VM is less than or equal to the suggested \\ cache sizes by URD.\end{tabular} \\ \hline
		
		$weightVM$     & Weight of VMs.                                                                                                                                                                                            \\ \hline
		$h[i]$     & Hit ratio of $VM_i$.                                                                                                                                                                                         \\ \hline
		$hsum$         & Sum of hit ratio of VMs.                                                                                                                                                                                  \\ \hline
		$diff$         & Difference between total SSD cache capacity and estimated cache sizes by ECI-Cache.                                                                                                                       \\ \hline
		$Obj$          & The objective variable.                                                                                                                                                                                   \\ \hline\hline
		\multicolumn{2}{|c|}{\textbf{Functions}}                                                                                                                                                                                                                                          \\ \hline\hline
		\multicolumn{1}{|l|}{\multirow{2}{*}{$calculateURD$}}          & Input: $VM$                                                                                                                                                                                                 \\ \cline{2-2} 
		\multicolumn{1}{|l|}{}                                       & Output: Useful Reuse Distance ($URD$) of each VM.                                                                                                                                                           \\ \hline
		\multicolumn{1}{|l|}{\multirow{2}{*}{$calculateURDbasedSize$}} & Input: $URD$                                                                                                                                                                                                \\ \cline{2-2} 
		\multicolumn{1}{|l|}{}                                       & Output: efficient cache size suggested by URD ($size_{URD}$)                                                                                                                                                   \\ \hline
		\multicolumn{1}{|l|}{\multirow{2}{*}{$calculateEffSize$}}      & \begin{tabular}[c]{@{}l@{}}Input: $size_{URD}[1..N]$, $C$\\ // This function is called only in infeasible states where the existing SSD cache size is \\ less than required cache sizes by VMs.\end{tabular}    \\ \cline{2-2} 
		 
		\multicolumn{1}{|l|}{}                                       & Output: efficient cache size for each VM ($c_{eff}[i]$)                                                                                                                                                    \\ \hline
		\multicolumn{1}{|l|}{$objectiveFunction$}                      & \begin{tabular}[c]{@{}l@{}}Input: ---\\ // This function is used within the $calculateEffSize$ function.\end{tabular}                                                                                           \\ \hline
			
	\end{tabular}
\end{table}
	\begin{algorithm*}[!htb]
	\tiny
	\DontPrintSemicolon
	\caption{{ECI-Cache size allocation algorithm (in more details).}}
	\label{alg:size_detail}
	\tcc{\textbf{Inputs:} Number of VMs: ($N$), SSD cache size: ($C$), HDD Delay: ($T_{HDD}$), SSD Delay: ($T_{SSD}$)}
	\tcc{\textbf{Output:} Efficient cache size for each VM: ($c_{eff}[1..N]$)}
		Sleep for $\Delta t$\; \label{lst:line:sleep}
		Extract the traces of the workloads running on the VMs including 1) destination address, 2) request size, and 3) request type\; \label{lst:line:delay} \label{lst:line:extract}
	\tcc{Here we \textbf{estimate} the efficient cache space for each VM by calculating URD of the running workloads.}
	\For {$i = 1$ to $N$}
	{
		$URD[i]=calculateURD(VM[i])$ \tcc{Here we find URD for each VM.} \label{lst:line:get_erd}
		$size_{urd}[i]=calculateURDbasedSize(URD[i])$ \tcc{Here we calculate the estimated cache size for each VM based on its URD.} \label{lst:line:calc_eff}
		$csum+=size_{urd}[i]$ \tcc{We keep the sum of estimated sizes in $csum$.} \label{lst:line:csum_d}
	}
	\tcc{In the following we check the feasibility of size estimation and minimize overall latency for estimated $size_{urd}[1..N]$}
	\If{csum $\leq C$}
	{\label{lst:line:elseif1_d}
		\tcc{If this condition is met, our estimation is feasible}
		$c_{eff}[1..N]=size_{urd}[1..N]$ \tcc{We assign the estimated sizes to the efficient sizes.} 
	}\ElseIf{csum $> C$} 
	{\label{lst:line:if1_d}
		\tcc{If this condition is met, the estimation is infeasible.}
		Update hit ratio function of VMs ($H_i(c)$) based on updated reuse distance of the workloads. \tcc{The structure of hit ratio function is provided in Algorithm \ref{alg:sample_hit_ratio}.}
		$c_{eff}[1..N]=calculateEffSize(size_{urd}[1..N], C)$ \tcc{We call $calculateEffSize$ to find the efficient sizes that fit in total SSD cache space.} \label{lst:line:ceff_utility}
	}
	$allocate(c_{eff}[1..N], VM[1..N])$ \tcc{This function allocates the calculated efficient cache spaces for each VM.} \label{lst:line:eff_allocate}
		\tcc{\newline \textbf{Function Declarations:}\newline calculateURD}
	\SetKwProg{Fn}{Function}{ is}{end}
	\Fn{$calculateURD$($VM$)}
	{
		\tcc{The purpose of this function is to find the URD‌ of the running workload on $VM$. This function calls PARDA \cite{parda}, which is modified to calculate URD (reuse distance only for RAR and RAW requests).}
		return $URD$\;
	}
		\tcc{\newline calculateURDbasedSize}
	\SetKwProg{Fn}{Function}{ is}{end}
	\Fn{$calculateURDbasedSize$($URD$)}
	{
		\tcc{The purpose of this function is to calculate URD based cache size of each VM.}
		$size_{urd} = URD \times cacheBlkSize$\;
		return $size_{urd}$\;
	}
		\tcc{\newline calculateEffSize}
	\SetKwProg{Fn}{Function}{ is}{end}
	\Fn{$calculateEffSize$($size_{urd}[1..N], C$)}
	{
		\tcc{This function is called in infeasible states and aims to minimize the overall latency (sum of VMs latencies).}
		$initialSize = \{c_{min}, ..., c_{min}\} $\tcc{We set $c_{min}$ as the initial cache space for each VM.} 
		$lowerBound = \{ c_{min}, ..., c_{min}\}$ \tcc{Here we set the minimum cache space for each VM equal to $c_{min}$ }
		$upperBound = \{ size_{urd}[1], ..., size_{urd}[N]\}$ \tcc{Here the maximum cache space for each VM is set.}
		$weightVM = \{1, ..., 1\}$ \tcc{We assume that the VMs are weighted identically.}
		\tcc{All abovementioned variables are the inputs of the $fmincon$ function. We pass $ObjectiveFunction$ to this function.} 
		$c_{eff}[1..N] = fmincon(ObjectiveFunction, initialSize, weightVM, C_{tot}, \{\}, \{\} , lowerBound, upperBound)$\;\label{lst:line:fmincon_d}
		return $c_{eff}[1..N]$\;
	}
			\tcc{\newline ObjectiveFunction}
	\SetKwProg{Fn}{Function}{ is}{end}
	\Fn{$ObjectiveFunction$()}
	{
		\tcc{This function is called until the condition of Eq. \ref{equ:conditions} is met by the estimated cache sizes ($c[1..N]$).}
		\For{$~i = 1$ to $N$}
		{
			$h[i]=H_i(c[i])$ \tcc{Here we calculate the hit ratio of each VM for $c[i]$ input. This function is updated in $\Delta t$ intervals.}
			$hsum+=h[i]$ \tcc{$hsum$ variable keeps the sum of hit ratios of the VMs}
			$csum+=c[i]$ \tcc{$csum$ variable keeps the sum of cache sizes of the VMs}
		}
		$diff = C - csum$ \tcc{$diff$ variable is used in maximizing the total estimated cache sizes}
		$Obj = diff + (hsum) \times T_{SSD}+(N - hsum) \times T_{HDD}$ \tcc{Objective: Maximizing estimated sizes and minimizing sum of VM latencies.}
		return $Obj$\;
	}
\end{algorithm*}

\end{appendices}

%
%
%
%

\end{document}